\documentclass[jhep,floats,superscriptaddress,floatfix,nofootinbib,preprintnumbers,notitlepage,eqsecnum]{revtex4-1}
\usepackage{amsmath}
\usepackage{amsfonts}
\usepackage{graphicx}
\usepackage{slashed}
\usepackage{dcolumn}
\usepackage{bm}
\usepackage{amssymb}
\usepackage{latexsym}
\usepackage{color}
\usepackage{url}

\newcommand{\pbrac}[1]{\left( #1 \right)}
\newcommand{\tbrac}[1]{\left[ #1 \right]}
\newcommand{\cbrac}[1]{\left\{ #1 \right\}}
\newcommand{\Tr}{\text{Tr}}

\newcommand{\bfrac}[2]{{\left(\frac{#1}{#2} \right)  }}

\def\Treh{T_{\rm reh}}

\begin{document}

\title{Classically Scale Invariant Inflation, Supermassive WIMPs, and Adimensional Gravity}

\author{Arsham Farzinnia}
\email[]{farzinnia@ibs.re.kr}
\affiliation{Center for Theoretical Physics of the Universe, Institute for Basic Science (IBS)\\Daejeon 34051, Republic of Korea}

\author{Seyen Kouwn}
\email[]{seyenkouwn@gmail.com}
\affiliation{Center for Theoretical Astronomy, Korea Astronomy and Space Science Institute (KASI)\\Daejeon 34055, Republic of Korea}

\preprint{CTPU-15-25}

\date{\today}

\begin{abstract}
We introduce a minimal and yet comprehensive framework with $CP$- and classical scale-symmetries, in order to simultaneously address the hierarchy problem, neutrino masses, dark matter, and inflation. One complex gauge singlet scalar and three flavors of the right-handed Majorana neutrinos are added to the standard model content, facilitating the see-saw mechanism, among others. An adimensional theory of gravity (Agravity) is employed, allowing for the trans-Planckian field excursions. The weak and Planck scales are induced by the Higgs portal and the scalar non-minimal couplings, respectively, once a Coleman-Weinberg dynamically-generated vacuum expectation value for the singlet scalar is obtained. All scales are free from any mutual quadratic destabilization. The $CP$-symmetry prevents a decay of the pseudoscalar singlet, rendering it a suitable WIMPzilla dark matter candidate with the correct observational relic abundance. Identifying the pseudo-Nambu-Goldstone boson of the (approximate) scale symmetry with the inflaton field, the model accommodates successful slow-roll inflation, compatible with the observational data. We reach the conclusion that a pseudo-Nambu-Goldstone inflaton, within a classically scale-symmetric framework, yields lighter WIMPzillas.
\end{abstract}

\maketitle

\section{Introduction}\label{intro}

Despite its undeniable success, the standard model (SM) leaves many questions unanswered within the modern study of particle physics and cosmology. The current treatment is devoted to simultaneously addressing a number of such open issues within a unified and consistent classically scale invariant framework. Specifically, an analysis of the slow-roll inflationary paradigm, dark matter, and the neutrino mass generation mechanism is presented, in which the various scales are free from a mutual quadratic sensitivity \cite{Bardeen}.

The model concerns the minimal addition of a complex gauge singlet to the SM content, in a scale- and $CP$-symmetric manner. It has been previously demonstrated \cite{Farzinnia:2013pga} that, within such an extended scalar sector, the Coleman-Weinberg mechanism \cite{Coleman:1973jx} may be successfully realized, while accommodating the discovered 125~GeV Higgs-like state \cite{LHCnew}, and, therefore, remedy the failure of the mechanism within the ordinary SM scalar sector. In particular, the dynamical generation of a nonzero vacuum expectation value (VEV) for the $CP$-even component of the complex singlet scalar via this mechanism can be transmitted to the electroweak sector via the Higgs portal operators, inducing the nonzero VEV for the SM Higgs boson, and, thereby, giving rise to a successful spontaneous breaking of the electroweak symmetry. The classical scale symmetry, subsequently, guarantees the absence of any quadratic sensitivity between the singlet and the weak scales. The $CP$-odd singlet component cannot decay due to the imposed $CP$-invariance, and constitutes a stable dark matter candidate. In addition, incorporating the see-saw mechanism \cite{seesaw} within the framework, by including three flavors of the right-handed Majorana neutrinos, yields nonzero masses for the SM neutrinos. The dark matter and collider phenomenology of the theory around the TeV scale were studied in \cite{Farzinnia:2014xia,Farzinnia:2015uma}, whereas the possibility for realizing a strongly first-order electroweak phase transition, important for the baryogenesis paradigm, was exhibited in \cite{Farzinnia:2014yqa}. Hence, this technically-natural (minimal and yet comprehensive) framework presents an extremely promising and economical route for theoretical and phenomenological exploration, from a model-building perspective.

In this analysis, we further investigate incorporating the slow-roll inflationary paradigm within the described framework, while additionally accounting for the gravitational effects by utilizing a renormalizable and scale symmetric theory of gravity, known as the Agravity \cite{Salvio:2014soa}. Within this fully scale- and $CP$-invariant framework, the dynamically-generated nonzero VEV of the $CP$-even singlet scalar induces the Planck scale via the scalar non-minimal couplings, in addition to the aforementioned generation of the weak scale via the Higgs portal operators. These induced scales are, then, shown to be free from quadratic divergences, and thus stable, as a consequence of the scale symmetry. The availability of a renormalizable theory of gravity, consistent with the symmetries of the framework, allows for the proper trans-Planckian field excursions, where the stability of the vacuum and the perturbativity of the couplings are satisfied to large trans-Planckian energies. Identifying the pseudo-Nambu-Goldstone boson of the (approximate) scale symmetry with the inflaton field, we explore the viability of the inflationary paradigm according to the latest observational values by the Planck Collaboration \cite{Ade:2015lrj},\footnote{Studies of the classically scale-invariant inflation, in various contexts, are available within the literature. See, e.g. \cite{SIinfl,Kannike:2015apa}.} and the reheating of the Universe due to the decay of the inflaton. Moreover, we demonstrate that the aforementioned pseudoscalar dark matter candidate constitutes a WIMPzilla \cite{Chung:1998rq}, satisfying the observed relic abundance \cite{Ade:2015lrj}. Hence, several important and pressing issues, faced by the contemporary physics, are captured within a single consistent framework.

In Sec.~\ref{formalism}, we present the overall theoretical setup of the minimal scale- and $CP$-invariant framework, introduce the employed notation, and determine the free input parameters. In Sec.~\ref{VST}, we proceed to check the consistency of the theory, by demanding the stability of the vacuum and the perturbativity of the couplings up to an energy scale of at least $250\bar{M}_\text{P}$ (all phenomena of interest, studied in the current treatment, occur below this energy range). We report all the relevant one-loop $\beta$-functions, solve the renormalization group equations, and determine the viable regions of the parameters space accommodating these requirements. Sec.~\ref{Infl} is devoted to the slow-roll inflation analysis, where the observational constraints on the model's free parameters are derived using two separate methods. The reheating of the Universe due to the decay of the inflaton, the WIMPzilla nature of the dark matter candidate, as well as its compatibility with the observed relic abundance, are exhibited in Sec.~\ref{DM}, and the results are displayed in various exclusion plots. In Sec.~\ref{Disc}, we combine the findings from all previous considerations within two benchmark exclusion plots, demonstrating that a pseudo-Nambu-Goldstone inflaton within a classically scale-symmetric framework results in lighter WIMPzillas. Finally, we conclude the study in Sec.~\ref{Concl}, providing a short summary.

\section{Formalism}\label{formalism}

We start by presenting the formalism of the theory and defining our notations. The model is characterized by adding one complex scalar to the SM content, which is a singlet under the full SM gauge group. The resulting scalar sector, containing the SM Higgs doublet $H$ and the additional complex singlet $S$, is assumed to be scale invariant as well as $CP$-symmetric. Moreover, three flavors of the gauge-singlet right-handed Majorana neutrinos are included to facilitate the seesaw mechanism \cite{seesaw}, by which nonzero masses for the ordinary SM neutrinos are generated. The right-handed Majorana neutrinos are conjectured to have $CP$-invariant Yukawa interactions with the complex singlet scalar, from which they obtain their masses. In order to account for the gravitational effects, the described model is, subsequently, embedded within a scale-symmetric renormalizable framework of gravity, the \textit{Adimensional gravity (Agravity)} \cite{Salvio:2014soa}, in which all scalars couple non-minimally to the curvature. Interestingly, the $CP$-symmetry of the action protects the pseudoscalar component of the complex singlet from decaying \cite{Farzinnia:2013pga}, rendering it a stable dark matter candidate.

\subsection{Action in the Jordan Frame and the Einstein Frame}

The Jordan frame action of the theory, invariant under the scale- and $CP$-symmetry, is defined according to
\begin{equation}\label{LJ}
\begin{split}
\sqrt{|\det g|}\, \mathcal L^{J}= \sqrt{|\det g|} \,\bigg\{& |D_\mu H|^2 + |\partial_\mu S|^2 - V^{(0)}(H,S) - \xi_H (H^\dagger H) R - \xi_S^1 |S|^2 R - \frac{1}{2} \xi_S^2 (S^2+S^{*2}) R \\
& + \frac{R^2}{6 f_0^2} + \frac{\frac{1}{3}R^2 - R_{\mu\nu}^2}{f_2^2} + \mathcal{L}_{\mathcal N} + \mathcal{L}_{\text{SM}}^{\text{rest}} \bigg\} \ ,
\end{split}
\end{equation}
where, $\mathcal{L}_{\text{SM}}^{\text{rest}}$ represents the (unaltered) SM gauge and fermionic sectors, while $H$ and $S$ contain the components
\begin{equation}\label{HS}
H= \frac{1}{\sqrt{2}}
\begin{bmatrix} \sqrt{2}\,\pi^+ \\ \pbrac{v_\phi+\phi}+i\pi^0 \end{bmatrix} \ , \qquad S =\frac{1}{\sqrt 2} \tbrac{\pbrac{v_\eta + \eta} + i\chi} \ .
\end{equation}
The nonzero VEVs of the $CP$-even scalars, $\phi$ and $\eta$, are generated dynamically via the Coleman-Weinberg mechanism \cite{Coleman:1973jx},\footnote{We assume that only the $CP$-even component of the complex singlet obtains a nonzero VEV; thereby, avoiding the possibility for a spontaneous violation of the $CP$-symmetry \cite{AlexanderNunneley:2010nw}.} and $\xi_{H}$ ($\xi_{S}^{1,2}$) in \eqref{LJ} denotes the non-minimal coupling of the $H$ ($S$)~scalar to the curvature (note that the singlet $S$ can couple non-minimally to the curvature via two $CP$-invariant operators).

The most general form of the scalar potential, satisfying the scale- and $CP$-symmetry, is composed of the following six operators \cite{Farzinnia:2013pga}
\begin{equation}\label{V0}
V^{(0)}(H,S) = \frac{\lambda_1}{6} \pbrac{H^\dagger H}^2 + \frac{\lambda_2}{6} |S|^4 + \lambda_3 \pbrac{H^\dagger H}|S|^2 + \frac{\lambda_4}{2} \pbrac{H^\dagger H}\pbrac{S^2 + S^{*2}} + \frac{\lambda_5}{12} \pbrac{S^2 + S^{*2}} |S|^2 + \frac{\lambda_6}{12} \pbrac{S^4 + S^{*4}} \ ,
\end{equation}
with all couplings being real and dimensionless. One observes that the ``Higgs portal'' terms, with the $\lambda_{3,4}$~coefficients, dynamically induce a mass term for the SM Higgs field, $\mu_{H}^{2}$, once the singlet obtains a nonzero VEV, $v_{\eta}$ (c.f. \eqref{HS}), by the Coleman-Weinberg mechanism,
\begin{equation}\label{muSM}
\mu_H^{2} = \frac{\lambda_{3}+\lambda_{4}}{2}\, v_{\eta}^{2} \equiv \frac{\lambda_{m}^{+}}{2}\, v_{\eta}^{2} \ .
\end{equation}
As we will show below, the $\lambda_{m}^{+}$~coupling (c.f. \eqref{couprel}) turns out to be negative; therefore, the spontaneous breaking of the electroweak symmetry is successfully triggered. Defining for convenience,
\begin{align}&\lambda_\phi \equiv \lambda_{1} \ , \quad \lambda_\eta \equiv \lambda_{2} + \lambda_{5} + \lambda_{6} \ , \quad \lambda_\chi \equiv \lambda_{2} - \lambda_{5} + \lambda_{6} \ , \quad \lambda_{\eta \chi} \equiv \frac{1}{3}\lambda_{2} - \lambda_{6} \ , \quad \lambda_m^\pm \equiv \lambda_{3} \pm \lambda_{4} \ , \label{couprel}\\
& \xi_\eta \equiv \xi_S^1 + \xi_S^2 \ , \quad \xi_\chi \equiv \xi_S^1 - \xi_S^2 \ , \label{nonminrel}
\end{align}
one may write explicitly for the interactions of the components (omitting the nonzero VEVs):
\begin{align}
& V^{(0)}_{\text{quartic}} = \frac{1}{24} \tbrac{ \lambda_\phi \,\phi^4 + \lambda_\eta \,\eta^4 + \lambda_\chi \,\chi^4 + \lambda_\phi\pbrac{\pi^0\pi^0 + 2 \pi^+ \pi^-}^2}+ \frac{1}{4}\tbrac{\lambda_m^+ \,\phi^2 \eta^2 + \lambda_m^- \,\phi^2 \chi^2 +\lambda_{\eta \chi} \,\eta^2 \chi^2} \notag
\\
&\qquad\qquad + \frac{1}{12} \tbrac{\lambda_{\phi} \,\phi^2 + 3\lambda_m^+\, \eta^2 + 3\lambda_m^- \,\chi^2}\pbrac{\pi^0\pi^0 + 2 \pi^+ \pi^-}\ , \label{V0quart}\\
&\xi_S^1 |S|^2 R + \frac{1}{2}\xi_S^2 (S^2+S^{*2}) R = \frac{1}{2} \xi_\eta\, \eta^2 R + \frac{1}{2} \xi_\chi\, \chi^2 R \ . \label{nonmincoupl}
\end{align}
Note that, according to the relation \eqref{nonmincoupl}, in general both the singlet scalar and pseudoscalar non-minimally couple to the curvature. Furthermore, the $CP$-symmetry of the Lagrangian \eqref{LJ} prohibits a decay of the $CP$-odd pseudoscalar, $\chi$, rendering it a stable dark matter candidate. The tree-level potential \eqref{V0} is bounded from below once the following conditions are satisfied \cite{Farzinnia:2014xia}
\begin{subequations}
\begin{align}
&
\lambda_\phi^{} > 0 \ , \qquad \lambda_\eta^{} > 0 \ , \qquad
\lambda_\chi^{} > 0 \,, \qquad
\lambda_{\eta\chi}^{} > -\frac{1}{3}\!\sqrt{\lambda_{\eta}^{}\lambda_{\chi}^{}} \ , \qquad
\lambda_m^+ > -\frac{1}{3} \sqrt{\lambda_\phi \lambda_\eta} \ , \qquad
\lambda_m^- > -\frac{1}{3} \sqrt{\lambda_\phi \lambda_\chi} \ ,
\label{stabtree1}
\\[2mm]
&
\lambda_{\eta\chi} \sqrt {\lambda_\phi} + \lambda_m^+ \sqrt{\lambda_{\chi}} + \lambda_m^- \sqrt{\lambda_{\eta}} > -\frac{1}{3} \tbrac{\sqrt{\lambda_{\phi} \lambda_{\eta} \lambda_{\chi}} + \sqrt{2\pbrac{3\lambda_{\eta\chi}+\sqrt{\lambda_{\eta} \lambda_{\chi}}}\pbrac{3\lambda_{m}^{+}+\sqrt{\lambda_{\phi} \lambda_{\eta}}}\pbrac{3\lambda_{m}^{-}+\sqrt{\lambda_{\phi} \lambda_{\chi}}}}} \ .
\label{stabtree2}
\end{align}
\end{subequations}

As mentioned, the three right-handed Majorana neutrinos are assumed to possess $CP$-invariant Yukawa interactions with the complex singlet scalar. Consequently, one obtains for the Lagrangian of this sector
\begin{equation}\label{LRHN}
\mathcal{L}_{\mathcal N} = \text{kin.} - \tbrac{Y^\nu_{ij}\, \bar{L}_{\ell}^{i} \tilde{H} \mathcal{N}^{j} + \text{h.c.}} -\frac{1}{2}y_{N} \pbrac{S + S^*} \bar{\mathcal{N}}^{i}\mathcal{N}^{i} \ ,
\end{equation}
where, the real Yukawa coupling $y_{N}$ is designated as flavor universal for simplicity ($i = 1,2,3$). In this Lagrangian, $\mathcal{N}_{i} = \mathcal{N}_{i}^{c}$ denotes the 4-component right-handed Majorana neutrino spinor, $\tilde{H} \equiv i \sigma^2 H^*$ (with $\sigma^2$ the second Pauli matrix), and $L_{\ell}^{i}$ represents the left-handed lepton doublet. In the following, one may ignore the complex Dirac neutrino Yukawa matrix, $Y^\nu_{ij}$, since its entities are negligibly small (of the order of the electron Yukawa coupling). According to \eqref{LRHN}, the flavor universal Yukawa coupling results in a (degenerate) mass term for the Majorana neutrino flavors, once the singlet obtains a dynamical nonzero VEV.

Finally, the pure scale-symmetric renormalizable gravity sector contains two (higher-derivative) operators with the dimensionless couplings $f_{0,2}^{2}$ (c.f. the second line in \eqref{LJ}). As discussed in \cite{Salvio:2014soa}, the operator parametrized by the $f_{0}^{2}$~coupling gives rise to a scalar graviton, whereas the operator with the $f_{2}^{2}$~coupling (a.k.a. the Weyl term) produces the usual massless spin-2 graviton together with its ``Lee-Wick''~(LW) partner \cite{LW};\footnote{These two degrees of freedom correspond to the two poles of the propagator in the higher-derivative theory. Ghosts may lead to the violation of the unitarity due to their negative norm; however, it has been shown that unitarity can be preserved at the expense of violating causality at ultra-high energies \cite{acausal}. The renormalizability of LW theories has been studied in \cite{Chivukula:2010kx}. See also the discussion and the related references provided in \cite{Salvio:2014soa,Salvio:2015gsi}.} i.e., a massive spin-2 ghost degree of freedom with a negative norm. It is instructive to make the massive scalar degree of freedom, $\Omega$, explicitly manifest in the action \cite{Salvio:2014soa}, by adding the vanishing term $-\dfrac{\pbrac{R + \frac{3}{2} f_0^2\, \Omega}^2}{6 f_0^2}$ to \eqref{LJ}, which subsequently yields
\begin{equation}\label{LJOmega}
\sqrt{|\det g|}\, \mathcal L^{J}= \sqrt{|\det g|} \,\bigg\{|D_\mu H|^2 + |\partial_\mu S|^2 - V^{(0)}(H,S) - \frac{3}{8} f_0^2\, \Omega^2 - \frac{F}{2}\, R + \frac{\frac{1}{3}R^2 - R_{\mu\nu}^2}{f_2^2} + \mathcal{L}_{\mathcal N} + \mathcal{L}_{\text{SM}}^{\text{rest}} \bigg\} \ , 
\end{equation}
where, we have defined
\begin{equation}\label{F}
F\equiv \tbrac{2\xi_H (H^\dagger H) + 2\xi_S^1 |S|^2 + \xi_S^2 (S^2+S^{*2}) + \Omega} \ .
\end{equation}
It is worth noting that in the Jordan frame Lagrangian \eqref{LJOmega}, the $\Omega$~scalar lacks a kinetic term. Furthermore, while the scale symmetry forbids the explicit presence of the Einstein-Hilbert term, $-\dfrac{\bar{M}_\text{P}^2}{2} R$, in the Lagrangian, one observes from \eqref{LJOmega} and \eqref{F} that the reduced Planck scale, $\bar{M}_\text{P} \equiv M_\text{P} / \sqrt{8\pi}$, is induced via the non-minimal interactions, once a nonzero scalar VEV, $v_{F}$ is obtained,
\begin{equation}\label{Mpl}
\bar{M}_\text{P}^2 = v_{F} \ .
\end{equation}

In order to explicitly reveal the canonical Einstein-Hilbert term, one may transform the Jordan frame scale- and $CP$-invariant action \eqref{LJOmega} into the Einstein frame, using the local Weyl transformation
\begin{equation}\label{Weyl}
g_{\mu\nu}^E \equiv \frac{F}{\bar{M}_\text{P}^2} \, g_{\mu\nu} \ , \qquad
\Phi^E \equiv \pbrac{\frac{\bar{M}_\text{P}^2}{F}}^{1/2} \Phi \ , \qquad \psi^E \equiv \pbrac{\frac{\bar{M}_\text{P}^2}{F}}^{3/4} \psi \ , \qquad A_\mu^E \equiv A_\mu \ ,
\end{equation}
where, $\Phi$, $\psi$, and $A_{\mu}$ correspond, respectively, to the real scalar, fermionic, and gauge vector degrees of freedom present within the Jordan frame Lagrangian. Under the local conformal transformation \eqref{Weyl}, all the non-derivative interactions, as well as the fermion and vector gauge boson kinetic terms remain invariant. The scalar kinetic terms are, in general, not invariant under such a transformation and additionally generate a kinetic term for the $F$~state \cite{Kaiser:2010ps}, which makes the latter a proper dynamical field in the Einstein frame. The Einstein frame action with the canonical Einstein-Hilbert term manifestly present, hence, takes the form
\begin{equation}\label{LE}
\sqrt{|\det g^E|}\, \mathcal L^E = \sqrt{|\det g^E|} \,\bigg\{ \mathcal L_\Phi^\text{kin} - V^{(0)E} -\frac{\bar{M}_\text{P}^2}{2} \, R^E + \frac{\frac{1}{3}(R^E)^{2} - (R_{\mu\nu}^{E})^{2}}{f_2^2} + \mathcal{L}^E_{\mathcal N} + \mathcal{L}_{\text{SM}}^{\text{rest}, \,E} \bigg\} \ ,
\end{equation}
where, the scalar sector is given by
\begin{equation}\label{scalarE}
\mathcal L_\Phi^\text{kin} \equiv \frac{\bar{M}_\text{P}^2}{F} \cbrac{|D_\mu H|^2 + |\partial_\mu S|^2 +\frac{3}{4F} (\partial_\mu F)^2} \ , \qquad V^{(0)E} \equiv \frac{\bar{M}_\text{P}^4}{F^2} \cbrac{ V^{(0)}(H,S) + \frac{3}{8} f_0^2 \, \Omega^2} \ .
\end{equation}
Defining the ``conformal" form of the $F$~scalar, according to
\begin{equation}\label{zeta}
F \equiv \frac{\zeta^2}{6} \ ,
\end{equation}
and using \eqref{F}, one deduces for the scalar kinetic and potential terms \eqref{scalarE} in the Einstein frame
\begin{align}
\mathcal L_\Phi^\text{kin} \equiv&\, \frac{6\bar{M}_\text{P}^2}{\zeta^2} \cbrac{|D_\mu H|^2 + |\partial_\mu S|^2 +\frac{1}{2} (\partial_\mu \zeta)^2} \ , \label{Lkinphi} \\
V^{(0)E}(H,S,\zeta) \equiv&\, \frac{36\bar{M}_\text{P}^4}{\zeta^4} \cbrac{ V^{(0)}(H,S) + \frac{3}{8} f_0^2 \tbrac{\frac{\zeta^2}{6} - 2\xi_H (H^\dagger H) - 2\xi_S^1 |S|^2 - \xi_S^2 (S^2+S^{*2})}^2} \label{VE} \ .
\end{align}

The presented framework, thus, contains three scalars which obtain nonzero VEVs; namely, $\phi$ and $\eta$ (c.f. \eqref{HS}), and $\zeta$. Using \eqref{Mpl} and \eqref{zeta}, one simply arrives at
\begin{equation}\label{vz}
v_{\zeta}^{2} = 6\bar{M}_\text{P}^2 \ .
\end{equation}
The Einstein-frame tree-level potential \eqref{VE} is bounded from below for the same conditions \eqref{stabtree1} and \eqref{stabtree2}.\footnote{We assume $f_{0}^{2} \geq 0$, which would otherwise lead to a tachyonic scalar graviton \cite{Salvio:2014soa}.} The scalar kinetic terms \eqref{Lkinphi} are, in general, non-canonical in the Einstein frame; however, as we will demonstrate, there is a particular flat direction in the field space of the potential, along which it is possible to define a single canonical scalar degree of freedom, which will additionally serve as the inflaton of the theory.

\subsection{Perturbative Minimization of the Potential and the Tree-level Mass Eigenstates}

In principle, in order to identify the true (dynamically-induced) vacuum of the system, one should compute and minimize the full one-loop effective potential, which may not be possible to perform analytically. In contrast, the perturbative minimization procedure, developed by Gildener and Weinberg \cite{Gildener:1976ih}, provides a convenient and economical method to achieve this goal. The perturbative minimization occurs in two steps, where initially only the tree-level potential is minimized with respect to its field content. This tree-level minimization happens at a definite energy scale, as a consequence of the running of the couplings as a function of the energy in the full quantum theory, and defines a flat direction among the scalar fields. Subsequently, one computes the one-loop corrections only along this flat direction, where they play the dominant role, remove the flatness, and specify the physical vacuum.

Since, in the Einstein frame, all of the scalar fields are manifestly dynamical, possessing (non-canonical) kinetic terms \eqref{Lkinphi}, we carry out the perturbative minimization of the potential in this frame using \eqref{VE}. Specifically, we demand
\begin{equation}\label{mins}
\frac{d \,V^{(0)E}}{d H} \Big|_{\phi = v_\phi} = \frac{d \,V^{(0)E}}{d S} \Big|_{\eta = v_\eta} = \frac{d \,V^{(0)E}}{d \zeta} \Big|_{\zeta = v_\zeta} = 0 \ .
\end{equation}
The conditions \eqref{mins} establish the flat direction of the tree-level potential, along which one derives the relations
\begin{align}
\lambda_m^+  (\Lambda)&= -\frac{\lambda_\phi (\Lambda)}{3}\, \frac{v_\phi^2}{v_\eta^2} + 3f_{0}^{2} (\Lambda) \, \xi_{H}(\Lambda)\, \frac{\bar{M}_\text{P}^{2} - \pbrac{\xi_{H}(\Lambda) v_{\phi}^{2} + \xi_{\eta}(\Lambda) v_{\eta}^{2}}}{v_\eta^2} \ , \label{lmp} \\
\lambda_\eta  (\Lambda)&= \lambda_\phi (\Lambda)\, \frac{v_\phi^4}{v_\eta^4} -9 f_{0}^{2} (\Lambda) \, \frac{\tbrac{\bar{M}_\text{P}^{2} - \pbrac{\xi_{H}(\Lambda) v_{\phi}^{2} + \xi_{\eta}(\Lambda) v_{\eta}^{2}}} \pbrac{\xi_{H}(\Lambda) v_{\phi}^{2} - \xi_{\eta}(\Lambda) v_{\eta}^{2}}}{v_\eta^4}  \ , \label{leta}
\end{align}
where, $v_{\zeta}$ has been eliminated using \eqref{vz}, and $\Lambda$ denotes the aforementioned minimization energy scale. These relations demonstrate instances of the dimensional transmutation phenomenon, where dimensionful quantities may be generated from the (combinations of) dimensionless couplings. The flat direction of the tree-level potential is identified by inserting the relations \eqref{lmp} and \eqref{leta} into \eqref{VE}, while keeping in mind the definite energy scale $\Lambda$ at which the flat direction is defined.

At this point, several observations are in order:
\begin{itemize}
\item The first term on the right-hand side of \eqref{lmp} exhibits the technical naturalness of the theory. In particular, it demonstrates that the weak scale is not destabilized by the quadratic contributions from the singlet scale, since the magnitude of the mixing between the two scales ($\lambda_{m}^{+}$) is proportional to their ratio; i.e., $\lambda_{m}^{+} v_{\eta}^{2} \sim v_{\phi}^{2}$ as a direct consequence of the scale symmetry.\footnote{We emphasize that no such relation exists within a general singlet-extended framework possessing no protective symmetry, where both the singlet VEV and the magnitude of its mixing with the electroweak sector appear as \textit{independent} parameters. In this case, a fine-tuning between these two free parameters is required, in order to prevent a quadratic destabilization of the weak scale due to the contributions from the singlet scale.} In addition, this terms is negative, resulting in the successful spontaneous electroweak symmetry breaking (c.f. \eqref{muSM}).
\item The second term on the right-hand side of \eqref{lmp} and \eqref{leta}, on the other hand, represents an additive correction, which is \textit{a priori} not under control by the scale symmetry, and may potentially jeopardize the technical naturalness of the theory. We shall demonstrate below (c.f. \eqref{V01min} and \eqref{Mplfin}) that this term is related to the cosmological constant problem, which is not explicitly incorporated within the current scale-invariant framework and remains an unresolved issue. Since the cosmological constant is currently not protected by the introduced scale symmetry, we have to ignore its additive contribution on the right-hand sides of \eqref{lmp} and \eqref{leta} throughout the remaining of this discussion.\footnote{It will be exhibited, in the forthcoming subsections, that these mentioned second terms are radiatively generated; nevertheless, they need not to be small as compared with the corresponding first terms, since their origin---the cosmological constant---is not protected by the scale symmetry within the current framework. A proper scale symmetric study of the cosmological constant (and more generally the dark energy) problem is outside the scope of the current treatment and will be presented elsewhere.}
\item The above notion of the technical naturalness is inherently based on the tree-level relation \eqref{lmp} and \eqref{leta} obtained at a particular energy scale $\Lambda$; therefore, one may (rightfully) wonder about the effects of the quantum corrections. These effects are, nonetheless, captured by the running of the relevant couplings as a function of the renormalization scale, which is only logarithmic in nature. The quadratic divergences, therefore, remain absent even at the quantum level, and the technical naturalness of the framework prevails.
\end{itemize}

The three scalars $\phi$, $\eta$, and $\zeta$ with nonzero VEVs exhibit quadratic mixings, manifest as off-diagonal terms in their mass matrix. These scalar masses may be diagonalized by a three-dimensional orthogonal rotation,\footnote{An example of a classically scale-invariant multi-Higgs portal model has been studied in \cite{Karam:2015jta}.} parametrized by the three angles $\omega_1$, $\omega_2$, and $\omega_3$, which is of the form
\begin{equation}\label{scaldiag}
\begin{pmatrix} \phi\\ \eta\\ \zeta \end{pmatrix}
= \mathcal R \begin{pmatrix} h \\ \sigma \\ \kappa \end{pmatrix} \ , \qquad \mathcal R = \begin{pmatrix} c_1c_3+s_1s_2s_3 & s_1c_2 & c_1s_3-s_1s_2c_3 \\ -s_1c_3+c_1s_2s_3 & c_1c_2 & -s_1s_3-c_1s_2c_3 \\ -c_2s_3 & s_2 & c_2c_3 \end{pmatrix} \ ,
\end{equation}
with $c_i \equiv \cos \omega_i$ and $s_i \equiv \sin \omega_i$ ($i=1,2,3$). The $h$, $\sigma$, and $\kappa$~degrees of freedom are the corresponding physical scalars in the mass basis. Note that for each pair of the angles being equal to zero (i.e. $\omega_i= \omega_j = 0$ for $i\neq j$), the orthogonal matrix $\mathcal R$ in \eqref{scaldiag} reduces to the ordinary two-dimensional rotation about the remaining axis.

Within the scale-symmetric framework, the mass matrix of the $\phi$, $\eta$, and $\zeta$~scalars possesses as one of its eigenvectors the direction along the (dynamically-generated) nonzero VEVs
\begin{equation}\label{eigenvec}
\frac{1}{\sqrt{v_\phi^2 + v_\eta^2+v_\zeta^2}} \begin{pmatrix} v_\phi\\ v_\eta\\ v_\zeta \end{pmatrix} \ .
\end{equation}
This direction corresponds to the pseudo-Nambu-Goldstone boson of the (approximate) scale symmetry, which is identified with the $\sigma$~boson in \eqref{scaldiag}. Hence, the eigenvector \eqref{eigenvec} may be inserted as the second column of the rotation matrix $\mathcal R$ in \eqref{scaldiag}, yielding
\begin{equation}\label{t1t2}
\tan\omega_1 = \frac{v_\phi}{v_\eta} \ , \qquad \tan\omega_2 = \frac{v_\zeta}{\sqrt{v_\phi^2 +v_\eta^2}} \ ,
\end{equation}
 along the flat direction. Inserting the relations \eqref{t1t2} into the scalar mass matrix and demanding all off-diagonal terms to be zero, we find for the remaining $\omega_{3}$~angle\footnote{In constructing \eqref{t3} and the mass expressions \eqref{treemass1} and \eqref{treemass2}, we have eliminated $\xi_{\eta}$ using \eqref{xieta} to illuminate the relations between the masses and their corresponding couplings. See also the discussion below \eqref{msig} for additional details.}
\begin{equation}\label{t3}
\begin{split}
&\tan\omega_3 = \mathcal C - \sqrt{1+\mathcal{C}^{2}} \ , \\
&\mathcal C \equiv -\frac{1}{2\tan\omega_1\sin\omega_2\tbrac{1 - 6 \xi_H \cot^2\omega_2}} + \frac{1}{2} \tan\omega_1\sin\omega_2\tbrac{1 - 6 \xi_H \cot^2\omega_2} + \frac{2\lambda_\phi}{f_0^2} \frac{\tan\omega_1\cos\omega_2\cot^3\omega_2}{1 - 6 \xi_H \cot^2\omega_2} \ .
\end{split}
\end{equation}
It is worth noting that in the limit of a singlet VEV much larger than the weak scale, $v_{\eta} \gg v_{\phi}$, one obtains $\tan \omega_1 \to 0$ and $\tan \omega_3 \to 0$. This, in turn, implies a negligible mixing of the electroweak sector with the singlet and the gravity sectors, as required by the technical naturalness of the theory as a consequence of the scale symmetry.

With the three mixing angles determined in \eqref{t1t2} and \eqref{t3}, the following diagonalized tree-level masses are obtained
\begin{equation}\label{treemass1}
\begin{split}
M_h^2 =&\, \frac{\lambda_\phi \, v_\phi^2}{3\cos^{2}\omega_1} \frac{1}{1+\tan\omega_1\sin\omega_2\tan\omega_3\tbrac{1-6\xi_H \cot^2\omega_2}} +\dots \ , \\
M_\kappa^2 =&\, \frac{f_0^2 \, \bar{M}_\text{P}^2}{2\cos^2\omega_2} \pbrac{1+\tan\omega_1 \sin\omega_2 \tan\omega_3\tbrac{1-6\xi_H \cot^2\omega_2}} +\dots \ , \\
M_\sigma^2 =&\, M_{\pi^{0,\pm}}^2 = 0 \ ,
\end{split}
\end{equation}
where, the ellipses represent the (ignored) contributions arising from the second term on the right-hand sides of \eqref{lmp} and \eqref{leta}. For convenience, we have expressed the mass of the $h$~scalar in terms of the electroweak parameters; whereas, the scalar graviton mass, $M_{\kappa}$, is expressed in terms of the reduced Planck scale. In this framework, we identify the $h$~scalar with the discovered Higgs-like boson at the LHC \cite{LHCnew}; i.e., $M_{h} = 125$~GeV. The pseudo-Nambu-Goldstone boson of the (approximate) scale symmetry, $\sigma$, is massless at tree-level; however, it obtains a radiatively-generated mass at one-loop (c.f.~\eqref{msig}). In contrast, the electroweak Nambu-Goldstone bosons, $\pi^{0,\pm}$, remain massless to all orders in perturbation theory. The remaining degrees of freedom---the dark matter pseudoscalar $\chi$, the right-handed Majorana neutrinos $\mathcal N$, and the LW~graviton $\theta$---possess diagonal mass terms. Expressing these, for convenience, in terms of the reduced Planck scale, we subsequently arrive at
\begin{equation}\label{treemass2}
\begin{split}
M_\chi^2 =&\, 3 \bar{M}_\text{P}^2 \cot^2\omega_2 \pbrac{\lambda_m^- \sin^2\omega_1 + \lambda_{\eta\chi}\cos^2 \omega_1} +\dots \ , \\
M_{\mathcal N} =&\, \sqrt2 \,y_{N} v_\eta = 2\sqrt3 \,y_{N} \bar{M}_\text{P} \cos\omega_1 \cot\omega_2 +\dots \ , \\
M_{\theta}^{2} = &\frac{1}{2} f_{2}^{2} \bar{M}_\text{P}^2 \ ,
\end{split}
\end{equation}
where, the ellipses represent, once more, the (ignored) contributions from the final term on the right-hand sides of \eqref{lmp} and \eqref{leta}.

\subsection{One-loop Effective Scalar Potential Along the Flat Direction}

As discussed in the previous subsection, the flat direction of the tree-level potential is characterized by the radial combination of the scalar fields with nonzero VEVs \eqref{eigenvec}, corresponding to the direction of the pseudo-Nambu-Goldstone boson of the (approximate) scale-symmetry, $\sigma$,
\begin{equation}\label{sigma}
\sigma^{2} = \phi^{2} + \eta^{2} + \zeta^{2} \ .
\end{equation}
On the one hand, using the relations \eqref{t1t2} between the fields along the flat direction, it is instructive to deduce from \eqref{sigma} the following equalities
\begin{equation}\label{sigmarel1}
\sigma = \frac{\phi}{\sin\omega_1\cos\omega_2} = \frac{\eta}{\cos\omega_1\cos\omega_2} = \frac{\zeta}{\sin\omega_2} \ .
\end{equation}
On the other hand, the $\sigma$~boson is expressed in terms of the field components according to the orthogonal rotation \eqref{scaldiag}
\begin{equation}\label{sigmarel2}
\sigma = \phi \, \sin\omega_1\cos\omega_2 + \eta \,\cos\omega_1\cos\omega_2 + \zeta \, \sin\omega_2 \ .
\end{equation}

The relation \eqref{sigma} (or equivalently \eqref{eigenvec}) implies that the kinetic term of the $\sigma$~scalar is non-canonical by virtue of \eqref{Lkinphi}, taking the form
\begin{equation}\label{skin}
\frac{6\bar{M}_\text{P}^2}{\zeta^2} \tbrac{\frac{1}{2} (\partial_\mu \sigma)^2} \ .
\end{equation}
Nevertheless, since this scalar constitutes the sole relevant degree of freedom along the flat direction, it's kinetic term can be brought into the canonical form by means of a field redefinition. Substituting $\zeta$ in \eqref{skin} by $\sigma$ using \eqref{sigmarel1}, one can perform the integration
\begin{equation}\label{sc}
\sigma_{c} - v_{\sigma_{c}} = \int_{v_{\sigma}}^{\sigma} \frac{\sqrt6 \bar{M}_\text{P}}{\sin\omega_2} \frac{d\sigma'}{\sigma'} = \frac{\sqrt6 \bar{M}_\text{P}}{\sin\omega_2} \log \frac{\sigma}{v_\sigma} \ ,
\end{equation}
where, $\sigma_{c}$ is the corresponding scalar with the canonical kinetic term, $\frac{1}{2} (\partial_\mu \sigma_{c})^2$.

The self-interactions of the $\sigma$~scalar (or equivalently its canonical redefinition $\sigma_{c}$) vanish in the tree-level potential \eqref{VE} with the flat direction conditions \eqref{lmp} and \eqref{leta} implemented. Hence, the leading-order self-interactions of this scalar along the flat direction arise at one-loop. According to \cite{Coleman:1973jx}, the one-loop corrections along the flat direction for the canonical field $\sigma_{c}$ at the scale $\Lambda$ may be written as
\begin{equation}\label{V1}
V^{(1)}(\sigma_{c}) = A\, \sigma_{c}^4 + B\, \sigma_{c}^4 \log \frac{\sigma_{c}^2}{\Lambda^2} \ ,
\end{equation}
with the coefficients $A$ and $B$ parametrizing the contributions from all the relevant degrees of freedom in the loop. Specifically, within the $\overline{\text{MS}}$~renormalization scheme,
\begin{equation}\label{AB}
\begin{split}
A &= \frac{1}{64\pi^{2}\, v_{\sigma_{c}}^{4}} \bigg\{\Tr \tbrac{M_{S}^{4}\pbrac{\log \frac{M_{S}^{2}}{v_{\sigma_{c}}^{2}} - \frac{3}{2}}} +3\Tr \tbrac{M_{V}^{4}\pbrac{\log \frac{M_{V}^{2}}{v_{\sigma_{c}}^{2}} - \frac{5}{6}}} -4\Tr \tbrac{M_{F}^{4}\pbrac{\log \frac{M_{F}^{2}}{v_{\sigma_{c}}^{2}} - 1}} \\
&\qquad \qquad\qquad + 5\Tr \tbrac{M_{T}^{4}\pbrac{\log \frac{M_{T}^{2}}{v_{\sigma_{c}}^{2}} - \frac{1}{4}}} \bigg\} \ , \\
B &=  \frac{1}{64\pi^{2}\, v_{\sigma_{c}}^{4}} \cbrac{\Tr M_{S}^{4} +3\, \Tr M_{V}^{4} -4\, \Tr M_{F}^{4} + 5\, \Tr M_{T}^{4}} \ ,
\end{split}
\end{equation}
with $M_{S}$, $M_{V}$, $M_{F}$, and $M_{T}$, the scalar, vector, fermion, and tensor masses in the loop, and the traces capturing the remaining internal degrees of freedom. In our model, we obtain for these coefficients
\begin{align}
A &= \frac{1}{64\pi^{2}\, v_{\sigma}^{4}} \bigg\{5 M_{\theta}^{4}\pbrac{\log \frac{M_{\theta}^{2}}{v_{\sigma}^{2}} - \frac{1}{4}} + M_{\kappa}^{4}\pbrac{\log \frac{M_{\kappa}^{2}}{v_{\sigma}^{2}} - \frac{3}{2}} + M_{\chi}^{4}\pbrac{\log \frac{M_{\chi}^{2}}{v_{\sigma}^{2}} - \frac{3}{2}} -6 M_{\mathcal N}^{4}\pbrac{\log \frac{M_{\mathcal N}^{2}}{v_{\sigma}^{2}} - 1} \label{Amodel} \\
&\qquad \qquad\qquad+ M_{h}^{4}\pbrac{\log \frac{M_{h}^{2}}{v_{\sigma}^{2}} - \frac{3}{2}} +6M_{W}^{4}\pbrac{\log \frac{M_{W}^{2}}{v_{\sigma}^{2}} - \frac{5}{6}} +3M_{Z}^{4}\pbrac{\log \frac{M_{Z}^{2}}{v_{\sigma}^{2}} - \frac{5}{6}} -12M_{t}^{4}\pbrac{\log \frac{M_{t}^{2}}{v_{\sigma}^{2}} - 1}  \bigg\} \ , \notag \\
B &= \frac{\mathcal M^{4}}{64\pi^{2}\, v_{\sigma}^{4}} \ , \qquad \mathcal M^{4}\equiv 5 M_{\theta}^{4}+ M_{\kappa}^{4}+ M_{\chi}^{4} -6 M_{\mathcal N}^{4}+ M_{h}^{4}+6M_{W}^{4} +3M_{Z}^{4} -12M_{t}^{4} \label{Bmodel}\ ,
\end{align}
where, the heavy SM degrees of freedom are included for completeness. In addition, since the kinetic term of the $\sigma$~boson becomes canonical at the minimum, we have set $v_{\sigma_{c}} = v_{\sigma}$.

The scale $\Lambda$ can be determined by minimizing \eqref{V1}, yielding
\begin{equation}\label{Lamb}
\frac{d V^{(1)}}{d \, \sigma_{c}} \Big |_{\sigma_{c} = v_{\sigma}} = 0 \qquad \Longrightarrow \qquad \Lambda = v_{\sigma} \exp\tbrac{\frac{A}{2B}+\frac{1}{4}} \ .
\end{equation}
Utilizing the expression for $\Lambda$, one can reconstruct the one-loop contribution \eqref{V1} entirely in terms of the $B$~coefficient \eqref{Bmodel},
\begin{equation}\label{V1fin}
V^{(1)}(\sigma_{c}) = \frac{\mathcal M^{4}}{64\pi^{2}\, v_{\sigma}^{4}} \, \sigma_{c}^4 \tbrac{\log \frac{\sigma_{c}^2}{v_{\sigma}^2} - \frac{1}{2}} \ .
\end{equation}
These one-loop corrections induce a radiatively-generated mass for the $\sigma$~scalar
\begin{equation}\label{msig}
m_\sigma^2 = \frac{d^{2} V^{(1)}}{d \, \sigma_{c}^{2}} \Big |_{\sigma_{c} = v_{\sigma}}  = 8B\, v_\sigma^2 = \frac{\mathcal M^{4}}{8\pi^{2}\, v_{\sigma}^{2}} \ .
\end{equation}

The value of the one-loop contribution \eqref{V1fin} is negative at the minimum. However, one can show that the tree-level potential \eqref{VE} is non-vanishing at the minimum along the flat direction, due to the gravitational contributions. In particular, one obtains
\begin{equation}\label{VEmin}
V^{(0)E}(v_{\sigma}) = \frac{3}{8}f_{0}^{2} \bar{M}_\text{P}^2 \tbrac{ \bar{M}_\text{P}^2- \pbrac{\xi_{H} v_{\phi}^{2} + \xi_{\eta} v_{\eta}^{2}}} \ , \qquad V^{(1)}(v_\sigma) = -\frac{\mathcal M^{4}}{128\pi^{2}} \ .
\end{equation}
At this point, we demand that the value of the full one-loop effective $\sigma$~potential (the sum of the tree-level and the one-loop contributions) to be zero at the minimum
\begin{equation}\label{V01min}
V(v_\sigma) \equiv V^{(0)E}(v_{\sigma}) + V^{(1)}(v_\sigma)  = 0 \ .
\end{equation}
This is equivalent to imposing the cosmological constant to be equal to zero at one-loop. Solving \eqref{V01min} for $\bar{M}_\text{P}^2$, we retrieve, up to the first-order, for the definition of the reduced Planck scale
\begin{equation}\label{Mplfin}
\bar{M}_\text{P}^2 \simeq \pbrac{\xi_{H} v_{\phi}^{2} + \xi_{\eta} v_{\eta}^{2}} \tbrac{1+ \frac{\mathcal M^{4}}{48\pi^{2} f_{0}^{2} \pbrac{\xi_{H} v_{\phi}^{2} + \xi_{\eta} v_{\eta}^{2}}^2}} \ ,
\end{equation}
which, includes a one-loop correction to the tree-level definition (c.f. \eqref{F} and \eqref{Mpl}).

Hence, the one-loop effective $\sigma$~potential along the flat direction takes the final form
\begin{equation}\label{V01}
V(\sigma_{c}) = \frac{\mathcal M^{4}}{128\pi^{2}} \tbrac{ \frac{\sin^{4}\omega_{2}}{36\, \bar{M}_\text{P}^4} \, \sigma_{c}^4 \pbrac{2\log \frac{\sigma_{c}^2\, \sin^{2}\omega_{2}}{6\bar{M}_\text{P}^2} - 1}+1} \ ,
\end{equation}
where, we have expressed $v_{\sigma}$ in terms of the reduced Planck scale, by means of \eqref{sigmarel1} and \eqref{vz},
\begin{equation}\label{vsig}
v_{\sigma}^{2} = \frac{v_{\zeta}^{2}}{\sin^{2}\omega_{2}} = \frac{6\bar{M}_\text{P}^2}{\sin^{2}\omega_{2}} \ .
\end{equation}
The final additive term inside the square brackets in \eqref{V01} is due to the non-vanishing gravitational contributions at the minimum of the tree-level potential (c.f. \eqref{VEmin}), which serve to cancel the negative one-loop contributions at the minimum (condition \eqref{V01min}). One observes that the one-loop effective potential \eqref{V01} is positive-definite and bounded from below for $\mathcal M^{4} > 0$, which implies a nontrivial relation between the masses (c.f. \eqref{Bmodel})
\begin{equation}\label{massrel}
5M_{\theta}^{4} +M_\kappa^4 + M_\chi^4 - 6M_{\mathcal N}^4 >  12M_t^4- M_h^4 - 6M_W^4 - 3M_Z^4 \simeq (300~\text{GeV})^{4}\ .
\end{equation}
This relation guarantees a non-tachyonic mass for the $\sigma$~boson by \eqref{msig}. Moreover, \eqref{massrel} implies that the right-handed Majorana neutrinos cannot constitute the heaviest state within the framework.

\subsection{Independent Free Parameters of the Framework}

The (\textit{ad hoc}) requirement of a vanishing cosmological constant at one-loop \eqref{V01min} provides an interesting insight into the nature of the additive contributions to the flat direction naturalness conditions \eqref{lmp} and \eqref{leta}. In particular, one notes that \eqref{Mplfin} compels the additive corrections on the right-hand sides of \eqref{lmp} and \eqref{leta} to be nonzero at one-loop and of the order $\mathcal O (\mathcal M^{4} / v_{\eta}^4)$. This is a direct consequence of the fact that our scale symmetry does not offer any protection against the cosmological constant, since the latter is not incorporated within the scale-invariant formalism of the current framework. Hence, we simply ignore the additive contribution of the cosmological constant to the tree-level relations.\footnote{Alternatively, one may ``fine-tune'' the difference between the bosonic and fermionic masses in order to obtain a very small $\mathcal M^{4}$ (c.f. \eqref{Bmodel}). For the additive corrections to be parametrically smaller than the leading term on the right-hand sides of \eqref{lmp} and \eqref{leta}, one needs $\mathcal M^{4} \lesssim v_{\phi}^{4}$, requiring the bosonic and fermionic masses to be extraordinary close to one another. Nevertheless, such a fine-tuning is fundamentally different than the one in the traditional sense, in which the \textit{couplings} would have been chosen to be unnaturally small to produce very light overall masses. Unfortunately, such a tiny $\mathcal M^{4}$ results in a very small amplitude for the one-loop effective potential \eqref{V01}, giving rise to an unacceptably suppressed primordial perturbation power spectrum (see Sec.~\ref{Infl}).}

At this stage, let us examine the independent free parameters of the theory. One can show that the model contains nine independent parameters, which, without loss of generality, may be conveniently taken as the following set
\begin{equation}\label{freepar}
\cbrac{\omega_{2}, M_{\theta}, M_{\kappa}, M_{\chi}, M_{\mathcal N}, \lambda_{m}^{-}, \lambda_{\chi}, \xi_{H}, \xi_{\chi}} \ .
\end{equation}
We recall that the mass of the $\sigma$~boson \eqref{msig} is radiatively generated at one-loop; therefore, it is fully determined in terms of the parameter set \eqref{freepar} and does not constitute an independent parameter,
\begin{equation}\label{msigfin}
m_{\sigma}^{2} = \frac{\sin^2\omega_2}{48\pi^2} \, \frac{\mathcal M^{4}}{\bar{M}_\text{P}^2}  \ ,
\end{equation}
with $\mathcal M^{4}$ defined in \eqref{Bmodel}. As a consequence, whenever convenient, any of the first five inputs in \eqref{freepar} may be replaced by $m_{\sigma}$ as a free parameter instead.

Fixing the values of the weak and the reduced Planck scales, the remaining parameters and couplings of the formalism may all be conveniently expressed in terms of the set \eqref{freepar}, where, as discussed, we neglect the one-loop (cosmological constant related) $\mathcal O (\mathcal M^{4} / v_{\eta}^4)$ corrections to the tree-level relations. In particular, defining the ratio of the two scales as
\begin{equation}\label{eps}
\varepsilon \equiv \frac{v_{\phi}}{\bar{M}_\text{P}} \sim 10^{-16} \ ,
\end{equation}
and using \eqref{vz} and \eqref{t1t2}, we obtain
\begin{equation}\label{veta}
v_{\eta}^{2} = \bar{M}_\text{P}^{2} \tbrac{6\cot^{2} \omega_{2}-\varepsilon^{2}} \ .
\end{equation}
This, in turn, results in an expression for the non-minimal coupling of the $\eta$~scalar via \eqref{Mplfin}
\begin{equation}\label{xieta}
\xi_{\eta} \simeq \frac{1-\xi_{H}\, \varepsilon^{2}}{6\cot^{2} \omega_{2}-\varepsilon^{2}} \ .
\end{equation}
Moreover, the mass definitions \eqref{treemass1} and \eqref{treemass2} may be utilized to convert their corresponding couplings in terms of the set \eqref{freepar}, according to
\begin{equation}\label{couplings}
\begin{split}
&\lambda_{\phi} \simeq \frac{3 M_{h}^{2}}{v_{\phi}^{2}} + \mathcal O(\varepsilon^{2}) \ , \qquad \lambda_{\eta \chi} =  \frac{1}{6\cot^{2} \omega_{2}-\varepsilon^{2}} \pbrac{\frac{2M_{\chi}^{2}}{\bar{M}_\text{P}^2} - \lambda_{m}^{-}\, \varepsilon^{2}} \ , \\
&f_{0}^{2} \simeq \frac{2M_{\kappa}^{2}}{\bar{M}_\text{P}^2} \cos^{2}\omega_{2}+ \mathcal O(\varepsilon^{2}) \ , \qquad f_{2}^{2} = \frac{2M_{\theta}^{2}}{\bar{M}_\text{P}^2} \ , \qquad y_{N} = \frac{M_{\mathcal N}}{\sqrt2 \bar{M}_\text{P} \sqrt{6\cot^{2} \omega_{2}-\varepsilon^{2}}} \ ,
\end{split}
\end{equation}
where, only the leading-order terms of the $\lambda_{\phi}$ and $f_{0}^{2}$~couplings are displayed, given the complicated form of their full expressions.\footnote{We retain the exact $\varepsilon$-dependent expressions of all the parameters and couplings in our numerical calculations.}

The $\omega_{3}$~mixing angle is defined in \eqref{t3}, with the relations \eqref{couplings} implied, whereas $\omega_{1}$ is obtained via its corresponding definition \eqref{t1t2}, and takes the final form
\begin{equation}\label{t1fin}
\tan\omega_{1} = \frac{\varepsilon}{\sqrt{6\cot^{2} \omega_{2}-\varepsilon^{2}}} \ .
\end{equation}
Finally, the above relations yield for the couplings \eqref{lmp} and \eqref{leta} 
\begin{equation}\label{flatcoupl}
\lambda_{m}^{+} \simeq -\frac{\lambda_{\phi}}{3} \tan^{2}\omega_{1} \ , \qquad \lambda_{\eta} \simeq \lambda_{\phi} \tan^{4}\omega_{1} \ ,
\end{equation}
where, as discussed, the one-loop induced cosmological constant contributions are ignored.
This concludes the introduction of the framework's formalism and notations. In the following sections, we proceed to analyze several phenomenological and cosmological implications of the presented minimal scale- and $CP$-symmetric model.

\section{Vacuum Stability and Perturbativity}\label{VST}

Gauge theories with an extended scalar sector may develop Landau poles at finite values of the renormalization scale \cite{GenBeta1,GenBeta2}, and violate the perturbativity of the theory; hence, one should examine the behavior of the running couplings as a function of the energy. As demonstrated in the following sections, the mixing angle range relevant to the inflation and the dark matter resides within $\tan\omega_{2} \gtrsim 0.01$. Using \eqref{veta}, one can deduce the corresponding energy range, $v_{\eta} \lesssim 250\bar{M}_\text{P}$. Therefore, within the framework under consideration, it is sufficient to demand that all gauge, fermionic, and scalar couplings remain perturbative up to at least an energy scale of $250\bar{M}_\text{P}$, where the internal consistency of the theory is guaranteed. As the indicator for a Landau pole, we choose a maximum running coupling value of $4\pi$.\footnote{One can check that larger indicator values will not significantly change the energy scale associated with the Landau pole.} In addition, the conditions for the vacuum stability of the scalar potential \eqref{stabtree1} and \eqref{stabtree2}, as well as the positivity of $f_{0,2}^{2}$, should be satisfied for the running couplings in the entire energy range up to the mentioned cutoff. The non-minimal coupling $\xi_{\eta}$ remains perturbative within the Agravity framework by enforcing $\xi_{\eta} \lesssim 1/f_{0,2}$ \cite{Salvio:2014soa}.

Within the current scale symmetric framework, the $\Lambda$~scale \eqref{Lamb}---at which the flat direction of the potential is defined---can be used as the characteristic energy for the starting point of the non-SM couplings. This energy scale is formally a function of the mixing angle and the masses in \eqref{freepar}. For consistency, we assume that this scale is smaller than the Planck scale, $\Lambda \lesssim M_\text{P}$, such that the non-SM couplings become active below the Planck scale.\footnote{For the purpose of the current analyses, we assume that the SM vacuum is stable up to $M_\text{P}$, which is a physical possibility, given the measured values of the Higgs boson and the top quark masses \cite{SMvac}. Alternatively, one may postulate beyond the SM physics (around the TeV scale), responsible for the stability of the vacuum up to the Planck scale (see \cite{Farzinnia:2013pga,Farzinnia:2015uma} for a classically scale invariant example), at the expense of a less minimal framework. For more recent discussions addressing the effects of the Planck scale physics on the SM vacuum stability, see \cite{StabPlanck}. Scale-symmetric theories satisfying Total Asymptotic Freedom are classified in \cite{Giudice:2014tma}.} Once the $\beta$-functions of the couplings are obtained, one may (numerically) solve the renormalization group~(RG) equation $\mu \, d \tau /d \mu = \beta_{\tau}$ ($\tau$ any of the running couplings and $\mu$ the renormalization scale), in order to determine the behavior of the couplings as a function of the energy. We shall report the one-loop $\beta$-functions of all relevant running couplings, calculated within the $\overline{\text{MS}}$-scheme. The SM couplings' $\beta$-functions are analytically known up to two-loops within the literature \cite{Schrempp:1996fb}, those of our extended scalar sector and the right-handed Majorana neutrinos were first calculated in \cite{Farzinnia:2013pga}, and all gravitational contributions\footnote{All the relevant $\beta$-functions are computed within the Jordan frame \cite{Salvio:2014soa}. It has been shown that, in the limit of the weak gravitational interactions, the one-loop $\beta$-functions in the Jordan and the Einstein frames are compatible with one another \cite{Kannike:2015apa}.} are deduced from the Agravity framework \cite{Salvio:2014soa}.\footnote{We note that some discrepancy exists between a number of the Agravity $\beta$-functions and the results previously reported within the literature (e.g. \cite{Buchbinder:1989jd}). See also the relevant discussion and the provided references in \cite{Salvio:2014soa} in this regard.}

Let us begin by discussing the gauge couplings' $\beta$-functions. Within the described model, the SM gauge interactions are unaffected and their usual $\beta$-function expressions are appropriate
\begin{equation} \label{SMgauge}
\pbrac{4\pi}^{2}\beta_{g} =-\frac{19}{6}g^{3} \ , \qquad \pbrac{4\pi}^{2}\beta_{g'} =+\frac{41}{6}g'^{\,3} \ , \qquad \pbrac{4\pi}^{2}\beta_{g_{c}} = -7g_{c}^{3} \ .
\end{equation}
We employ a hypercharge coupling normalization according to $g' = \sqrt{3/5} \, g_{1}$, where $g_{1}$ is the corresponding coupling with the GUT normalization.\footnote{Within the SM, the hypercharge coupling develops a Landau pole around $\mu \sim 10^{41}$~GeV.} Furthermore, the $\beta$-functions of the Agravitational interactions $f_{0}^{2}$ and $f_{2}^{2}$ are given by
\begin{equation} \label{agravity}
\begin{split}
\pbrac{4\pi}^{2}\beta_{f_{0}^{2}} =&\,\frac{5}{3}f_{2}^{4} + 5f_{0}^{2}f_{2}^{2} + \frac{f_{0}^{4}}{12} \tbrac{ 10 + 4\pbrac{1+6\xi_{H}}^{2} + \pbrac{1+6\xi_{\eta}}^{2} + \pbrac{1+6\xi_{\chi}}^{2}} \ , \\
\pbrac{4\pi}^{2}\beta_{f_{2}^{2}} =&\,-f_{2}^{4} \tbrac{ \frac{133}{10} + \frac{12}{5} + \frac{45+3}{20} + \frac{4+3}{60}} \ ,
\end{split}
\end{equation}
where, $\beta_{f_{2}^{2}}$ contains, in addition to the SM contributions, those of the three right-handed Majorana neutrino flavors, as well as the additional three (real) scalars, $\eta,\zeta,\chi$. Interestingly, one observes that the $f_{2}^{2}$ coupling is asymptotically free.

The dominant Yukawa interactions of the model are represented by those of the top quark and the flavor universal right-handed Majorana neutrinos. The $\beta$-functions of these fermions receive an additional Agravitational correction from $f_{2}^{2}$. Moreover, the right-handed Majorana neutrinos are SM gauge singlets and do not mix with the ordinary SM fermions. Hence, one obtains
\begin{equation}\label{Yukawa}
\pbrac{4\pi}^{2}\beta_{y_{t}} =y_{t} \tbrac{ - \frac{9}{4} g^{2} - \frac{17}{12} g'^{\,2} -8 g_{c}^{2}+\frac{9}{2} y_{t}^{2} +\frac{15}{8} f_{2}^{2}}\ , \qquad \pbrac{4\pi}^{2}\beta_{y_{N}} =y_{N} \tbrac{ 9 y_{N}^{2}+\frac{15}{8} f_{2}^{2} } \ .
\end{equation}

The scalar non-minimal couplings also run as a function of the renormalization scale, and their $\beta$-functions read
\begin{equation} \label{non-min}
\begin{split}
\pbrac{4\pi}^{2}\beta_{\xi_{H}} =&\, \xi_{H}\tbrac{-\frac{5}{3} \frac{f_{2}^{4}}{f_{0}^{2}} + f_{0}^{2}\pbrac{\frac{2}{3} + \xi_{H}} \pbrac{1+6\xi_{H}}} + \pbrac{1+6\xi_{H}} \tbrac{\frac{\lambda_{\phi}}{3} + 2y_{t}^{2} - \frac{1}{4} g'^{2} - \frac{3}{4} g^{2}} \\
&+ \pbrac{1+6\xi_{\eta}} \frac{\lambda_{m}^{+}}{6} + \pbrac{1+6\xi_{\chi}} \frac{\lambda_{m}^{-}}{6} \ , \\
\pbrac{4\pi}^{2}\beta_{\xi_{\eta}} =&\, \xi_{\eta}\tbrac{-\frac{5}{3} \frac{f_{2}^{4}}{f_{0}^{2}} + f_{0}^{2}\pbrac{\frac{2}{3} + \xi_{\eta}} \pbrac{1+6\xi_{\eta}}} + \pbrac{1+6\xi_{\eta}} \tbrac{\frac{\lambda_{\eta}}{6} + 4y_{N}^{2}}+ \pbrac{1+6\xi_{H}} \frac{2\lambda_{m}^{+}}{3} + \pbrac{1+6\xi_{\chi}} \frac{\lambda_{\eta\chi}}{6} \ , \\
\pbrac{4\pi}^{2}\beta_{\xi_{\chi}} =&\, \xi_{\chi}\tbrac{-\frac{5}{3} \frac{f_{2}^{4}}{f_{0}^{2}} + f_{0}^{2}\pbrac{\frac{2}{3} + \xi_{\chi}} \pbrac{1+6\xi_{\chi}}} + \pbrac{1+6\xi_{\chi}} \frac{\lambda_{\chi}}{6} + \pbrac{1+6\xi_{H}} \frac{2\lambda_{m}^{-}}{3} + \pbrac{1+6\xi_{\eta}} \frac{\lambda_{\eta\chi}}{6} \ . \\
\end{split}
\end{equation}
Note that, according to \eqref{LRHN}, the $CP$-symmetry only allows for a right-handed Majorana neutrino Yukawa interaction with the $\eta$~scalar. Finally, taking into account the coupling normalizations of the potential \eqref{V0}, the scalar quartic coupling $\beta$-functions may be expressed as
\begin{equation}\label{scalar}
\begin{split}
\pbrac{4\pi}^{2}\beta_{\lambda_{\phi}} =&\, 4\lambda_{\phi}^{2} + 3\tbrac{ \lambda_{m}^{+\, 2}+  \lambda_{m}^{-\, 2}} +3\lambda_{\phi} \tbrac{4 y_{t}^{2} - 3g^{2} - g'^{2}}- \frac{9}{4} \tbrac{16 y_{t}^{4} - 2 g^{4} - (g^{2} + g'^{2})^{2}} \\
& +\lambda_{\phi} \tbrac{5f_{2}^{2} + f_{0}^{2}\pbrac{1+6\xi_{H}}^{2}} +3\xi_{H}^{2} \tbrac{5f_{2}^{4} + f_{0}^{4}\pbrac{1+6\xi_{H}}^{2}} \ , \\
\pbrac{4\pi}^{2}\beta_{\lambda_{\eta}} =&\, 3\lambda_{\eta}^{2} + 12 \lambda_{m}^{+\, 2}+ 3 \lambda_{\eta\chi}^{2} +24\lambda_{\eta} y_{N}^{2}- 288 y_{N}^{4}+\lambda_{\eta} \tbrac{5f_{2}^{2} + f_{0}^{2}\pbrac{1+6\xi_{\eta}}^{2}}  \\
&+3\xi_{\eta}^{2} \tbrac{5f_{2}^{4} + f_{0}^{4}\pbrac{1+6\xi_{\eta}}^{2}} \ , \\
\pbrac{4\pi}^{2}\beta_{\lambda_{\chi}} =&\, 3\lambda_{\chi}^{2} + 12 \lambda_{m}^{-\, 2}+ 3 \lambda_{\eta\chi}^{2} +\lambda_{\chi} \tbrac{5f_{2}^{2} + f_{0}^{2}\pbrac{1+6\xi_{\chi}}^{2}} +3\xi_{\chi}^{2} \tbrac{5f_{2}^{4} + f_{0}^{4}\pbrac{1+6\xi_{\chi}}^{2}} \ , \\
\pbrac{4\pi}^{2}\beta_{\lambda_{m}^{+}} =&\, 4\lambda_{m}^{+\, 2} + \lambda_{m}^{+} \tbrac{2\lambda_{\phi} + \lambda_{\eta}} + \lambda_{\eta\chi} \lambda_{m}^{-} +\frac{3}{2}\lambda_{m}^{+} \tbrac{4 \pbrac{y_{t}^{2}+2y_{N}^{2}} - 3g^{2} - g'^{2}} \\
& +\lambda_{m}^{+} \tbrac{5f_{2}^{2} + \frac{f_{0}^{2}}{6}\tbrac{\pbrac{1+6\xi_{H}}^{2} + \pbrac{1+6\xi_{\eta}}^{2} + 4 \pbrac{1+6\xi_{H}}\pbrac{1+6\xi_{\eta}}}}\\
& + \frac{1}{2}\xi_{H}\xi_{\eta} \tbrac{5f_{2}^{4} + f_{0}^{4}\pbrac{1+6\xi_{H}}\pbrac{1+6\xi_{\eta}}}\ , \\
\pbrac{4\pi}^{2}\beta_{\lambda_{m}^{-}} =&\, 4\lambda_{m}^{-\, 2} + \lambda_{m}^{-} \tbrac{2\lambda_{\phi} + \lambda_{\chi}} + \lambda_{\eta\chi} \lambda_{m}^{+} +\frac{3}{2}\lambda_{m}^{-} \tbrac{4 y_{t}^{2} - 3g^{2} - g'^{2}} \\
& +\lambda_{m}^{-} \tbrac{5f_{2}^{2} + \frac{f_{0}^{2}}{6}\tbrac{\pbrac{1+6\xi_{H}}^{2} + \pbrac{1+6\xi_{\chi}}^{2} + 4 \pbrac{1+6\xi_{H}}\pbrac{1+6\xi_{\chi}}}} \\
&+ \frac{1}{2}\xi_{H}\xi_{\chi} \tbrac{5f_{2}^{4} + f_{0}^{4}\pbrac{1+6\xi_{H}}\pbrac{1+6\xi_{\chi}}} \ ,\\
\pbrac{4\pi}^{2}\beta_{\lambda_{\eta\chi}} =&\, 4\lambda_{\eta\chi}^{2} + \lambda_{\eta\chi} \tbrac{\lambda_{\eta} + \lambda_{\chi}} + 4\lambda_{m}^{+} \lambda_{m}^{-} +12\lambda_{\eta\chi}y_{N}^{2}+ \frac{1}{2}\xi_{\chi}\xi_{\eta} \tbrac{5f_{2}^{4} + f_{0}^{4}\pbrac{1+6\xi_{\eta}}\pbrac{1+6\xi_{\chi}}} \\
& +\lambda_{\eta\chi} \tbrac{5f_{2}^{2} + \frac{f_{0}^{2}}{6}\tbrac{\pbrac{1+6\xi_{\eta}}^{2} + \pbrac{1+6\xi_{\chi}}^{2} + 4 \pbrac{1+6\xi_{\eta}}\pbrac{1+6\xi_{\chi}}}}\ .
\end{split}
\end{equation}

As mentioned, the non-SM couplings become effectively active at the flat direction energy scale $\Lambda \lesssim M_\text{P}$. At this scale, the starting values of four of the couplings directly constitute as independent input parameters; namely, those of $\lambda_{m}^{-}, \lambda_{\chi}, \xi_{H}, \xi_{\chi}$ (c.f \eqref{freepar}). The matching value of $\lambda_{\phi}$ as well as the starting values of the remaining couplings, at this same energy scale, are defined in \eqref{xieta}, \eqref{couplings}, and \eqref{flatcoupl} in terms of the input parameter set. Armed with the theory's $\beta$-functions \eqref{SMgauge}-\eqref{scalar}, one may (numerically) solve the relevant (coupled) RG~equations, determine the behavior of the running couplings as a function of the renormalization scale $\mu$, and explore the regions of the parameter space \eqref{freepar} where the perturbativity and the vacuum stability conditions are satisfied.

\begin{figure}
\includegraphics[width=.4\textwidth]{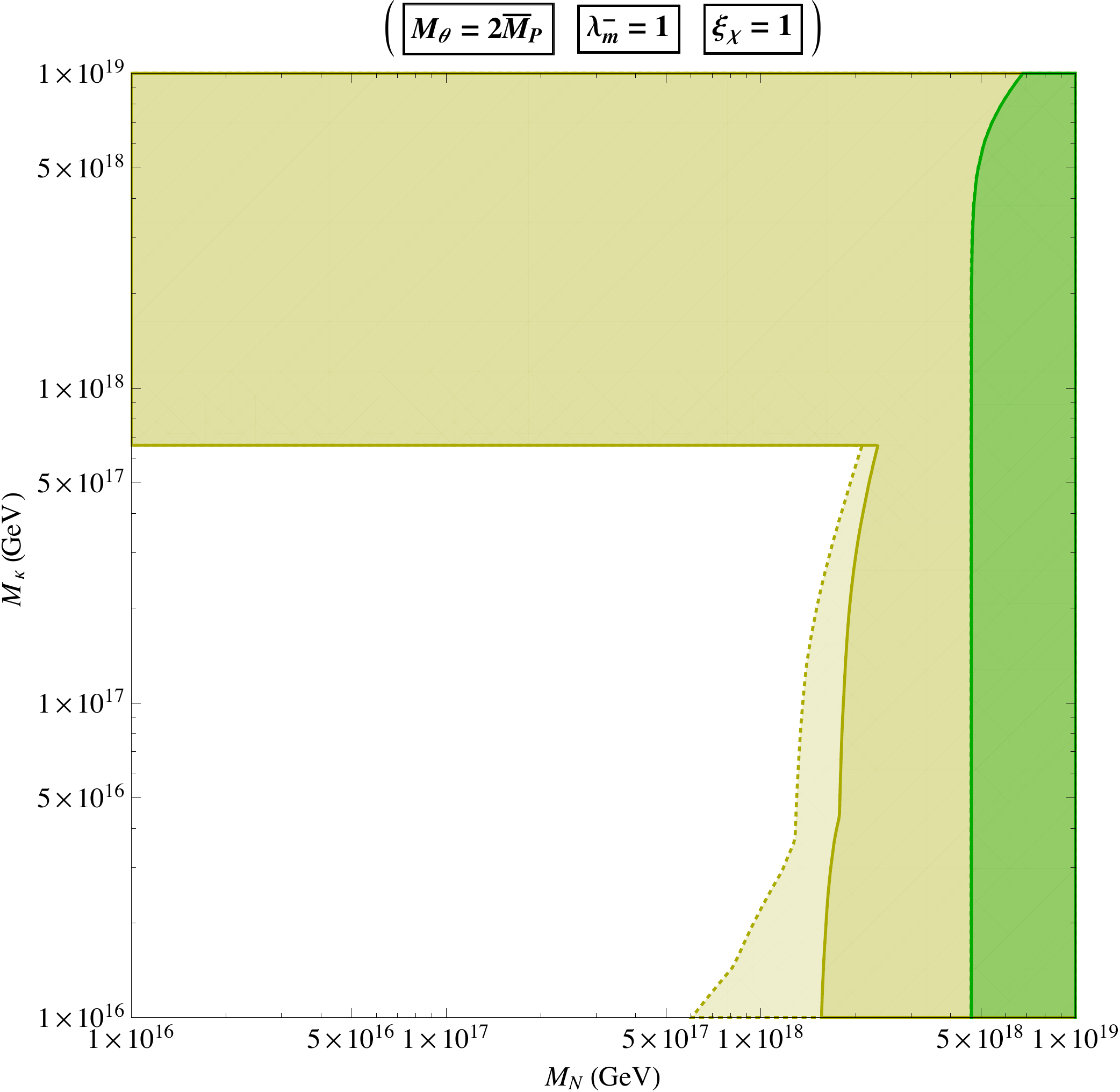}
\includegraphics[width=.4\textwidth]{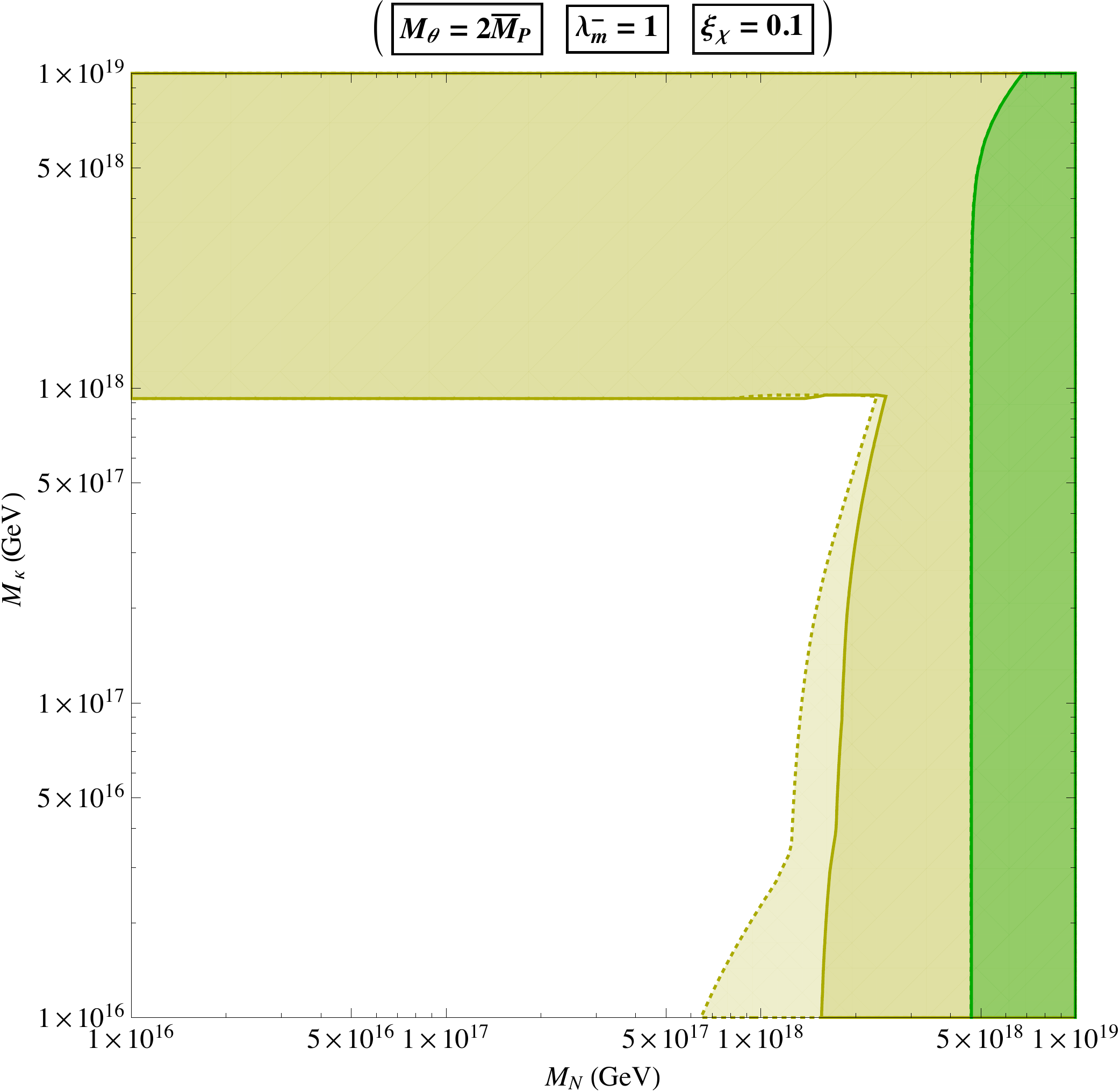}
\includegraphics[width=.4\textwidth]{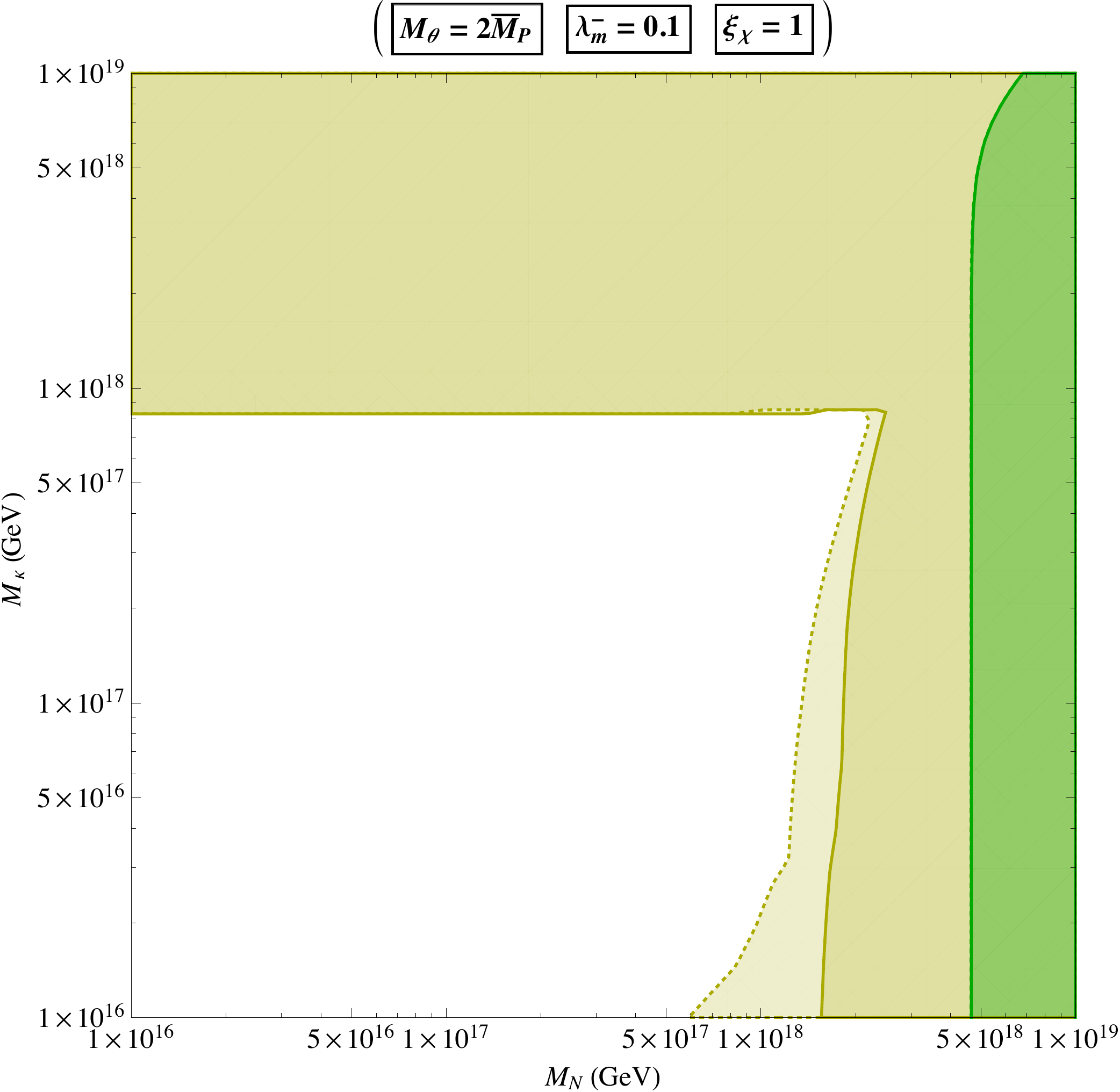}
\includegraphics[width=.4\textwidth]{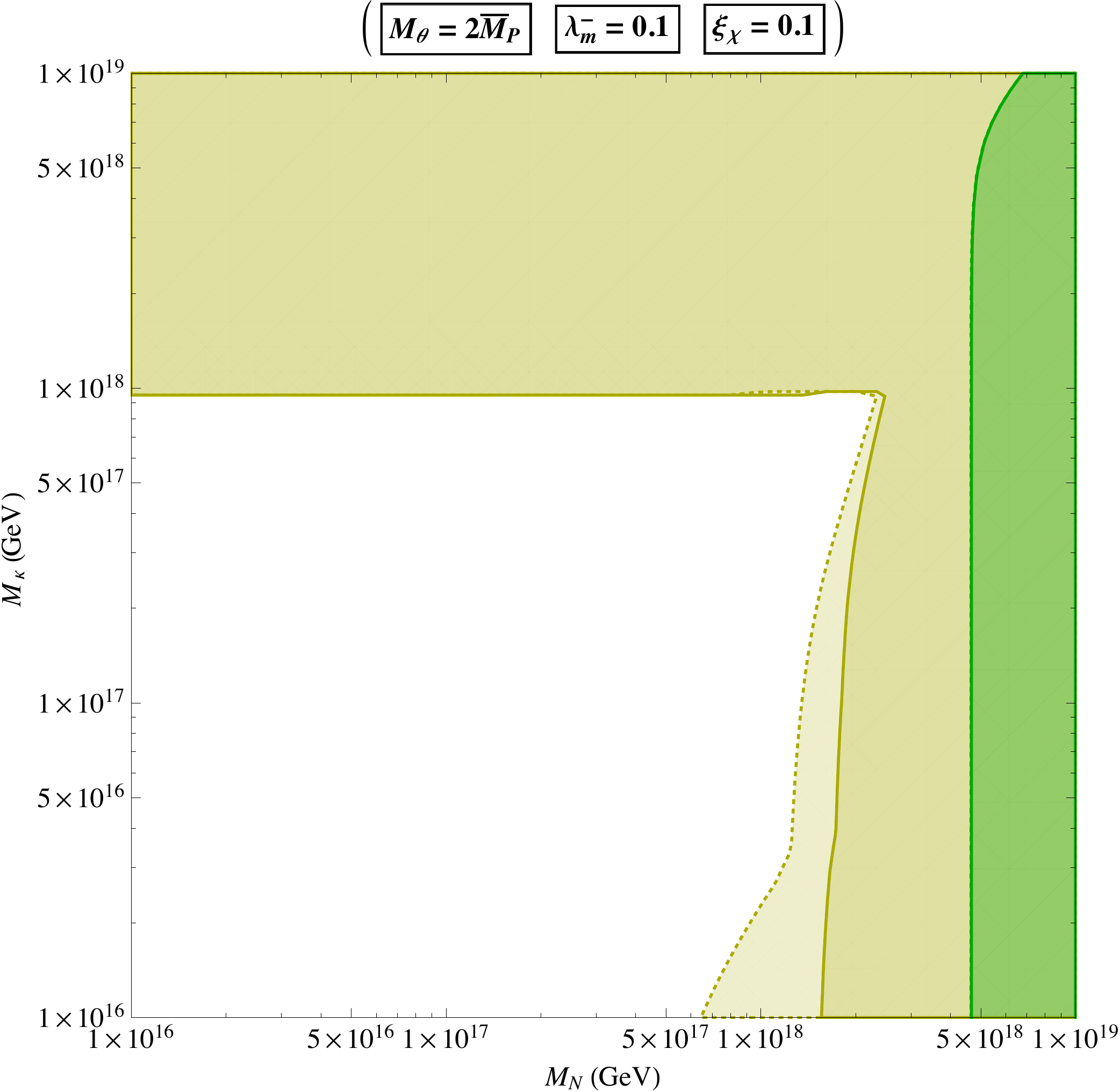}
\caption{Vacuum stability and perturbativity constraints (yellow region), displayed within the $M_{\mathcal N}-M_{\kappa}$~plane, for a cutoff energy $\mu \sim 250\bar{M}_\text{P}$, and the benchmark values $\lambda_{\chi} = \xi_{H} = 1$ and $M_{\theta} = 2\bar{M}_\text{P}$. All colored regions are excluded, and the panels exhibit the dependence on the varying $\lambda_{m}^{-}$ and $\xi_{\chi}$ from large to small. The solid lines correspond to the dark matter mass $M_{\chi} = \bar{M}_\text{P}$, whereas the dotted lines signify an additional exclusion for the lighter $M_{\chi} = 0.1\bar{M}_\text{P}$. The upper bound on $M_{\kappa}$ is determined by $\lambda_{\phi}$ ($\lambda_{\chi}$) developing a Landau pole for $\xi_{\chi} = 0.1$ ($\xi_{\chi} = 1$), and is virtually independent of the dark matter mass. At larger right-handed Majorana neutrino masses, the vacuum stability is violated below the cutoff due to the large negative fermionic contributions to the $\beta$-functions; hence, a lower bound on $M_{\kappa}$ is developed, since an adequate positive bosonic contribution is necessary for compensation. Note that this lower bound becomes more stringent for the lighter (bosonic) dark matter masses. The green region to the right side of the panels indicates the one-loop bounded from below constraint \eqref{massrel}, and is almost identical for both dark matter masses, given the dominant $M_{\theta}$ contribution. A universal value of the mixing angle $\tan\omega_{2} = 0.1$ has been selected for illustration; nevertheless, a dependence on the value of this parameter is relatively mild for most of its range.}
\label{ST1a}
\end{figure}

\begin{figure}
\includegraphics[width=.4\textwidth]{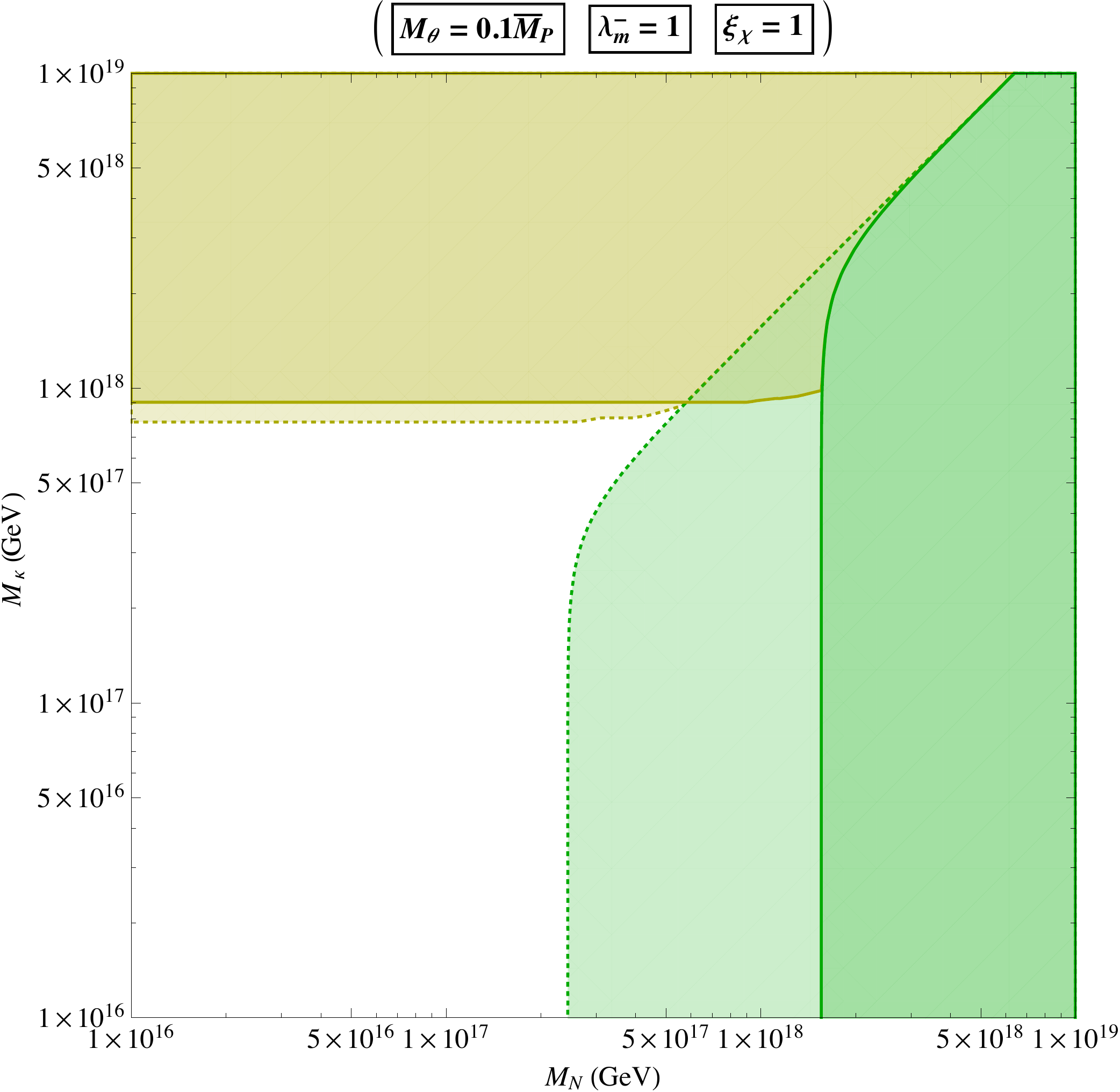}
\includegraphics[width=.4\textwidth]{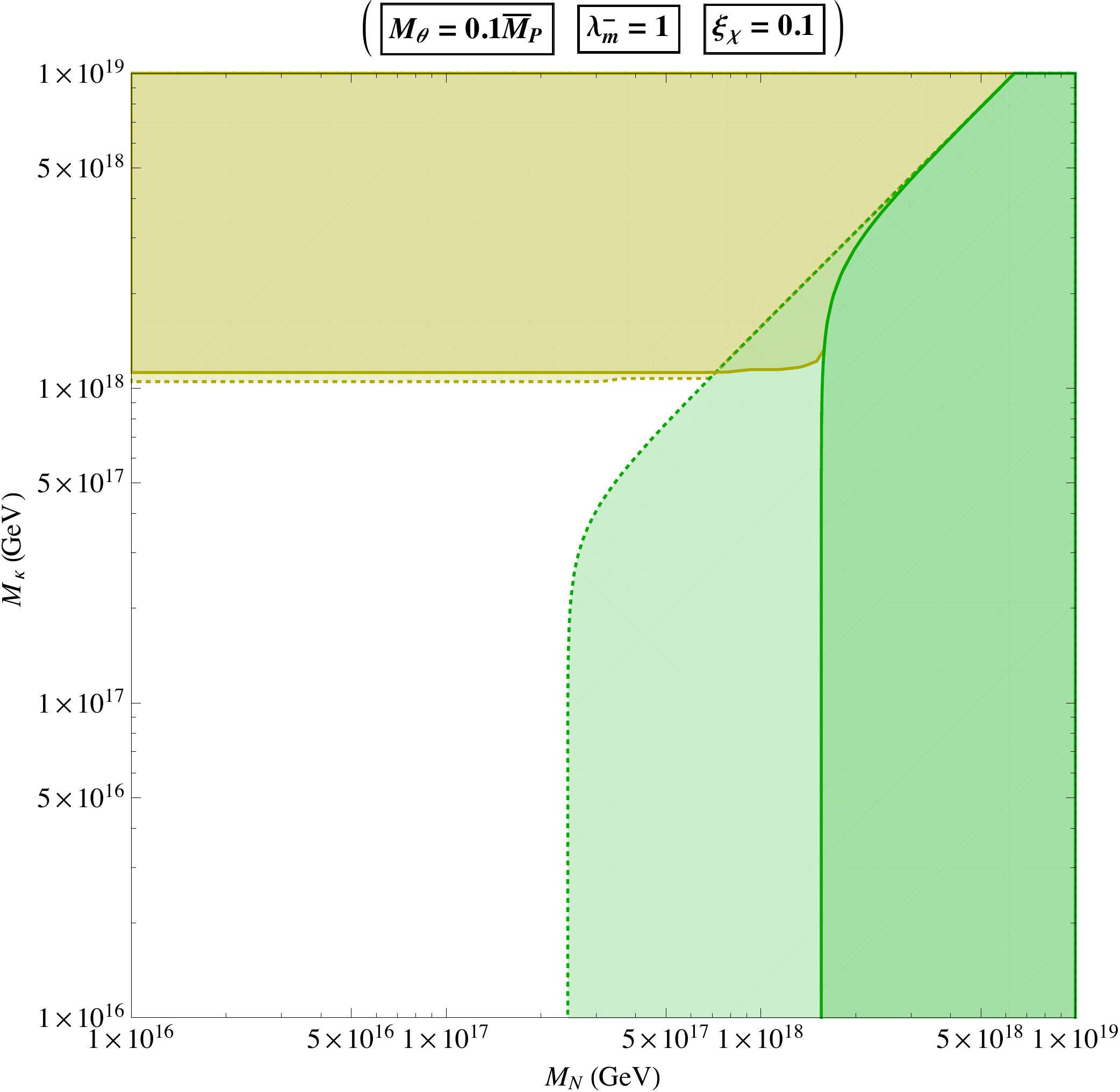}
\includegraphics[width=.4\textwidth]{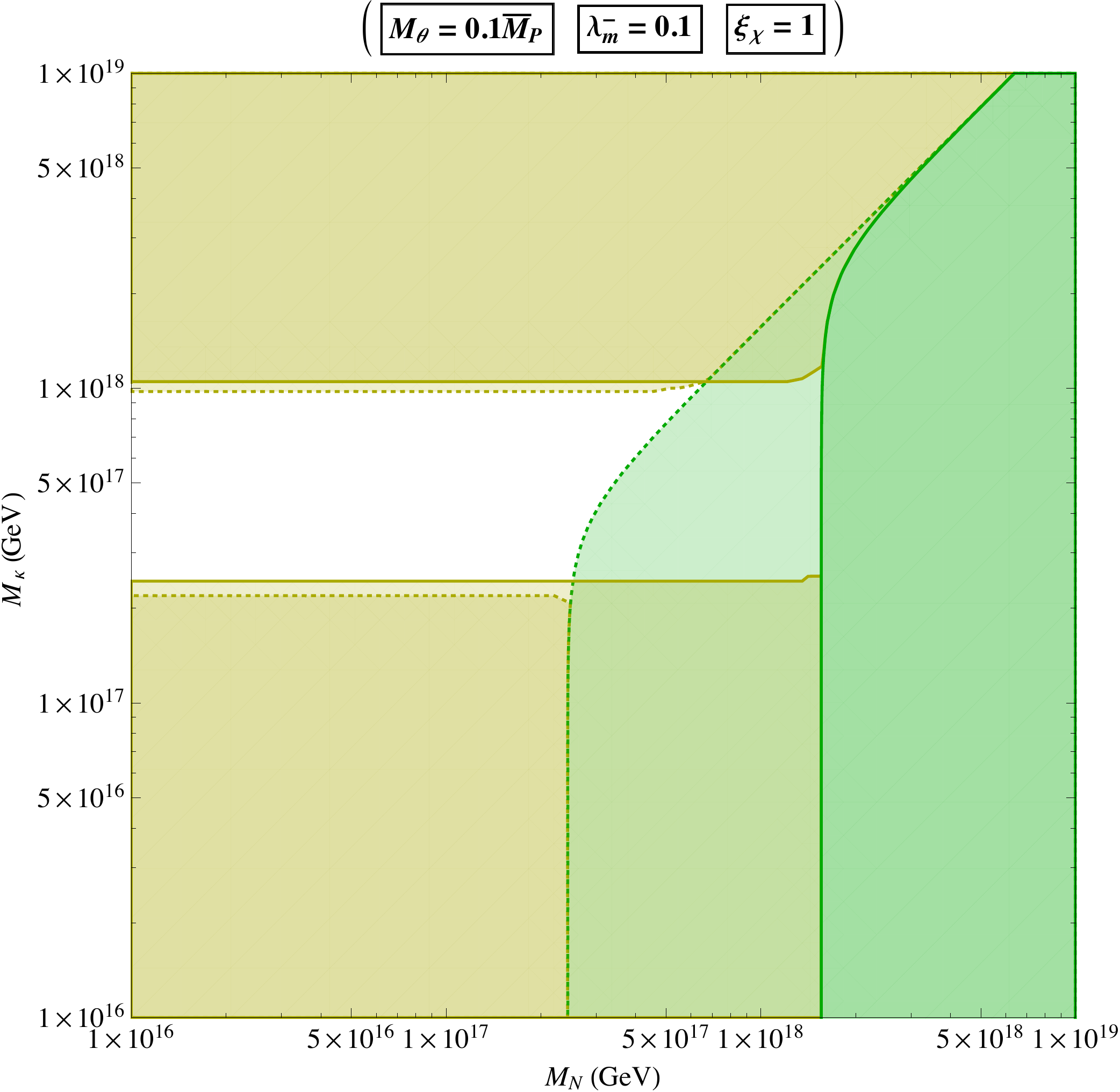}
\includegraphics[width=.4\textwidth]{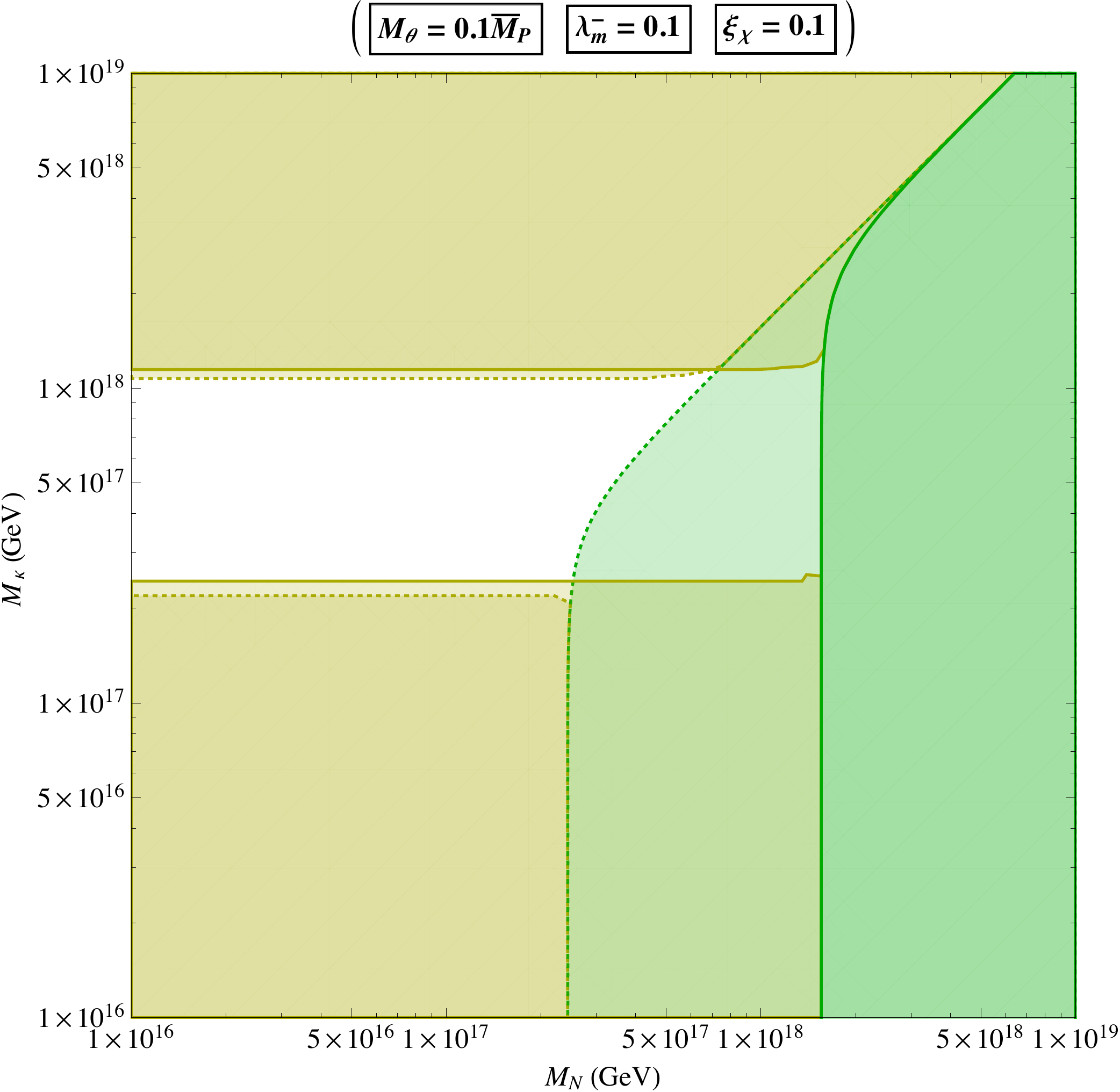}
\caption{Vacuum stability and perturbativity constraints (yellow region), displayed within the $M_{\mathcal N}-M_{\kappa}$~plane, for a cutoff energy $\mu \sim 250\bar{M}_\text{P}$, and the benchmark values $\lambda_{\chi} = \xi_{H} = 1$ and $M_{\theta} = 0.1\bar{M}_\text{P}$. All colored regions are excluded, and the panels exhibit the dependence on the varying $\lambda_{m}^{-}$ and $\xi_{\chi}$ from large to small. The solid lines correspond to the dark matter mass $M_{\chi} = \bar{M}_\text{P}$, whereas the dotted lines signify an additional exclusion for the lighter $M_{\chi} = 0.1\bar{M}_\text{P}$. The green region to the right side of the panels indicates the one-loop bounded from below constraint \eqref{massrel}. A universal value of the mixing angle $\tan\omega_{2} = 0.1$ has been selected for illustration.}
\label{ST1b}
\end{figure}

\begin{figure}
\includegraphics[width=.4\textwidth]{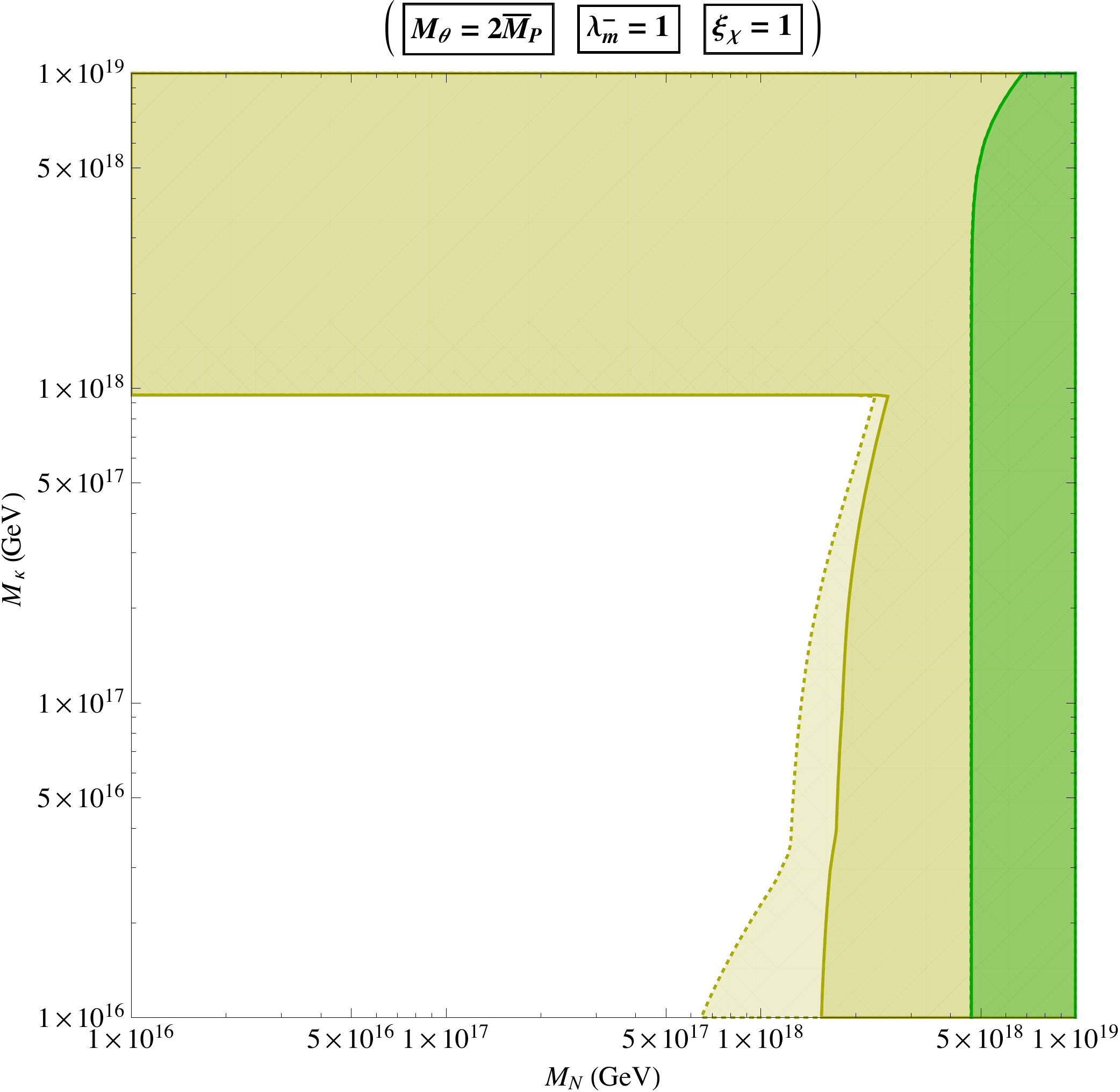}
\includegraphics[width=.4\textwidth]{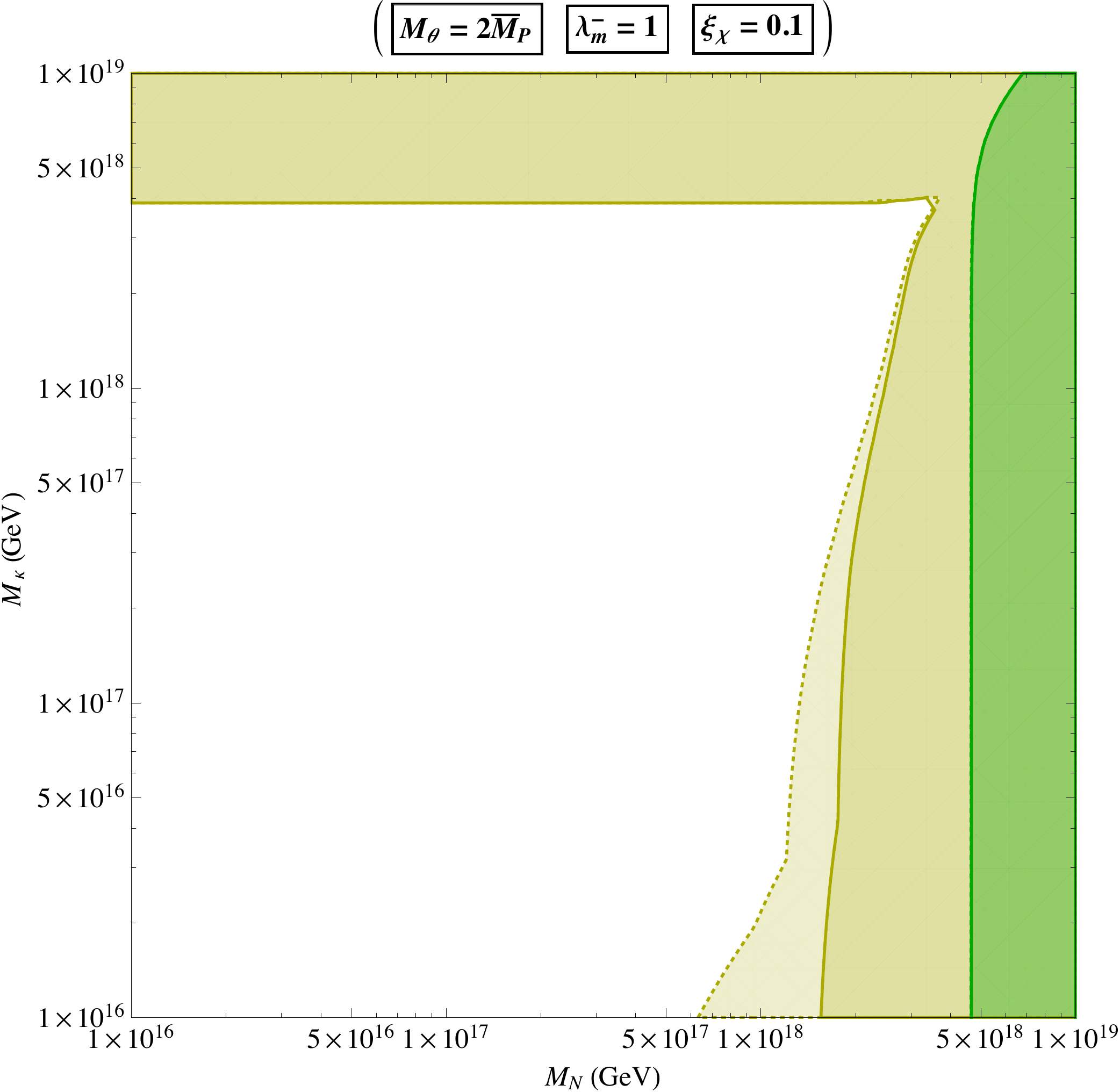}
\includegraphics[width=.4\textwidth]{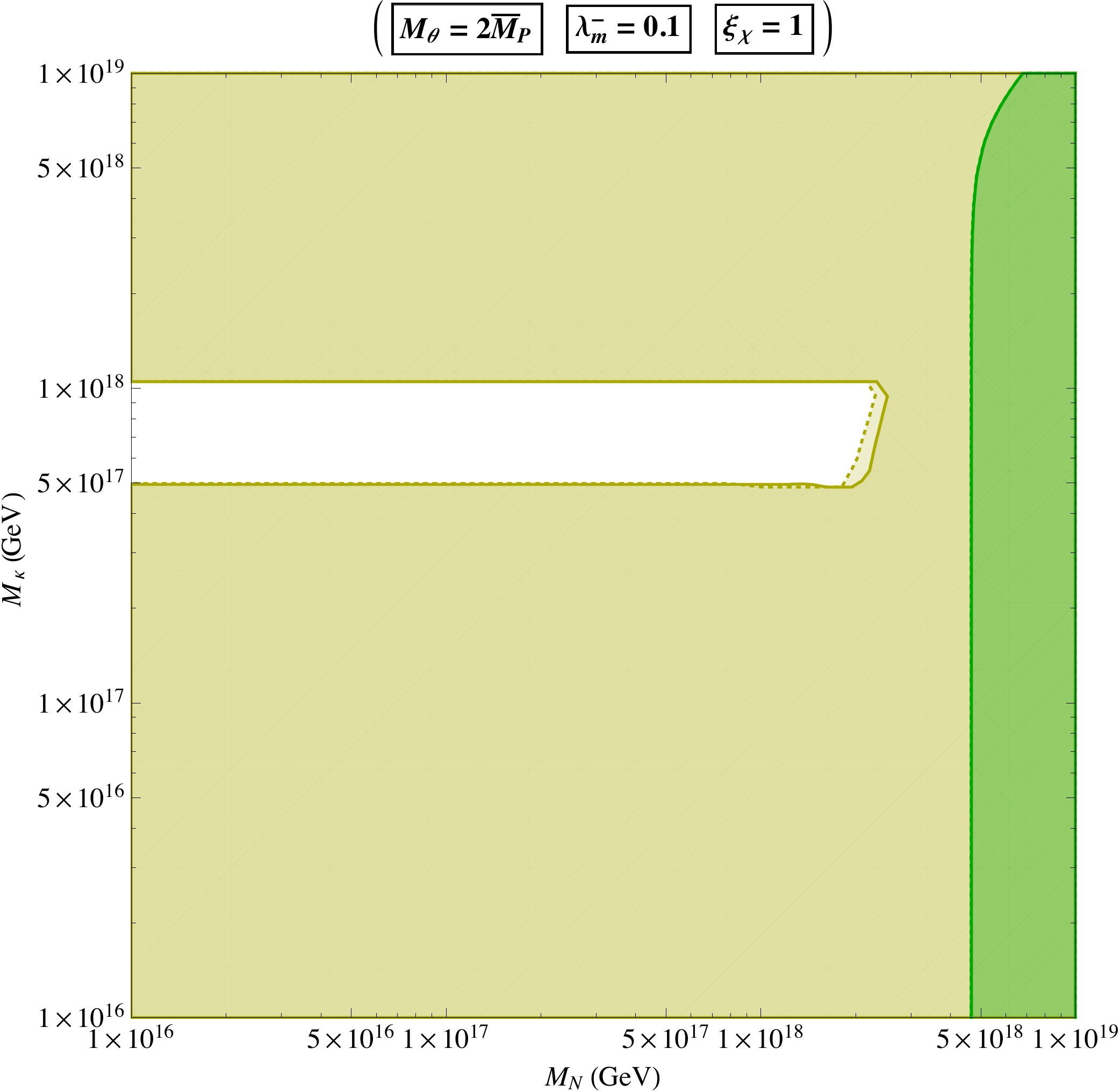}
\includegraphics[width=.4\textwidth]{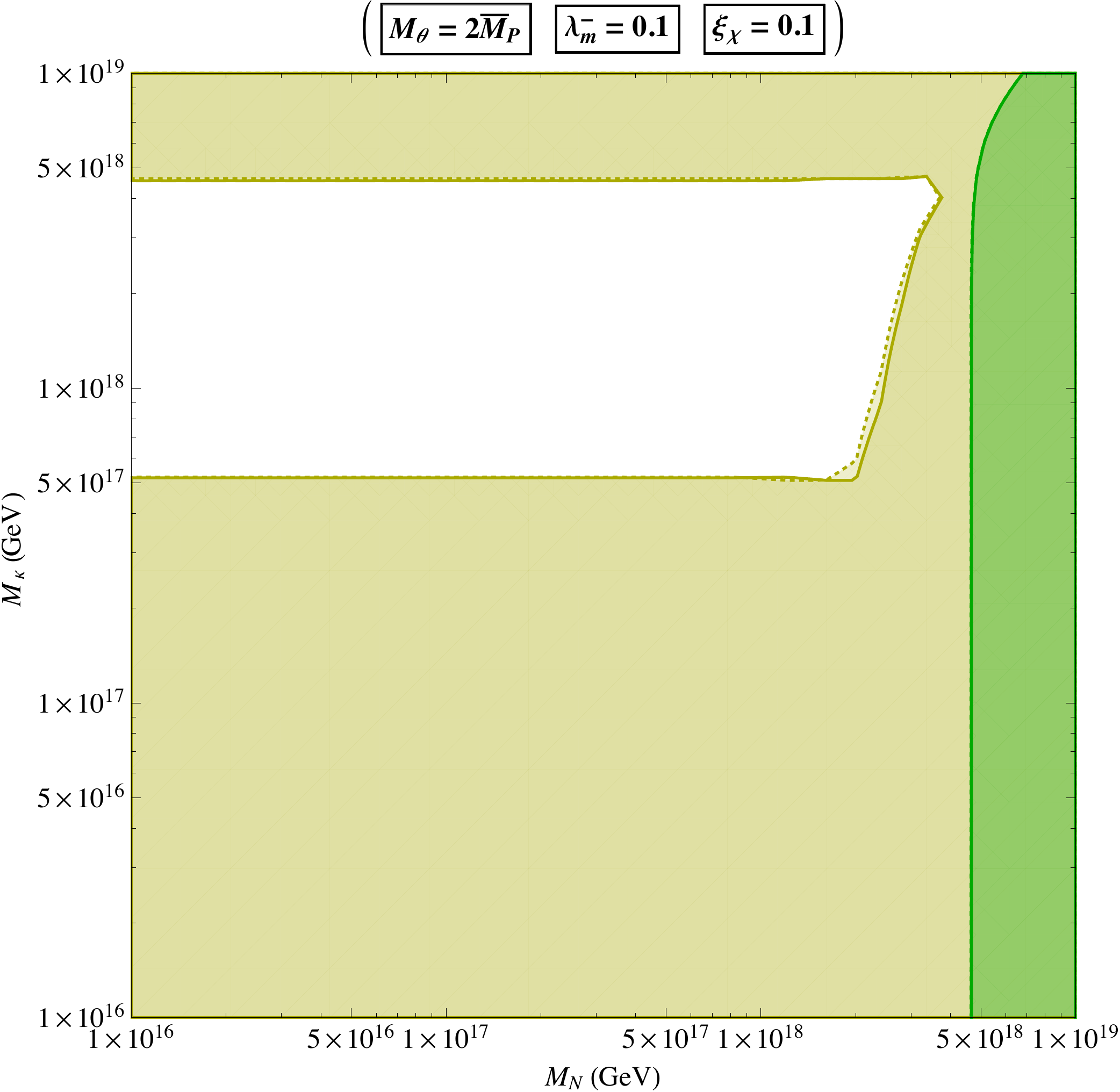}
\caption{Vacuum stability and perturbativity constraints (yellow region), displayed within the $M_{\mathcal N}-M_{\kappa}$~plane, for a cutoff energy $\mu \sim 250\bar{M}_\text{P}$, and the benchmark values $\lambda_{\chi} = \xi_{H} = 0.1$ and $M_{\theta} = 2\bar{M}_\text{P}$. All colored regions are excluded, and the panels exhibit the dependence on the varying $\lambda_{m}^{-}$ and $\xi_{\chi}$ from large to small. The solid lines correspond to the dark matter mass $M_{\chi} = \bar{M}_\text{P}$, whereas the dotted lines signify an additional exclusion for the lighter $M_{\chi} = 0.1\bar{M}_\text{P}$. The green region to the right side of the panels indicates the one-loop bounded from below constraint \eqref{massrel}, and is almost identical for both dark matter masses, given the dominant $M_{\theta}$ contribution. A universal value of the mixing angle $\tan\omega_{2} = 0.1$ has been selected for illustration.}
\label{ST2a}
\end{figure}

\begin{figure}
\includegraphics[width=.4\textwidth]{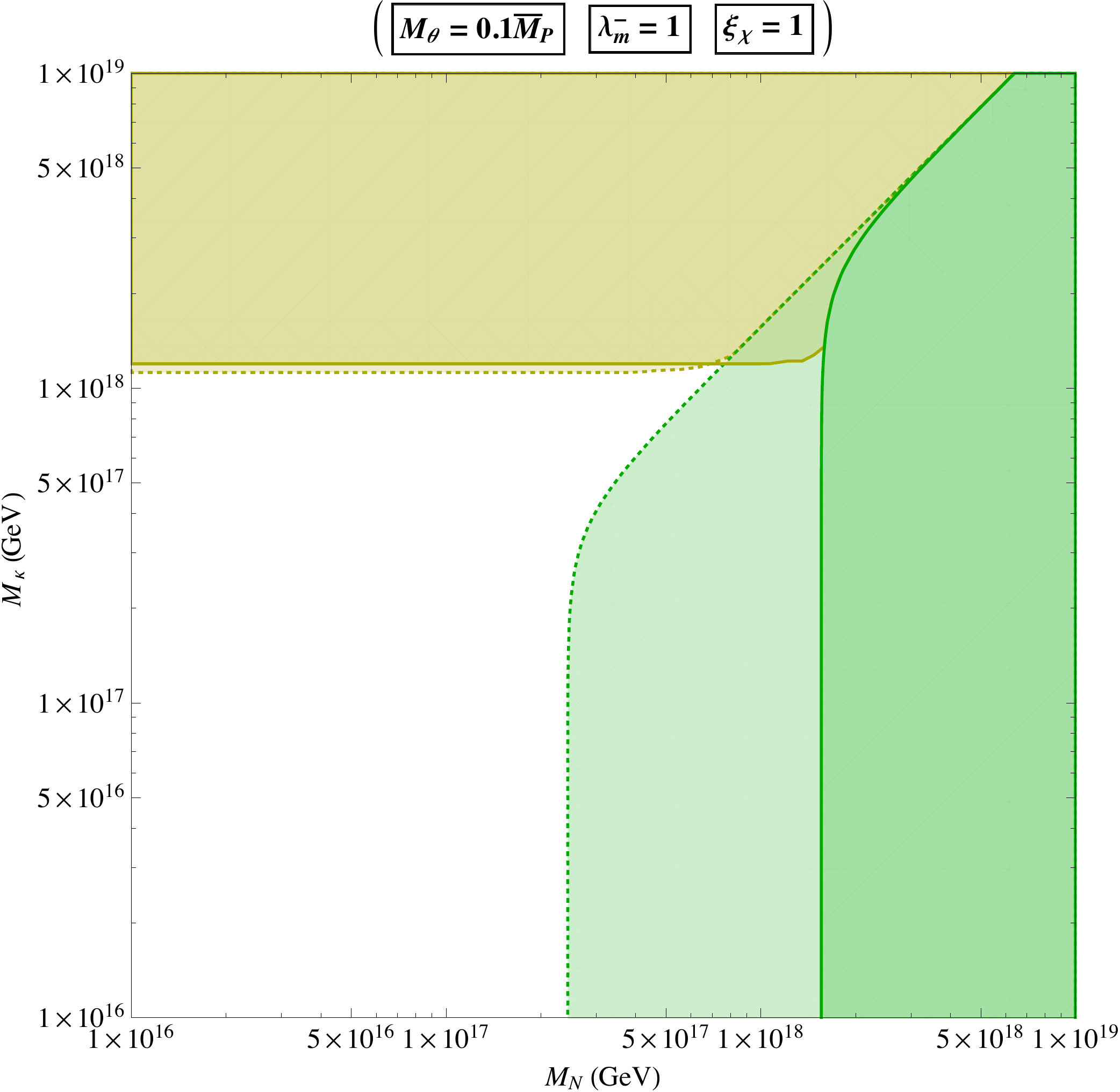}
\includegraphics[width=.4\textwidth]{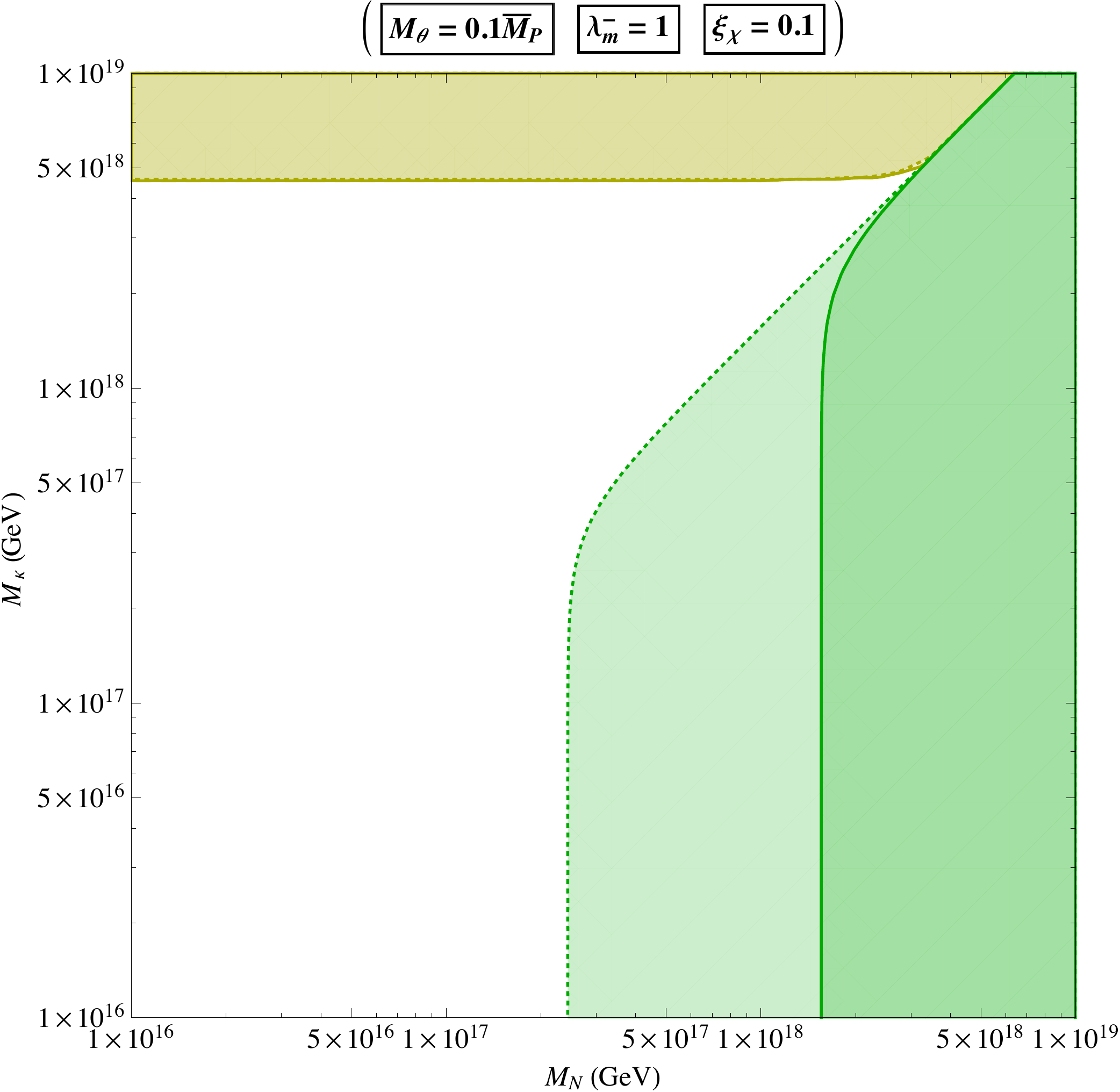}
\includegraphics[width=.4\textwidth]{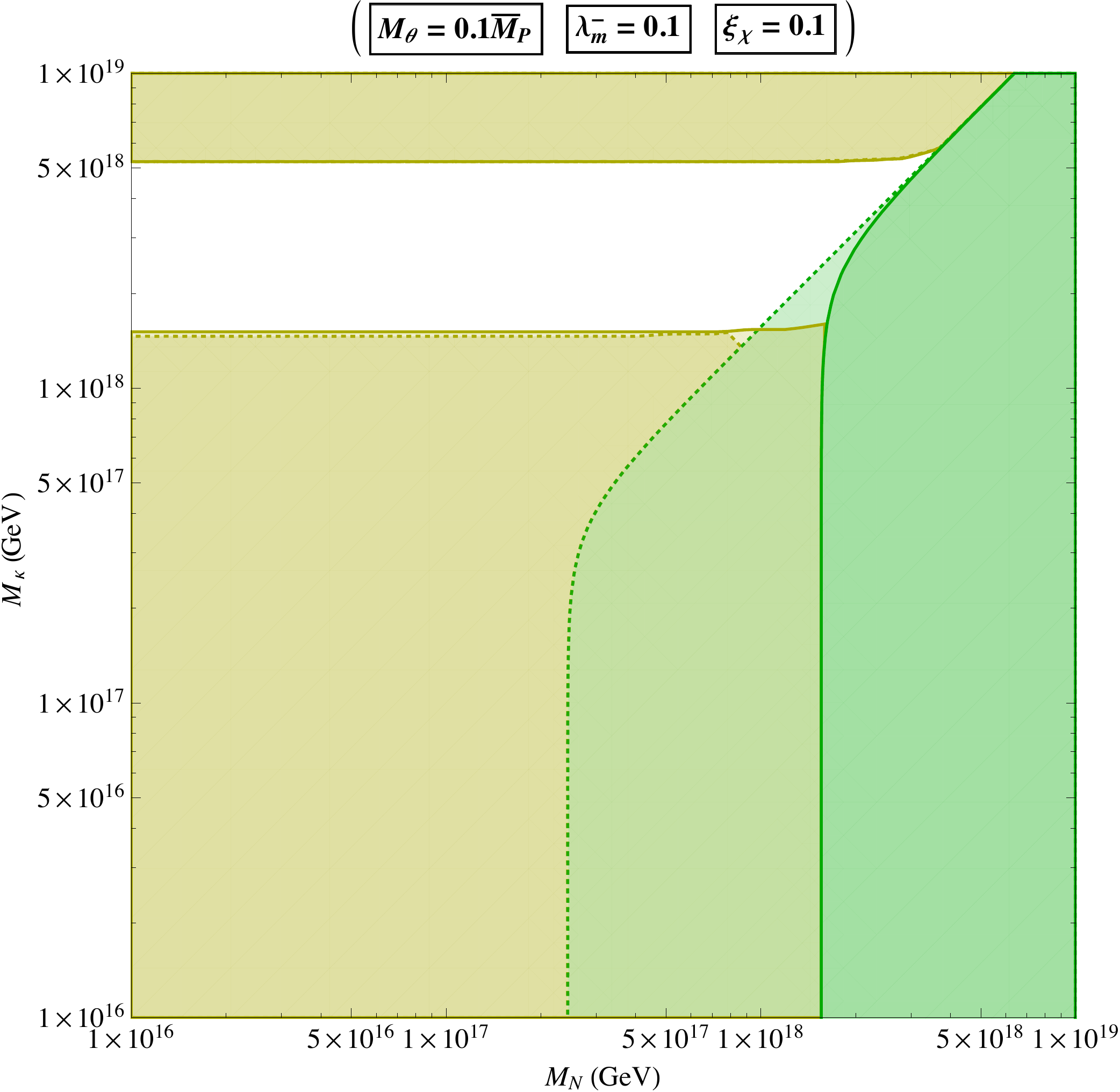}
\caption{Vacuum stability and perturbativity constraints (yellow region), displayed within the $M_{\mathcal N}-M_{\kappa}$~plane, for a cutoff energy $\mu \sim 250\bar{M}_\text{P}$, and the benchmark values $\lambda_{\chi} = \xi_{H} = 0.1$ and $M_{\theta} = 0.1\bar{M}_\text{P}$. All colored regions are excluded, and the panels exhibit the dependence on the varying $\lambda_{m}^{-}$ and $\xi_{\chi}$ from large to small (the panel for $\lambda_{m}^{-}=0.1$ and $\xi_{\chi}=1$ is entirely excluded, and hence, not shown). The solid lines correspond to the dark matter mass $M_{\chi} = \bar{M}_\text{P}$, whereas the dotted lines signify an additional exclusion for the lighter $M_{\chi} = 0.1\bar{M}_\text{P}$. The green region to the right side of the panels indicates the one-loop bounded from below constraint \eqref{massrel}. A universal value of the mixing angle $\tan\omega_{2} = 0.1$ has been selected for illustration.}
\label{ST2b}
\end{figure}

The constraints arising from demanding the stability of the vacuum and the perturbativity of the couplings up to at least the energy $\mu \sim 250\bar{M}_\text{P}$, as well as the one-loop bounded from below condition \eqref{massrel}, are plotted in Figs.~\ref{ST1a}-\ref{ST2b} within the $M_{\mathcal N}-M_{\kappa}$~plane, for various (large and small) benchmark choices of the remaining free parameters. A universal value of the mixing angle, $\tan\omega_{2} = 0.1$, has been applied throughout for illustration. A dependence on the exact value of this parameter is relatively mild for most of its range, with larger mixing angles allowing for heavier $M_{\kappa}$.

Fig.~\ref{ST1a} depicts the mentioned constraints for the larger benchmarks $\lambda_{\chi} = \xi_{H} = 1$ and $M_{\theta} = 2\bar{M}_\text{P}$. The four panels correspond to the varying $\lambda_{m}^{-}$ and $\xi_{\chi}$ from large to small. A heavy $M_{\kappa}$ drives the scalar couplings swiftly toward a Landau pole before the cutoff is reached; therefore, the perturbativity condition implies an upper limit on this mass parameter. For the larger $\xi_{\chi}$ (left column), it is the $\lambda_{\chi}$~quartic coupling which encounters a Landau pole, whereas, for the smaller $\xi_{\chi}$ (right column), this occurs for $\lambda_{\phi}$. On the other hand, heavier right-handed Majorana neutrinos result in a destabilization of the potential below the cutoff, via their large negative fermionic contributions to the $\beta$-functions. For a given $M_{\theta}$ and $M_{\chi}$, this necessitates a large enough positive bosonic contribution from the $\kappa$~scalar, in order to counterbalance the destabilizing fermionic effects. As a result, the vacuum stability condition imposes a lower limit on $M_{\kappa}$. Although the $M_{\kappa}$~upper limit is practically independent of the dark matter mass, $M_{\chi}$, its lower limit is noticeably affected by this parameter; in particular, a lighter $M_{\chi}$ provides a less positive bosonic contribution to the $\beta$-functions, resulting in a more stringent lower bound on $M_{\kappa}$ for a heavier $M_{\mathcal N}$. Moreover, given the large $M_{\theta}$, as well as $\lambda_{\chi}$ and $\xi_{H}$ in this figure, the $\beta$-functions are essentially dominated by the gravitational contribution of $f_{2}^{2}$, leaving a moderate dependence on the remaining parameters. The same observation is also valid for the depicted constraints from the one-loop bounded from below condition \eqref{massrel}.

In contrast, the effect of lowering $M_{\theta}$ to $ 0.1\bar{M}_\text{P}$, while keeping the remaining aforementioned benchmarks unaltered, is exhibited in Fig.~\ref{ST1b}, where, once more, the panels correspond to the varying $\lambda_{m}^{-}$ and $\xi_{\chi}$ from large to small. In this case, given the less prominent contribution of the LW~graviton, a dark matter mass dependence of the condition \eqref{massrel} is far more pronounced, as evident from the panels. This condition immediately excludes large values of $M_{\mathcal N}$, taming their negative fermionic contributions to the scalar couplings' $\beta$-functions. A large $\lambda_{m}^{-}$ (top row) provides, then, an adequate positive counterbalance within the $\beta$-functions, guaranteeing the stability of the vacuum up to the cutoff. Consequently, the $M_{\kappa}$~lower bound resides at much smaller values, outside the displayed region. Nonetheless, for a small $\lambda_{m}^{-}$ (bottom row), a heavy enough $M_{\kappa}$ becomes, once more, necessary and a lower limit on this mass parameter is imposed by the stability of the vacuum; thereby, significantly shrinking the viable region. As in Fig.~\ref{ST1a}, the $M_{\kappa}$~upper limits are determined by the perturbativity of $\lambda_{\chi}$ (for the large $\xi_{\chi}$) and $\lambda_{\phi}$ (for the small $\xi_{\chi}$). A dependence on the dark matter mass is virtually negligible for these $M_{\kappa}$~upper and lower limits.

In Fig.~\ref{ST2a}, we examine the viable regions of the parameter space for the smaller benchmarks $\lambda_{\chi} = \xi_{H} = 0.1$ and the larger $M_{\theta} = 2\bar{M}_\text{P}$, with the panels corresponding to the varying $\lambda_{m}^{-}$ and $\xi_{\chi}$ as before. In contrast with the case plotted in Fig.~\ref{ST1a}, the smaller values of $\lambda_{\chi}$ and $\xi_{H}$ largely tame the gravitational contributions of $f_{2}^{2}$ within the scalar couplings' $\beta$-functions, rendering a dependence on the remaining parameters more relevant. In particular, due to the smaller $\xi_{H}$, the running of the $\lambda_{\phi}$~coupling is much less prominently affected by the gravitational contributions, and the $M_{\kappa}$~upper limit in all panels is exclusively determined by the $\lambda_{\chi}$~Landau pole. As in Fig.~\ref{ST1b}, a larger $\lambda_{m}^{-}$ (upper row) provides adequate positive contributions to counterbalance the negative fermionic effects of the smaller values of $M_{\mathcal N}$, whereas for the heavier right-handed Majorana neutrinos additional bosonic contributions from the scalar graviton become essential. This leads to a $M_{\kappa}$~lower limit for larger values of $M_{\mathcal N}$, with a noticeable dependence on the dark matter mass, in analogy with Fig.~\ref{ST1a}. On the other hand, a smaller $\lambda_{m}^{-}$ (lower row) cannot, on its own, compensate for the destabilizing fermionic contributions even for the lighter right-handed Majorana neutrinos, and, as in Fig.~\ref{ST1b}, the vacuum stability implies a lower bound on $M_{\kappa}$, significantly reducing the viable parameter space, practically independent of the mass of the dark matter.

Finally, in Fig.~\ref{ST2b}, we demonstrate the effect of lowering $M_{\theta}$ to $ 0.1\bar{M}_\text{P}$, while keeping the remaining benchmarks of Fig.~\ref{ST2a} unaltered, and varying $\lambda_{m}^{-}$ and $\xi_{\chi}$ from large to small as before. Once more, the reduced LW~graviton mass results in a more pronounced dependence of the condition \eqref{massrel} on the dark matter mass. In this case the bosonic gravitational $f_{2}^{2}$~contributions are subdominant within the $\beta$-functions due to the small $M_{\theta}$, as well as the suppression by the small $\lambda_{\chi}$ and $\xi_{H}$. The upper and lower bounds on $M_{\kappa}$ essentially follow the previous discussions pertaining to Figs.~\ref{ST1b}~and~\ref{ST2a}; however, with the noticeable difference that the parameter space for the small $\lambda_{m}^{-}$ and large $\xi_{\chi}$ is now completely excluded. This is attributed to the lack of adequate positive bosonic contributions within the $\beta$-functions of the scalar couplings to properly compensate for the negative contributions of the Majorana fermions. This results in the $M_{\kappa}$~lower limit (due to the vacuum stability) to surpass its upper limit (due to the perturbativity), therefore, excluding the entire region of the parameter space.

The aforementioned discussions in this section lead to the general observation that, for the cases where the gravitational contributions are sufficiently suppressed within the current framework, a smaller $\lambda_{m}^{-}$ significantly reduces the viable regions of the parameter space, mainly due to the vacuum stability requirement.

\section{Inflation}\label{Infl}

In our developed model, the sole relevant scalar degree of freedom along the flat direction with a non-vanishing (radiatively-induced) potential is the $\sigma$~boson. In this section we analyze the consequences of identifying this degree of freedom, in its canonical form, with the inflaton, whose potential is given by \eqref{V01}, and confront our predictions with the available cosmological data.

For this purpose, we consider the slow-roll inflation paradigm, and demand the inflaton potential, $V$ given in \eqref{V01}, to satisfy the slow-roll approximations,
\begin{equation} \label{sr_approx}
\dot{\sigma}_c^2/2 \ll V \ ,
\qquad \ddot{\sigma}_c \ll 3H\dot{\sigma}_c \ .
\end{equation}
To reflect these approximations, we introduce the slow-roll parameters
\begin{equation}\label{slowpa}
\epsilon \equiv \frac{\bar{M}_\text{P}^2}{2}\bfrac{V_{\sigma_c}}{V}^2  \ , \qquad
\eta \equiv \bar{M}_\text{P}^2 \, \frac{V_{\sigma_c\sigma_c}}{V} \ , \qquad
\xi^2 \equiv \bar{M}_\text{P}^4 \, {V_{\sigma_c} V_{\sigma_c\sigma_c\sigma_c}\over V^2} \ , 
\end{equation}
where, the field subscripts denote taking the appropriate derivative(s) of the potential with respect to the argument.
With the approximations in \eqref{sr_approx},
the background equations of motion are given by
\begin{equation}\label{sr-bgeq2}
3 \bar{M}_\text{P}^2 \, H^2 \simeq V \ , \qquad
3H\dot{\sigma}_c +V_{\sigma_c} \simeq 0 \ .
\end{equation}

As evident from the potential \eqref{V01}, the background dynamics of the $\sigma_c$~field is entirely governed by two parameters; namely, the mixing angle $\sin \omega_2$, and the overall mass combination $\mathcal M$ \eqref{Bmodel}. The number of $e$-foldings may, subsequently, be computed using the slow-roll equations of motion \eqref{sr-bgeq2},
\begin{equation}\label{efdN}
N = \int_{t}^{t_e} Hdt = -\int_{\sigma_{c,e}}^{\sigma_{c,i}}
 \frac{H}{\dot{\sigma_c}} d{\sigma_c}
\simeq \frac1{\bar{M}_\text{P}^2} \int_{\sigma_{c,e}}^{\sigma_{c,i}}
\frac{V}{V_{\sigma_c} }\,
d{\sigma_c} \ ,
\end{equation}
where, $\sigma_{c,i}$ and  $\sigma_{c,e}$ represent
the values of the $\sigma_{c}$~inflaton at the beginning and
at the end of the inflation, respectively. Given the inflaton potential \eqref{V01}, the integration in \eqref{efdN} may be performed analytically, yielding
\begin{equation}\label{N}
N 
= \cbrac{
\frac{3}{8 \sin^2 \omega_2}
\left[
{\rm Ei}\left(-\log \frac{\sin^2 \omega_2\,\sigma_{c}^2}{6 \bar{M}_\text{P}^2}\right)
-{\rm li}\left(\frac{\sin^2 \omega_2\,\sigma_{c}^2}{6 \bar{M}_\text{P}^2}\right)
\right]
+\frac{\sigma_{c}^2}{8 \bar{M}_\text{P}^2}} \Bigg|_{\sigma_{c,e}}^{\sigma_{c,i}} \ ,
\end{equation}
with ${\rm Ei}(z)$ the exponential integral (${\rm Ei}(z)=-\int_{-z}^{\infty}{e^{-t}\over t}dt$), and ${\rm li}(z)$ the logarithmic integral (${\rm li}(z)=\int_{0}^{z}\frac{1}{\log t} dt$).
Furthermore, the end of the inflation is characterized by the usual condition: $\epsilon|_{\sigma_c=\sigma_{c,e}} \sim 1$.

Other relevant inflationary quantities, such as the amplitude of the scalar perturbations, $A_s$,
the scalar spectral index, $n_s$, 
the tensor-to-scalar ratio, $r$, 
along with its running, $\alpha_s = d n_s / d \log k$,
are defined in the usual way
\begin{equation}\label{inflobs}
A_s =\frac{V_*}{24\pi^2\bar{M}_\text{P}^4 \,  \epsilon_*}\ , \qquad n_s-1 = -6\epsilon_*+2\eta_*\ , \qquad
r =16\epsilon_* \ , \qquad \alpha_s = 16 \epsilon_* \eta_* -24 \epsilon_*^2 -2 \xi_*^2 \ ,
\end{equation}
where, the subscripted asterisk indicates the value at the horizon crossing point, $\sigma_c=\sigma_{c,i}$.
Using the inflaton potential \eqref{V01} and the slow-roll parameter definitions \eqref{slowpa}, the above observables can be calculated in a straightforward manner, as functions of $\sin \omega_2$, $\mathcal M$, and the field value of $\sigma_{c,i}$ at the horizon exit.

As mentioned, the end of the inflation is characterized by requiring $\epsilon|_{\sigma_c=\sigma_{c,e}} \sim 1$. From this condition, one may compute the field value $\sigma_{c,e}$ at the end of the inflation as a function of the mixing angle, $\sin\omega_{2}$.\footnote{Note that within the definitions of the slow-roll parameters \eqref{slowpa}, the amplitude $\mathcal M$ of the potential cancels.} Using the analytical expression for $N$ \eqref{N}, the corresponding initial field value, $\sigma_{c,i}$, is subsequently determined, as a function of the mixing angle, for a given $e$-folding number. It is noteworthy that the initial and the final field values, as connected by the number of  $e$-foldings, may reside on either side of the VEV \eqref{vsig}, corresponding to the small and the large field inflation scenarios. Once the initial field value, $\sigma_{c,i}$, for either scenario is obtained, it may be utilized to compute the inflationary observables \eqref{inflobs}. A comparison of the model's predictions for these inflationary quantities with their observational values, thus, results in constraints on the framework's relevant free parameters.

\begin{figure}
\includegraphics[width=.7\textwidth]{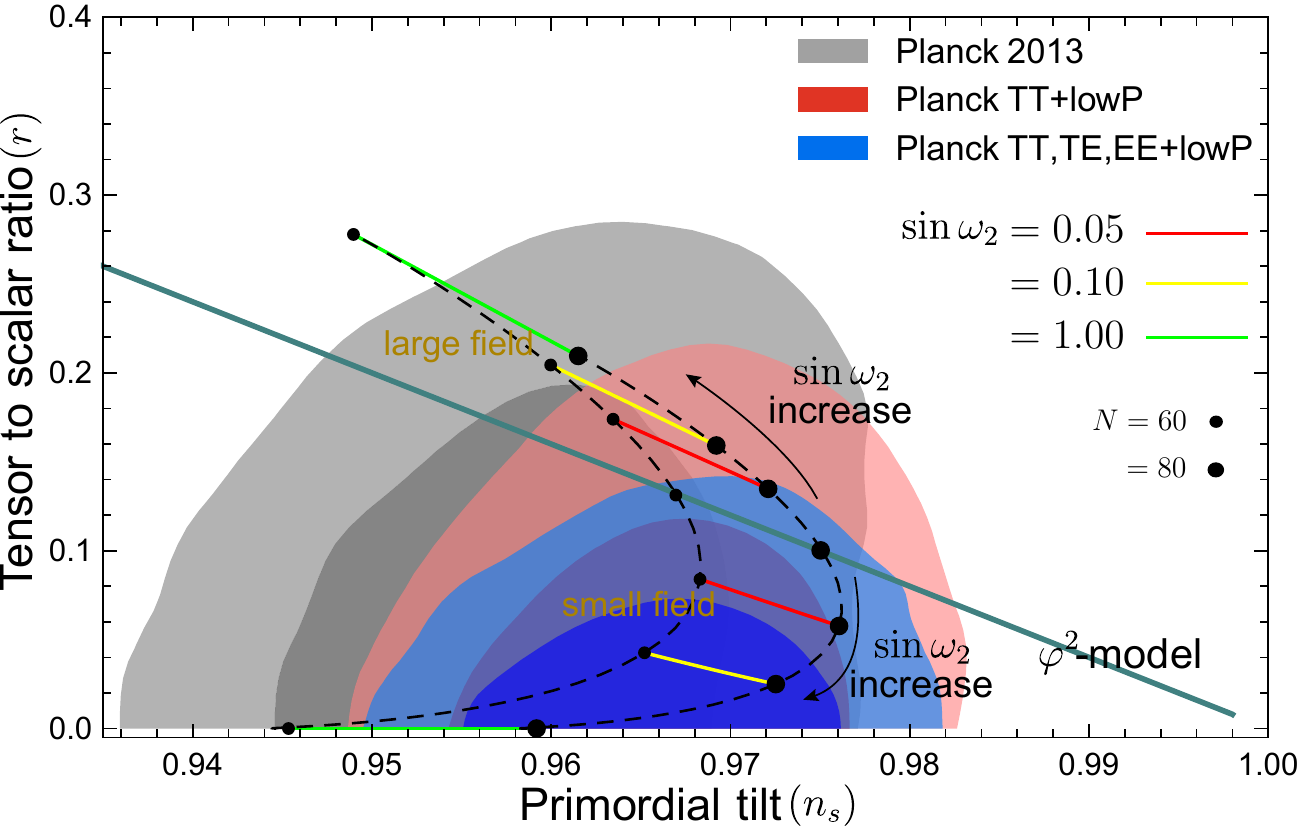}
\caption{The $n_s - r$~plane, incorporating several observational data sets by the Planck Collaboration \cite{Ade:2015lrj} at 68\%~C.L. (darker region) and at 95\%~C.L. (lighter region). Model's predictions for both the small and large field inflation scenarios are also depicted (dashed lines) for the $e$-folding numbers, $N=60,80$, and the full range of the mixing angle, $\sin\omega_{2}$. Near the $\sin\omega_{2} \to 0$ limit, the model reduces to the ordinary chaotic inflation scenario, indicated by the diagonal line labeled as ``$\varphi^{2}$-model'' within the plot.}
\label{fig:nsr}
\end{figure}

The constraints from several cosmological data sets, as published by the Planck Collaboration \cite{Ade:2015lrj}, are plotted within the $n_s - r$~panel, displayed in Fig.~\ref{fig:nsr}. In addition, the model's predictions for the small and large field inflation scenarios are superimposed for two given $e$-folding numbers, $N=60,80$, while covering the full range of the mixing angle, $\sin\omega_{2}$. It is interesting to observe that, in the limit $\sin\omega_{2} \to 0$, the inflaton potential \eqref{V01} near its minimum \eqref{vsig} reduces to
\begin{equation}
V(\sigma_c) \simeq 
{{\mathcal M}^4 \sin^2 \omega_2 \over 96 \pi^2 \bar{M}_\text{P}^2}
\left(\sigma_c-{\sqrt{6} \bar{M}_\text{P} \over \sin \omega_2}\right)^2 \ .
\end{equation}
This is the usual potential for the chaotic inflation scenario, from which one reads for the spectral index and the tensor-to-scalar ratio,
\begin{equation} 
(n_s, r)\Big |_{\sin\omega_{2}\to 0} \simeq \left(1-{4\over 1+ 2N}, {16\over 1+ 2N}\right) \ .
\end{equation}
Hence, one obtains the values $(n_s,r) = (0.967,0.132)$ for $N=60$, and $(n_s,r) = (0.975,0.099)$ for $N=80$ in this limit.
In the opposite limit, $\sin\omega_{2} \to 1$, the inflation can formally be driven at either side of the VEV, for small or large field values. For larger values of the mixing angle, it is, however, increasingly more difficult to reconcile the large field inflation scenario with the observational data, as the potential becomes too steep. Conversely, within the small field inflation scenario, the potential assumes a relatively flat shape, and the corresponding slow-roll parameter $\epsilon$ is extremely small; therefore, the spectral index (c.f.~\eqref{inflobs}) is mainly determined by $\eta$.
In the large mixing angle limit of this scenario, the leading behavior of the observables takes, subsequently, the approximate form
\begin{equation} 
(n_s, r)\Big |_{\sin\omega_{2}\to 1} \simeq \left(1-{3\over N}, 0\right) \ .
\end{equation}

Furthermore, we perform a statistical study of the model, utilizing the $\chi^2$~analysis.
For this purpose, we have employed the MCMC chains from the Planck Legacy archive data 
to obtain constraints on $(\log A_s, n_s, \alpha, r)$. This information can, subsequently, be employed to derive statistical constraints on the parameter space of the model.
To this end, we use the following two observational data sets
\begin{itemize}
\item Planck TT+lowP: Planck full mission temperature data, and
\item Planck TT, TE, EE+lowP: Planck full mission temperature data, combined with the Planck high-$\ell$ polarization.
\end{itemize}
The set of Planck TT+lowP data has the following mean values:
\begin{equation}\label{meanTT}
\left<\log A_s\right> = 3.12\,, \quad
\left<n_s\right> = 0.967\,, \quad
\left<\alpha\right> = -0.013\,, \quad
\left<r\right> = 0.063\ ,
\end{equation}
and the inverse covariance matrix of $(\log A_s, n_s, \alpha, r)$ is given by
\begin{equation}
\begin{pmatrix}\label{icovTT}
 874.956 & -2369.85 & 2312.25 & 156.566 \\
 -2369.85 & 29862. & -6773.72 & -885.52 \\
 2312.25 & -6773.72 & 20217.5 & 1359.07 \\
 156.566 & -885.52 & 1359.07 & 455.688 \\
\end{pmatrix} \, .
\end{equation}
For the Planck TT, TE, EE+lowP data set, the mean values of  $(\log A_s, n_s, \alpha, r)$
are:
\begin{equation}\label{meanTTTEEE}
\left<\log A_s\right> = 3.10\,, \quad
\left<n_s\right> = 0.964\,, \quad
\left<\alpha\right> = -0.008\,, \quad
\left<r\right> = 0.056\ ,
\end{equation}
with the inverse covariance matrix
\begin{equation}
\begin{pmatrix}\label{icovTTTEEE}
 1077.1 & -2892.56 & 2273.95 & 151.824 \\
 -2892.56 & 51718.1 & -12064.4 & -1162. \\
 2273.95 & -12064.4 & 25074.3 & 1517.58 \\
 151.824 & -1162. & 1517.58 & 570.099 \\
\end{pmatrix} \,.
\end{equation}

\begin{table*}
\begin{center}
\caption{Fitting results for the relevant parameters, with the $1\sigma$~C.L., for two Planck data sets.}
\label{table:result}
\begin{tabular}{cccc}
\hline\hline 
& Planck TT + lowP  && Planck TT,TE,EE + lowP\\
\hline
$\log_{10} (\mathcal M / \bar{M}_\text{P})$   & $-1.37^{+0.35}_{-0.92}$ && $-1.39^{+0.29}_{-0.87}$ \\

$\log_{10} \sin \omega_2$ & $-0.82^{+0.80}_{-0.63}$ && $-0.90^{+0.81}_{-0.41}$  \\

$N$ & $72.04^{+34.11}_{-16.31}$ && $69.70^{+33.12}_{-11.00}$ \\

\hline 
$\chi^{2}_{\rm min}$  & 1.78 & &  2.57  \\
\hline
\end{tabular}
\end{center}
\end{table*}

\begin{figure}
\scalebox{0.45}[0.45]{
\includegraphics{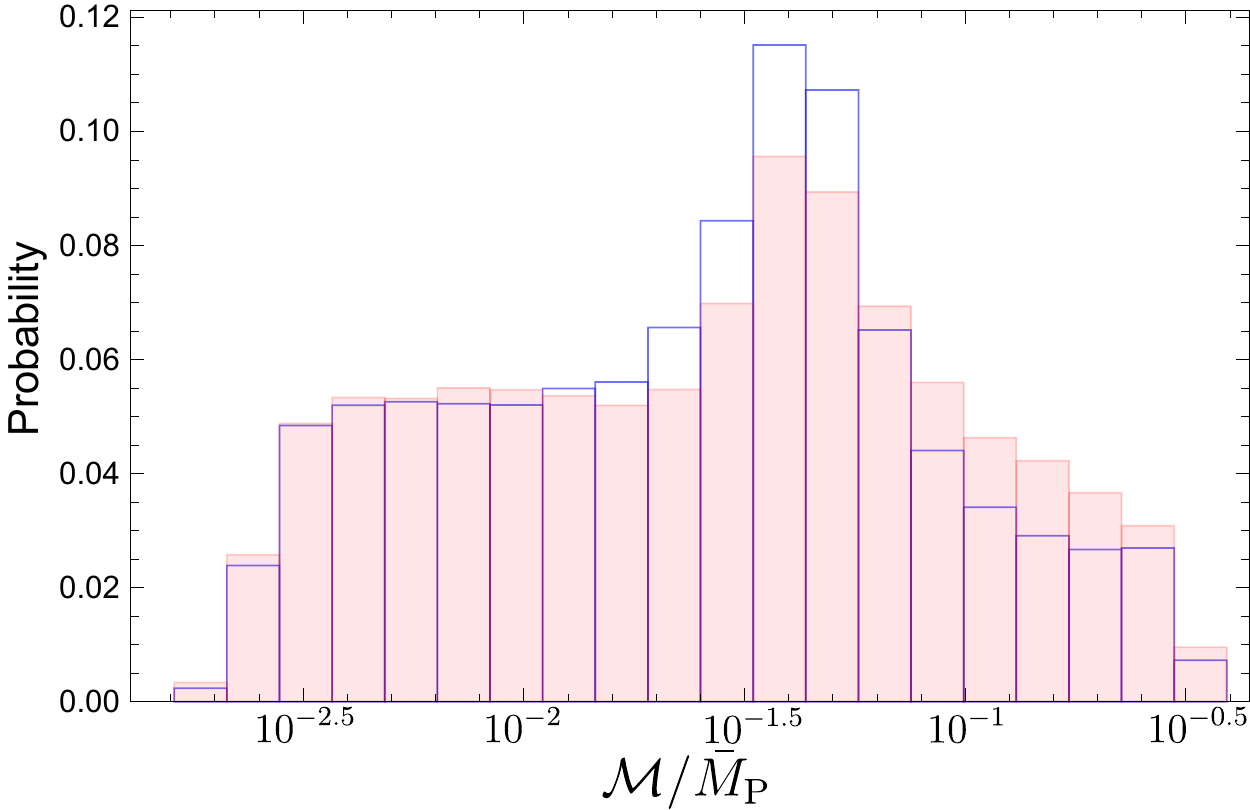}
\includegraphics{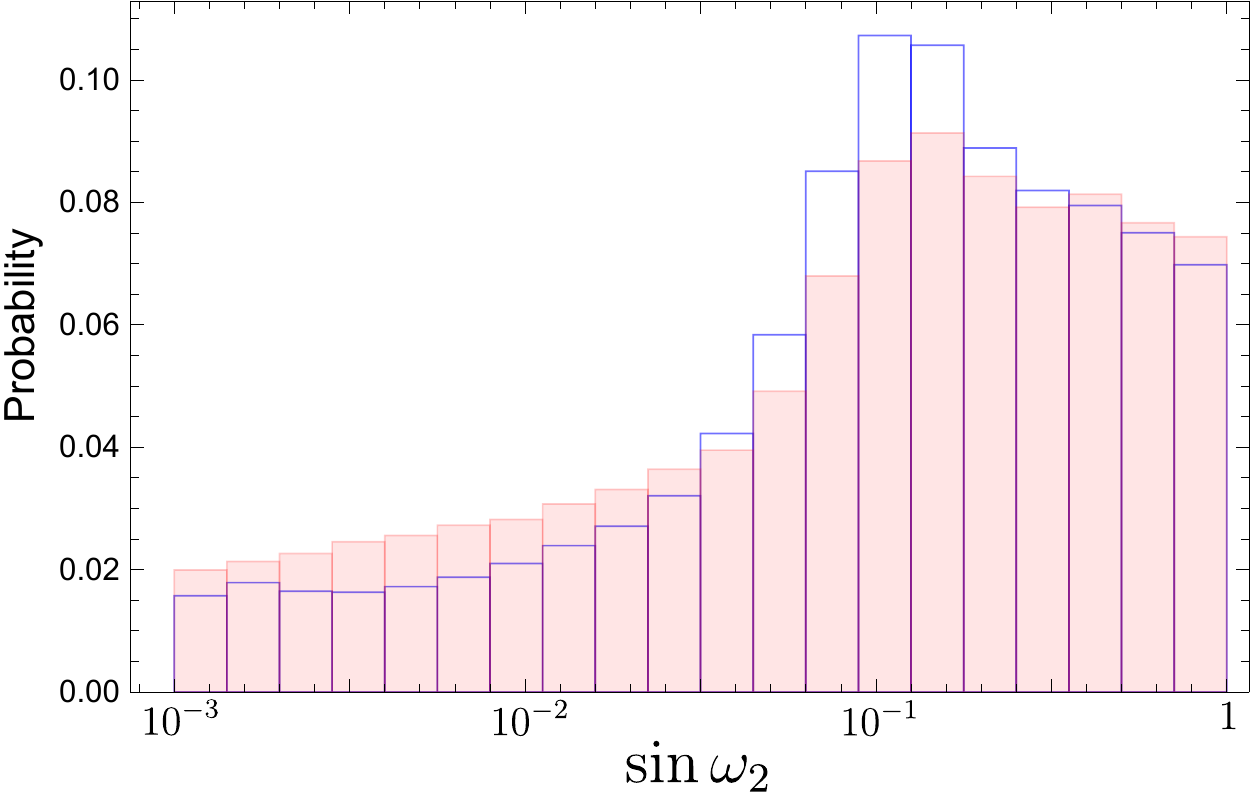}
\includegraphics{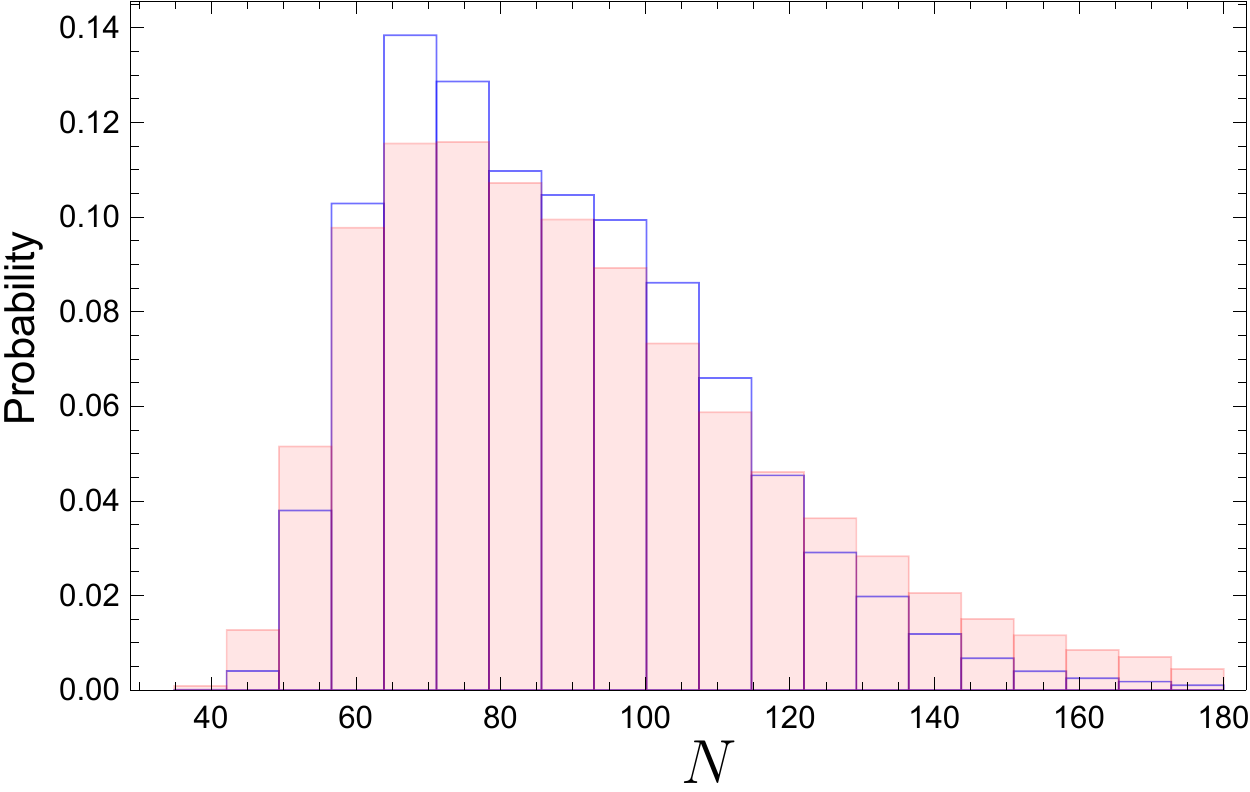}
}
\caption{The one dimensional marginalized probability distributions of
${\cal M}/\bar{M}_\text{P}$ (left), $\sin \omega_2$ (middle), and $N$ (right),
for Planck TT+lowP~(red) and Planck TT, TE, EE+lowP~(blue) data.
}
\label{fig:like}
\end{figure}

It is straightforward to derive statistical constraints on the model parameters
by computing the corresponding $\chi^2$ for each data set.
Defining $x_1 = \log A_s$, $x_2 = n_s$, $x_3 = \alpha$ and $x_4 = r$, the $\chi^2$~parameters are given according to
\begin{align}
& \chi^2 = \Delta x_i \, {\rm Cov}^{-1}(x_i,x_j) \, \Delta x_j \,, \\
& \Delta x_i = x_i - x_i^{\rm data} \,,
\end{align}
where, ${\rm Cov}^{-1}$ is the inverse of the covariance matrix, given in \eqref{icovTT} and \eqref{icovTTTEEE},
and $x_i^{\rm data}$ are the mean values from \eqref{meanTT} and \eqref{meanTTTEEE}.
Starting with the input parameters 
$\left(\log_{10} ({\cal M}/\bar{M}_\text{P}),\, \log_{10} \sin \omega_2,\, N\right)$, 
one can calculate the $\chi^2$ value in this 3-parameter space.
The input parameters are taken in the following priors:
$\log_{10} ({\cal M}/\bar{M}_\text{P}) = [-3, 0]$, $\log_{10} \sin \omega_2 = [-3,0]$
and $N = [20, 180]$.
To obtain the likelihood distributions for the model parameters, we use the
Markov chain Monte Carlo method, based on the Metropolis-Hastings
algorithm, which allows for a random exploration of the parameter space favored by the
observational data \cite{MCMC}.
The method makes decisions for accepting or rejecting
a randomly chosen chain element via the probability function
$P \propto \exp(-\chi^2/2)$.
Additionally, during the MCMC analysis, we use a simple diagnostic to test the convergence
of the MCMC chain; i.e., the means estimated from the first (after the burning process)
and the last 10\% of the chain should be approximately equal to one another,
if the chain has converged (see e.g. Appendix~B in \cite{Abrahamse-etal-2008}).

The results are displayed in Tab.~\ref{table:result} for a summary of the parameter constraints 
with the mean, as well as the $1\sigma$~C.L., 
and in Fig.~\ref{fig:like} for the marginalized one-dimensional probability distributions of the individual parameters. One observes that the second data set (temperature data combined with the polarization data) yields more restricted results as compared with the first data set (temperature data alone). This is perhaps not surprising, given the more stringent nature of the Planck TT, TE, EE+lowP data set. 
The best-fit locations in the parameter space are determined as
\begin{align}
(\log_{10} ({\cal M}/\bar{M}_\text{P}),\, \log_{10} \sin \omega_2,\, N)
=(-1.43, -0.99, 62.95) \,,
\end{align}
for the Planck TT+lowP data set, and
\begin{align}
(\log_{10} ({\cal M}/\bar{M}_\text{P}),\, \log_{10} \sin \omega_2,\, N)
=(-1.42, -1.01, 65.05) \,,
\end{align}
for the Planck TT, TE, EE+lowP data set.
One deduces from this statistical analysis that the most likely value of the mixing angle compatible with the observation is near $\sin \omega_2\sim 0.1$, whereas the corresponding most favorable value of the mass combination lies around $\mathcal M \sim10^{-1.4} \bar{M}_\text{P} \sim 10^{17}$~GeV, with a most likely $e$-folding number in the vicinity of $N\sim 65$.

\begin{figure}
\includegraphics[width=.48\textwidth]{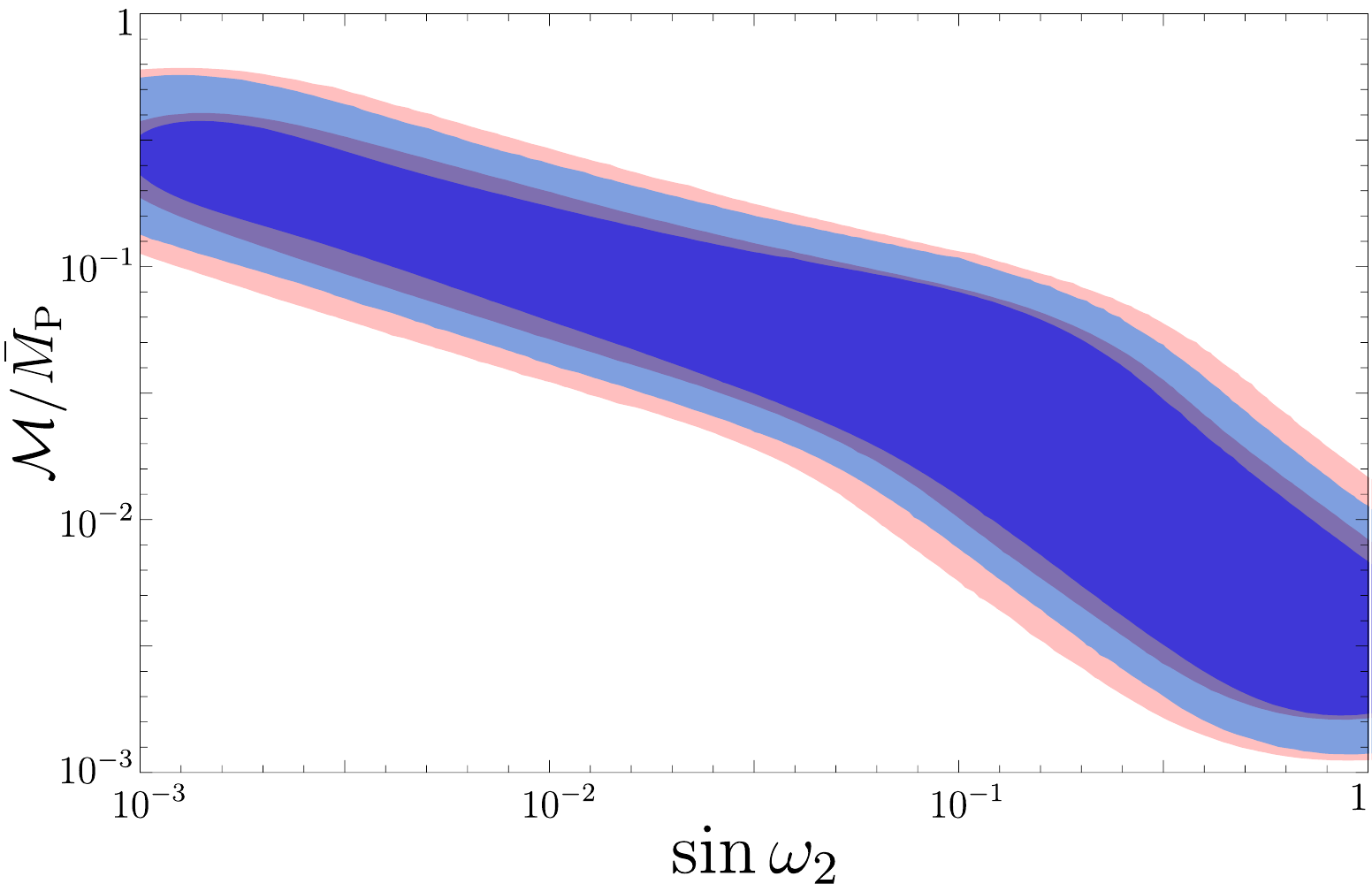}
\includegraphics[width=.5\textwidth]{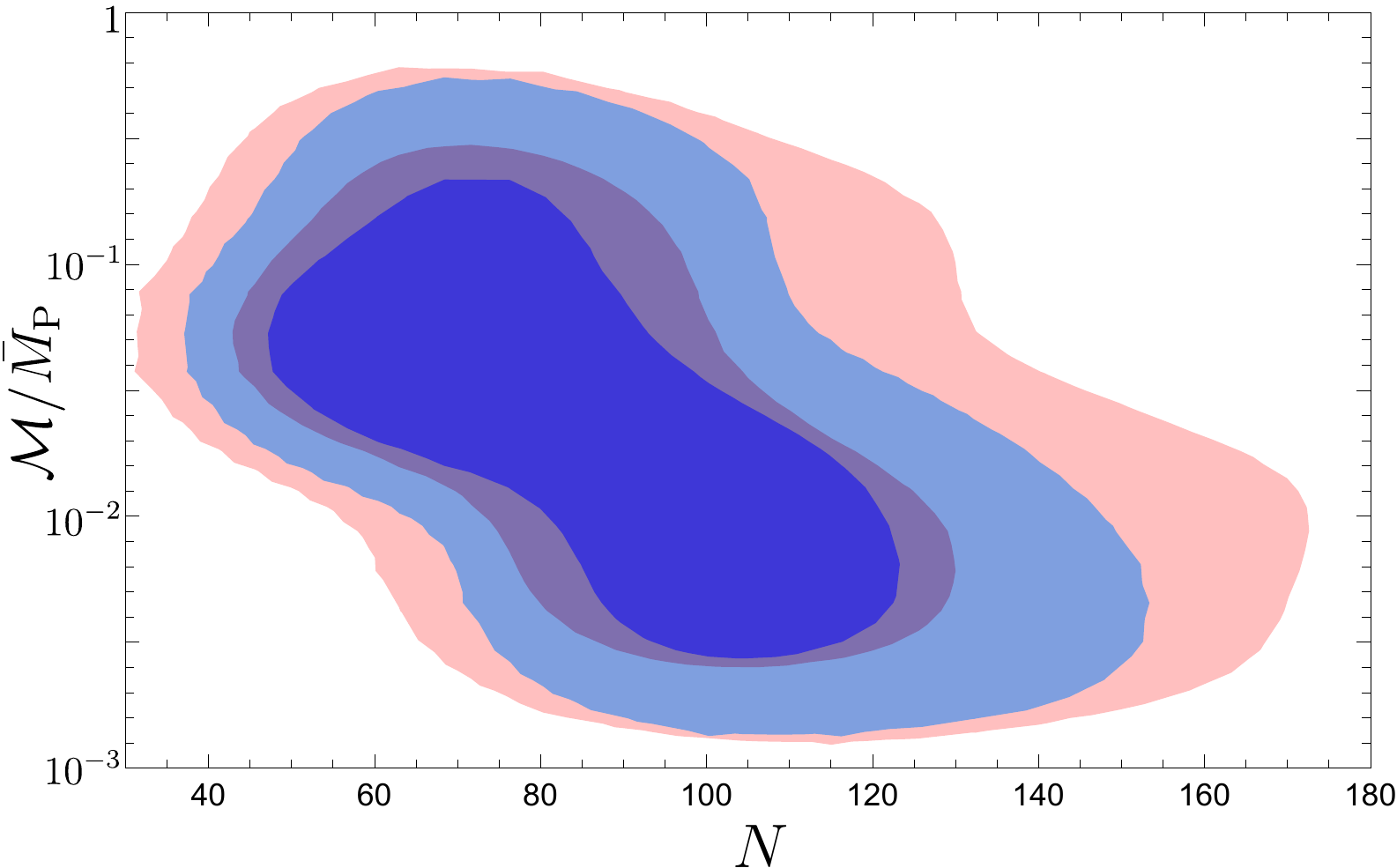}
\caption{Marginalized probability distributions displayed within the $\sin \omega_2- \mathcal M /\bar{M}_\text{P}$~plane (left) and the $N- \mathcal M /\bar{M}_\text{P}$~plane (right), with 68\%~C.L. (darker inner region) and 95\%~C.L. (lighter outer region), obtained by the parameter estimation using the Planck TT+lowP~(red) and Planck TT, TE, EE+lowP~(blue) data sets.}
\label{fig:cnt1}
\end{figure}

Thus, based on the aforementioned discussion, we conclude that the small field inflation scenario is fully compatible with the observational data within the current framework. This fact is further illustrated in the panels of Fig.~\ref{fig:cnt1}, where the marginalized probability distributions are shown within the $\sin \omega_2- \mathcal M /\bar{M}_\text{P}$ and the $N- \mathcal M /\bar{M}_\text{P}$~planes, for the Planck TT+lowP and Planck TT, TE, EE+lowP data sets. The contours constrain the range of the relevant parameter space at 68\%~C.L. and 95\%~C.L.; specifically, one observes that the mass combination \eqref{Bmodel} is confined to $0.001 \lesssim \mathcal M /\bar{M}_\text{P} \lesssim 1$. This range is predominantly dictated by the amplitude of the scalar perturbations, $A_s$, which is a function of $\mathcal M$ (c.f.~\eqref{inflobs}). It is evident that the observational data favor smaller values of $\mathcal M$ for the larger mixing angles and $e$-folding numbers. 

\section{Reheating and dark matter}\label{DM}

Toward the end of the inflation, the inflaton field, $\sigma_{c}$, oscillates around the minimum of the potential \eqref{V01}, and quickly settles in its nonzero VEV \eqref{vsig}. During this oscillation, the Universe is dominated by the matter-like phase. The inflaton field finally decays into the lighter particles and reheats the Universe. Assuming that the reheating is mainly due to the perturbative decay of the inflaton field, the dominant decay occurs when the inflaton decay rate is comparable in magnitude to the Hubble expansion rate, $\Gamma_{\sigma_{c}} \sim H (\Treh)$. Using the expression for the Hubble expansion rate as a function of temperature \cite{DMrelic,Kolb:book}
\begin{equation}\label{HT}
H(T) = \frac{\pi}{3}\sqrt{\frac{g_{\text{rad}}}{10}} \, \frac{T^{2}}{\bar{M}_\text{P}} \ ,
\end{equation}
one may, subsequently, compute the reheating temperature, $\Treh$, in terms of the inflaton decay rate $\Gamma_{\sigma_{c}}$
\begin{equation} \label{Treh}
\Treh \simeq \bfrac{90}{\pi^2 \, g_{\text{rad}}}^{1/4}\sqrt{\bar{M}_\text{P}\, \Gamma_{\sigma_{c}}} \ .
\end{equation}
Here, $g_{\text{rad}} = 106.75 + N_{s}$ represents the effective number of thermalized relativistic degrees of freedom at $\Treh$, with $N_{s}$ the number of the non-SM scalar contributions. The pseudoscalar dark matter, $\chi$, constitutes a WIMPzilla candidate \cite{Chung:1998rq}, with a mass much larger than the reheating temperature, which becomes non-relativistic at the time of the reheating. As the inflaton mass is also larger than the reheating temperature, hence, only a sufficiently light scalar graviton, $\kappa$, can thermalize and contribute to $g_{\text{rad}}$; i.e., $g_{\text{rad}} = 107.75$ for $M_{\kappa} \ll \Treh$.

In order to calculate the total decay rate of the inflaton, $\Gamma_{\sigma_{c}}$, one must identify the various (kinematically) allowed decay channels within the current framework. This can be achieved by re-expressing the field basis potential along the flat direction (potential \eqref{VE} with the conditions \eqref{lmp} and \eqref{leta} implemented) in terms of the physical scalar mass eigenstates \eqref{scaldiag}. Using the mass definitions \eqref{treemass1} and \eqref{treemass2}, one then obtains the following scalar trilinear couplings\footnote{As the inflaton resides in the minimum of the potential during the decay, its non-canonical definition coincides with the canonical definition.}
\begin{equation}\label{Infltriscal}
i\lambda_{\sigma \chi \chi} = i\sqrt{\frac{2}{3}} \frac{M_{\chi}^{2}}{\bar{M}_\text{P}} \sin \omega_{2} \ , \qquad i\lambda_{\sigma \kappa \kappa} = i\sqrt{\frac{2}{3}} \frac{M_{\kappa}^{2}}{\bar{M}_\text{P}} \sin \omega_{2} \ , \qquad i\lambda_{\sigma h h} = 2i \frac{M_{h}^{2}}{v_{\phi}} \sin \omega_{1} \cos \omega_{2} \ .
\end{equation}
The $\sigma \to \chi \chi$ and $\sigma\to \kappa \kappa$ decay channels are kinematically open for the masses $M_{\chi},M_{\kappa} \leq m_{\sigma}/2$, respectively.

Since the inflaton field has a component along the SM Higgs boson $\phi$ (c.f. \eqref{sigmarel2}), its coupling to the SM vector bosons and fermions is proportional to the usual SM values, albeit with a mixing angle suppression factor
\begin{equation}\label{InflSM}
i\lambda_{\sigma i j} = i\lambda_{\phi i j} \,  \sin \omega_{1} \cos \omega_{2} \ ,
\end{equation}
where, $i\lambda_{\phi i j}$ denotes the SM value for the coupling of the SM Higgs boson, $\phi$, to the $i$ and $j$~fermions or vector bosons. For our purposes, it is sufficient to consider only the heavy SM states; i.e., $ij \in \cbrac{\bar{t}t, W^{+}W^{-}, ZZ}$.

Finally, the inflaton field may decay into a pair of right-handed Majorana neutrinos. This coupling is proportional to the corresponding Yukawa coupling with the $\eta$~scalar \eqref{LRHN}, suppressed by the appropriate mixing angle factor from \eqref{sigmarel2},
\begin{equation}\label{InflRHN}
i\lambda_{\sigma \mathcal N \mathcal N} = i\lambda_{\eta \mathcal N \mathcal N} \,  \cos \omega_{1} \cos \omega_{2} = -\frac{i}{\sqrt 6} \frac{M_{\mathcal N}}{\bar{M}_\text{P}} \sin \omega_{2} \ .
\end{equation}
In analogy with the $\chi$ and $\kappa$~states, the $\sigma \to \mathcal N \mathcal N$~decay channel may be kinematically available depending on the mass $M_{\mathcal N}$; i.e., for $M_{\mathcal N} \leq m_{\sigma}/2$.

Having determined all relevant decay channels, the inflaton decay rate into pairs of each particle species can be computed using the standard methods, and we quote the results for completeness
\begin{equation}\label{decays}
\begin{split}
\Gamma_{S} =&\, \frac{f_{S}}{16\pi \, m_{\sigma}}\left(1-\frac{4 M_S^2}{m_\sigma^2}\right)^{1/2} |\lambda_{S}|^{2}  \ , \qquad \Gamma_{F} = \frac{f_{F} \, m_{\sigma}}{8\pi}\left(1-\frac{4 M_F^2}{m_\sigma^2}\right)^{3/2} |\lambda_{F}|^{2}  \ , \\
&\Gamma_{V} = \frac{f_{V}}{16\pi \, m_{\sigma}}\left(1-\frac{4 M_V^2}{m_\sigma^2}\right)^{1/2} \tbrac{2+\pbrac{1-\frac{m_{\sigma}^2}{2M_V^2}}^2} |\lambda_{V}|^{2}  \ ,
\end{split}
\end{equation}
with $S,F,V$ denoting the scalar, fermion, and vector boson species, respectively, and the couplings $\lambda_{S,F,V}$ as defined in \eqref{Infltriscal}--\eqref{InflRHN}. The statistical factors take into account the remaining internal degrees of freedom associated with each particle species, as well as a double-counting for identical particles in the final state; specifically, for identical scalars in the final state $f_{S}= 1/2$, colored top quarks $f_{t} = 3$, three flavors of the right-handed Majorana neutrinos $f_{\mathcal N} = 3/2$, and the vector bosons $f_{W} = 1$ and $f_{Z} = 1/2$. The total (dominant) decay rate of the inflaton is, thus, obtained by the sum of the aforementioned decay rates
\begin{equation}\label{decaytot}
\Gamma_{\sigma_{c}} = \Gamma_{\sigma\chi\chi} + \Gamma_{\sigma\kappa\kappa} + \Gamma_{\sigma\mathcal N \mathcal N} + \Gamma_{\sigma hh} + \Gamma_{\sigma \bar t t} + \Gamma_{\sigma W^{+}W^{-}} + \Gamma_{\sigma ZZ} \ ,
\end{equation}
which, is a function of the first five free parameters in \eqref{freepar}; i.e., the mixing angle $\omega_{2}$, and the four masses. Substituting \eqref{decaytot} into \eqref{Treh}, consequently, yields the reheating temperature as a function of these five input parameters.

Within the current framework, the stable supermassive $\chi$~pseudoscalar can serve as a WIMPzilla dark matter candidate. Such supermassive particles may be pair-produced early on during the reheating epoch, when the maximum temperature is much higher than the characteristic reheating temperature \eqref{Treh}. As a consequence, WIMPzillas with $\mathcal O (2000\, \Treh)$~masses may be produced by the scattering of the thermalized decay products of the inflaton, while exhibiting the correct dark matter relic abundance \cite{Chung:1998rq}. Hence, such WIMPzillas are highly non-relativistic once the characteristic reheating temperature, $\Treh$, is eventually reached.

An analytical estimate for the relic abundance of these supermassive dark matter candidates, in the slow reheating process, reads \cite{Chung:1998rq}
\begin{equation}\label{relic}
\Omega_{\chi} h^{2} \simeq M_{\chi}^{2} \, \langle\sigma v \rangle_{\text{ann}} \pbrac{\frac{g_{\text{rad}}}{200}}^{-3/2} \pbrac{\frac{2000\, \Treh}{M_{\chi}}}^{7} \ ,
\end{equation}
with $\langle\sigma v \rangle_{\text{ann}}$ representing the thermally averaged total cross section for the pair-annihilation of WIMPzillas, taking into account the M$\o$ller flux factor. For heavy non-relativistic WIMPzillas, the center of mass energy of the pair-annihilation process simply amounts to $s_{\text{CM}} \simeq 4M_{\chi}^{2}$; therefore, one may approximate the leading order thermally averaged total cross section as \cite{DMrelic,Kolb:book}
\begin{equation}\label{tacs}
\langle\sigma v \rangle_{\text{ann}} \simeq \left.\frac{E_1E_2\, v_{12}\, \sigma_\text{ann}}{M_\chi^2}\right|_{s_{\text{CM}}=4M_\chi^2} \ ,
\end{equation}
where, $E_{1,2}$ denotes the energy of the annihilating $\chi$~pseudoscalars, $v_{12}$ is their relative velocity, and $\sigma_\text{ann}$ represents the total $2\to 2$ scattering cross section.

\begin{figure}
\includegraphics[width=.9\textwidth]{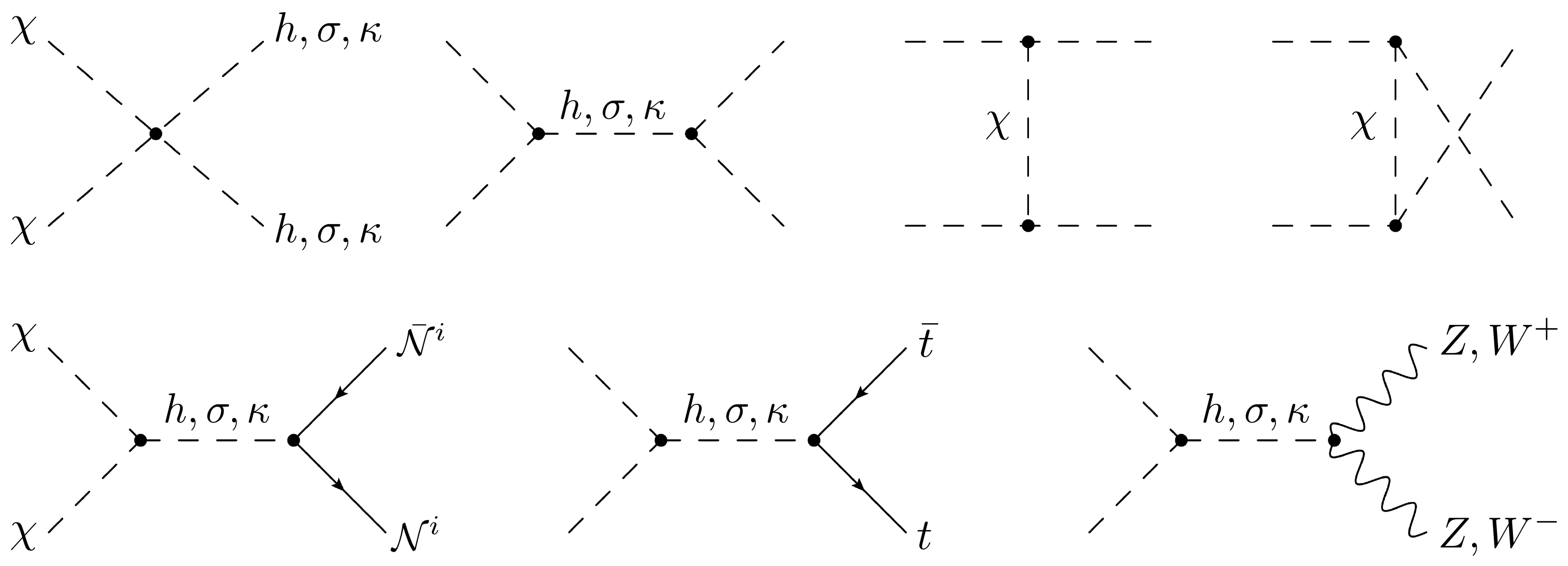}
\caption{Dominant final state products of the $\chi$~WIMPzilla pair-annihilation. Various scalar final states (including non-identical pairs) are obtained via the contact interactions, as well as the $s,t,u$-channel processes (diagrams in the top row from left to right), whereas the fermionic and vector boson final states only proceed through the $s$-channel (bottom row). The relevant mediator(s) are indicated within each diagram. Note that the (tree-level) $\sigma\sigma h$, $\sigma\sigma\kappa$, $\sigma\sigma\sigma$, and $\sigma\kappa h$~couplings are absent due to the classical scale symmetry.}
\label{DMann}
\end{figure}

Fig.~\ref{DMann} displays the relevant diagrams for the (kinematically available) dominant final state products of the WIMPzilla pair annihilation, where, once more only the heavy SM degrees of freedom are taken into account, in addition to the non-SM states. All relevant trilinear and quartic scalar couplings can be extracted from the potential along the flat direction, as previously elaborated (c.f. the explanation above \eqref{Infltriscal}), in a straightforward manner.\footnote{Given the relatively complicated form of some of these couplings in the mass eigenstate basis, we avoid explicitly quoting them for brevity.} The couplings of the physical scalars to the fermions and gauge bosons are proportional to those of their corresponding field basis definitions with the appropriate mixing angle suppression factors. Computing the total $2\to 2$ scattering cross section using the standard methods, one arrives at the thermally averaged total cross section \eqref{tacs}, which is formally a function of all of the model's input parameters \eqref{freepar}, except for the dark matter quartic coupling, $\lambda_{\chi}$. Inserting this expression into \eqref{relic}, one can exploit the available inputs, in order to determine the parameter space giving rise to the observed value of the dark matter relic abundance, $\Omega_{\chi} h^{2} = 0.1199 \pm 0.0022$ \cite{Ade:2015lrj}.

\begin{figure}
\includegraphics[width=.49\textwidth]{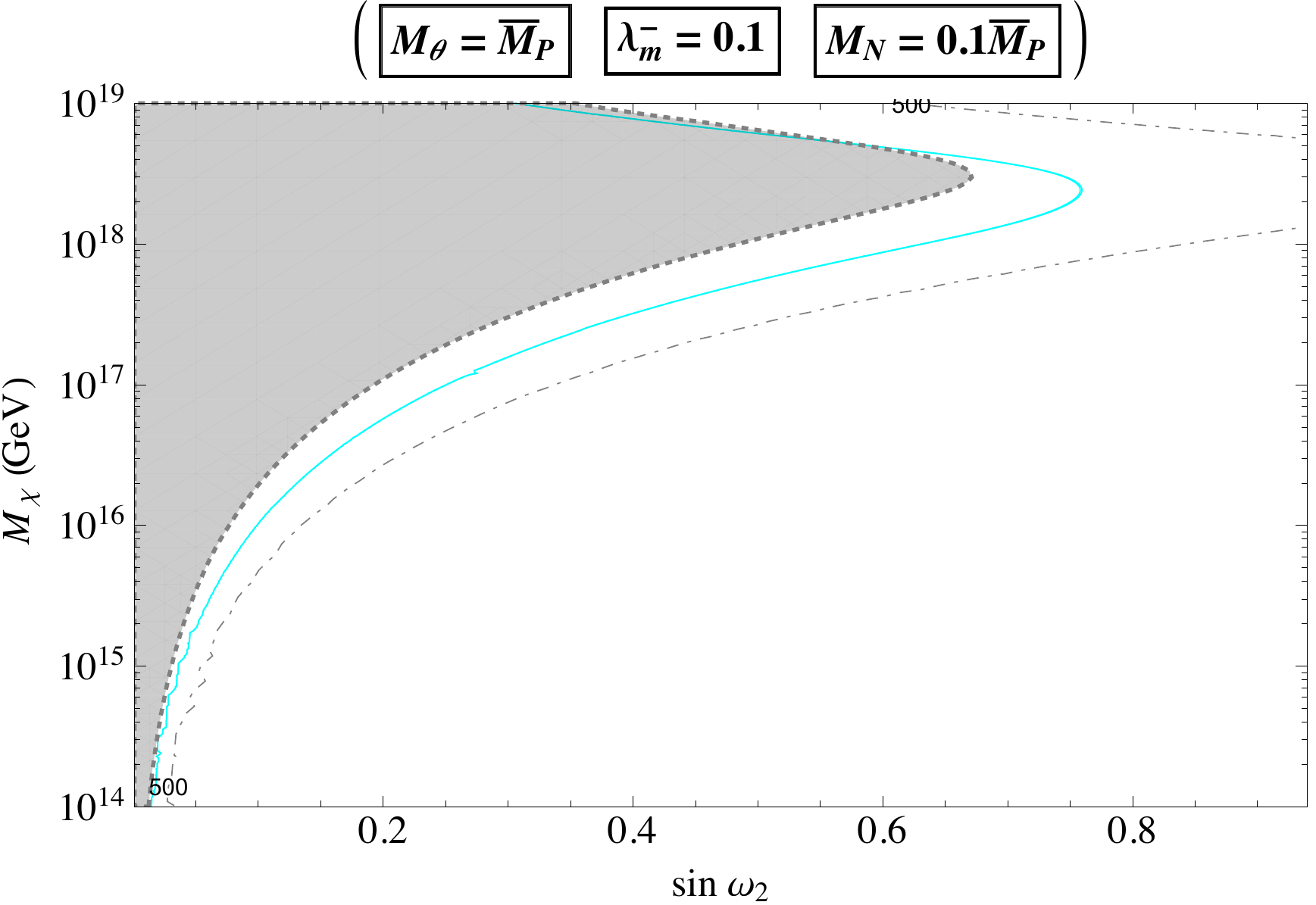}
\includegraphics[width=.49\textwidth]{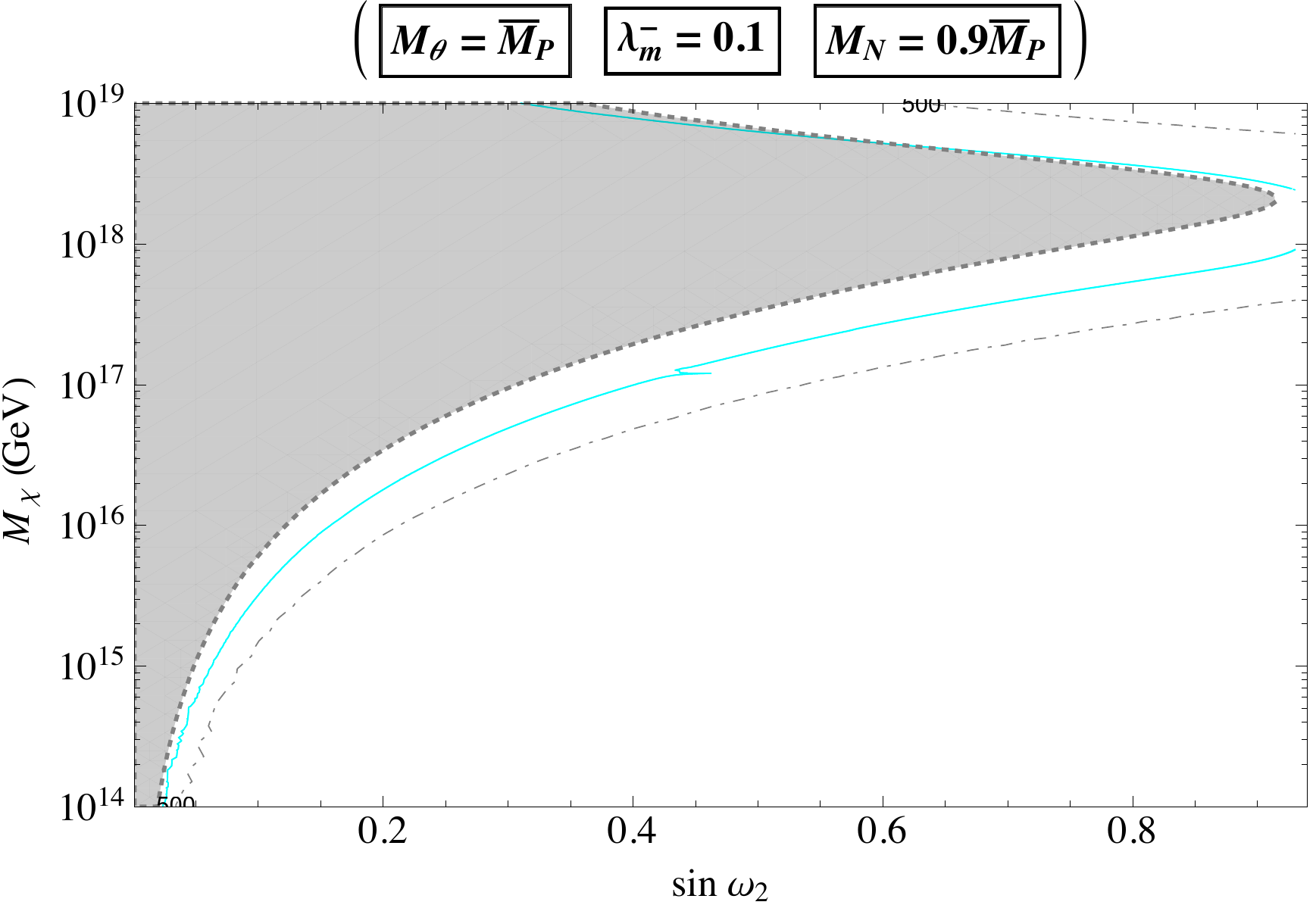}
\includegraphics[width=.49\textwidth]{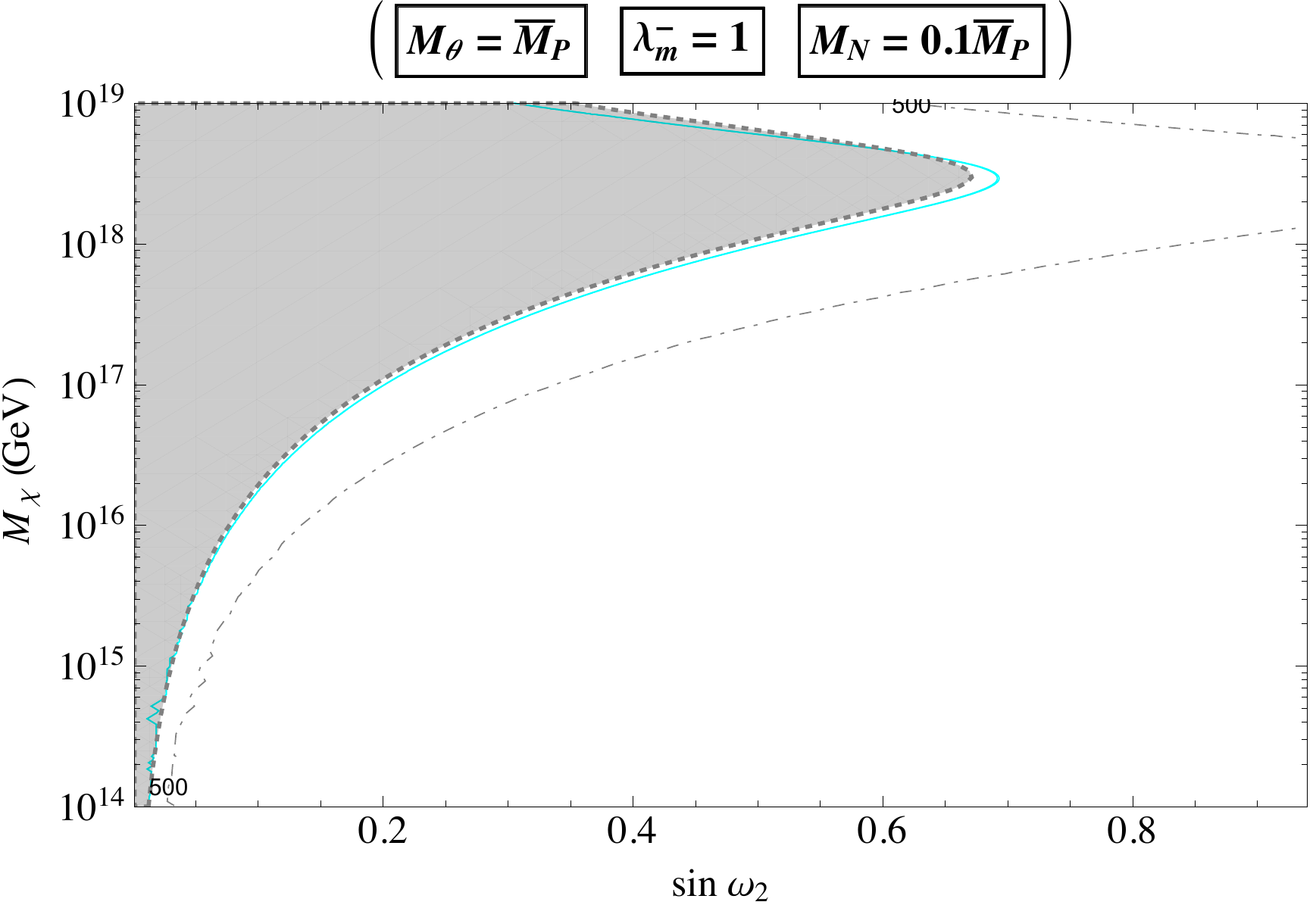}
\includegraphics[width=.49\textwidth]{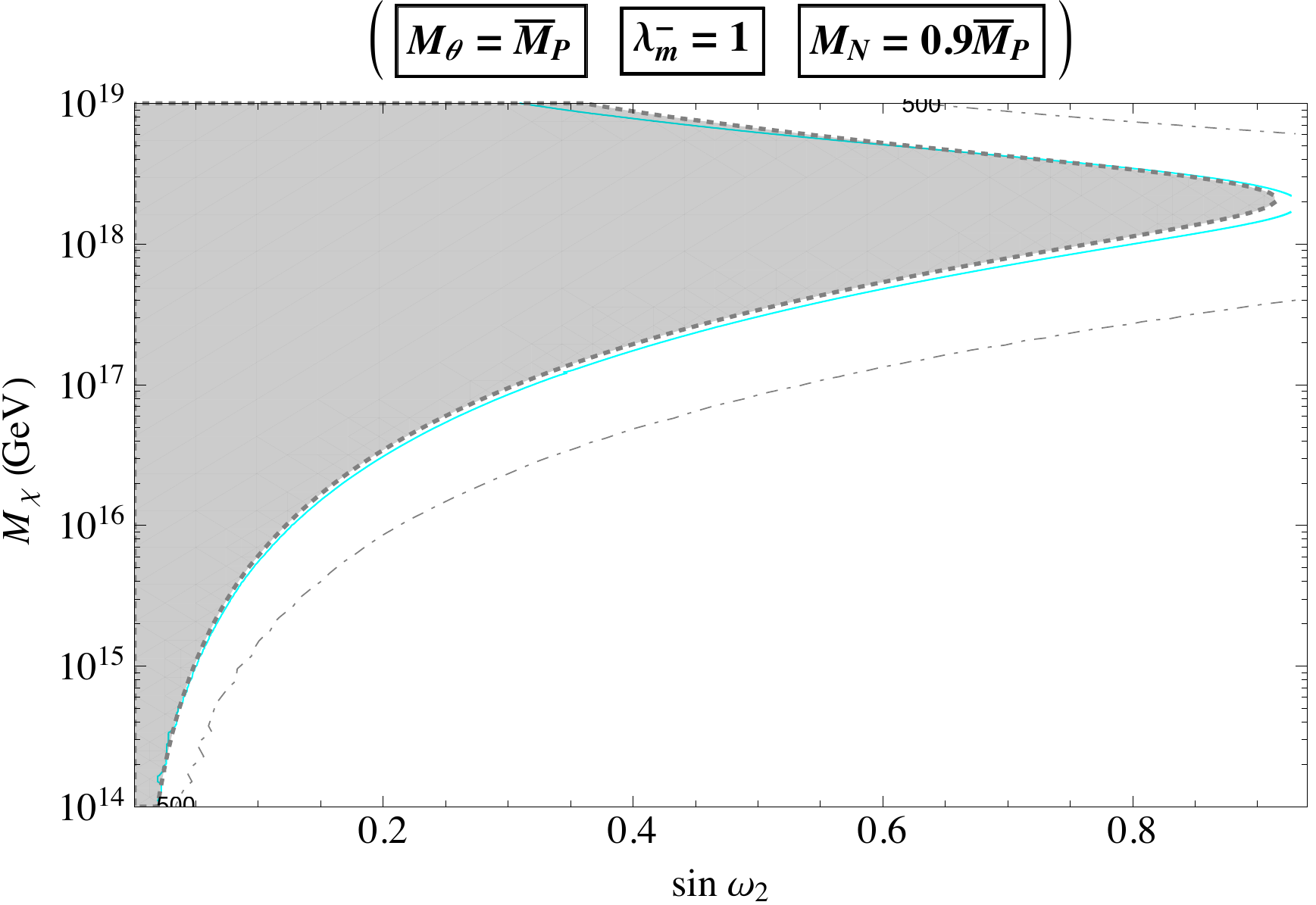}
\caption{Dark matter relic abundance curve (solid blue) as reported by the Planck Collaboration \cite{Ade:2015lrj}, with a $1\sigma$~uncertainty thickness, in the $\sin\omega_{2} - M_{\chi}$~plane, for $M_{\theta} = \bar{M}_\text{P}$, and several (small and large) values of the free parameters $\lambda_{m}^{-}$ and $M_{\mathcal N}$. The gray-shaded region (dotted line) corresponds to $M_{\chi} / \Treh > 2000$, and is considered to be ruled out within the WIMPzilla paradigm. The dot-dashed line illustrates $M_{\chi} / \Treh = 500$, signifying the fact that the non-relativistic approximation remains valid for the WIMPzilla relic abundance curve, which always remain far above this value. Universal values of the remaining free parameters $\xi_{H} = \xi_{\chi} =1$ and $M_{\kappa} = 0.1 \bar{M}_\text{P}$ are selected for illustration (the ``kink'' of the relic abundance curve near $M_{\chi} \sim 10^{17}$~GeV is attributed to the $\kappa$~threshold). While (slightly) larger values of the $\kappa$~mass pushes the relic abundance curve into the excluded gray region, a dependence on the exact values of these three parameters within the allowed region is negligible.}
\label{relicplots1}
\end{figure}

\begin{figure}
\includegraphics[width=.49\textwidth]{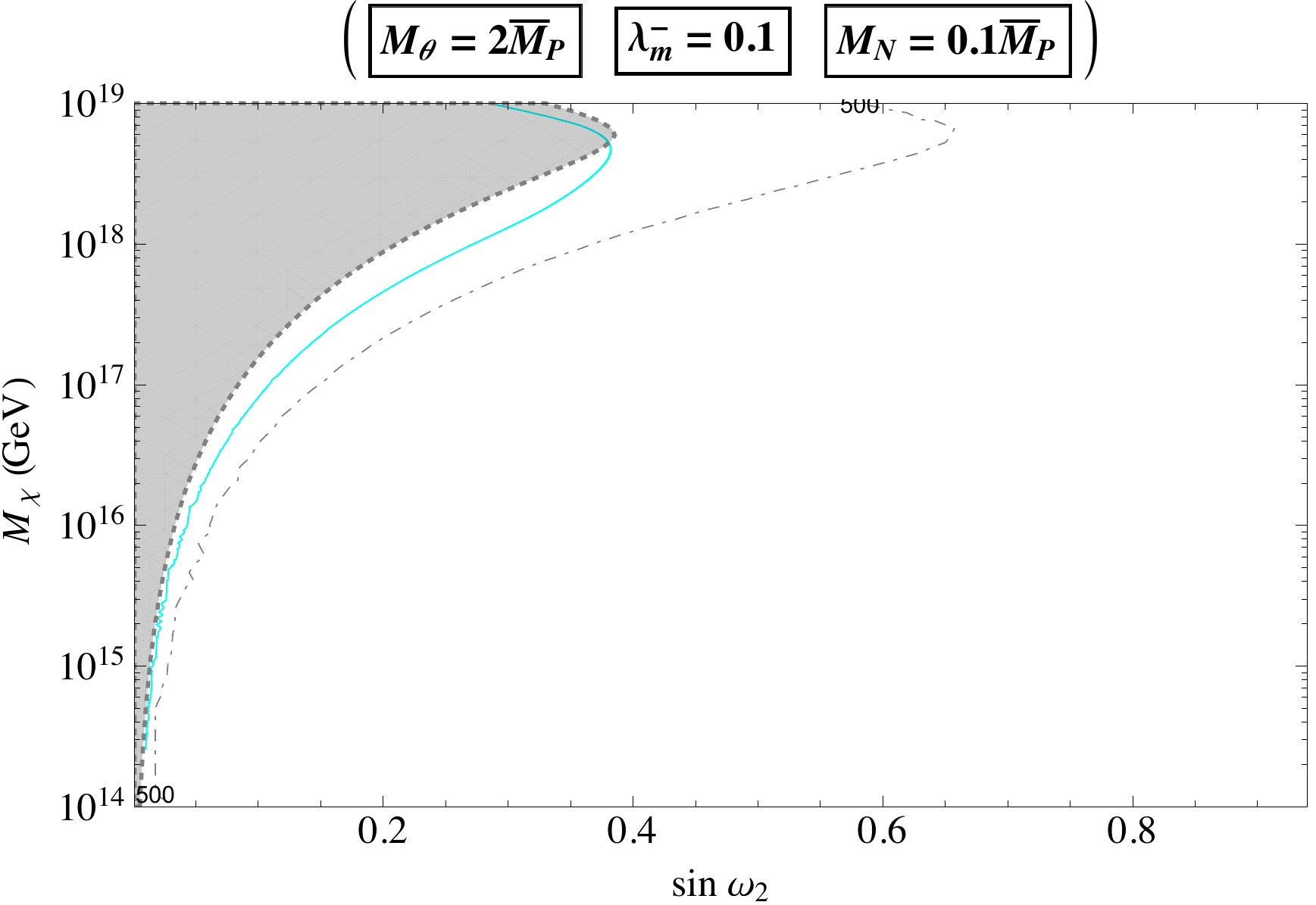}
\includegraphics[width=.49\textwidth]{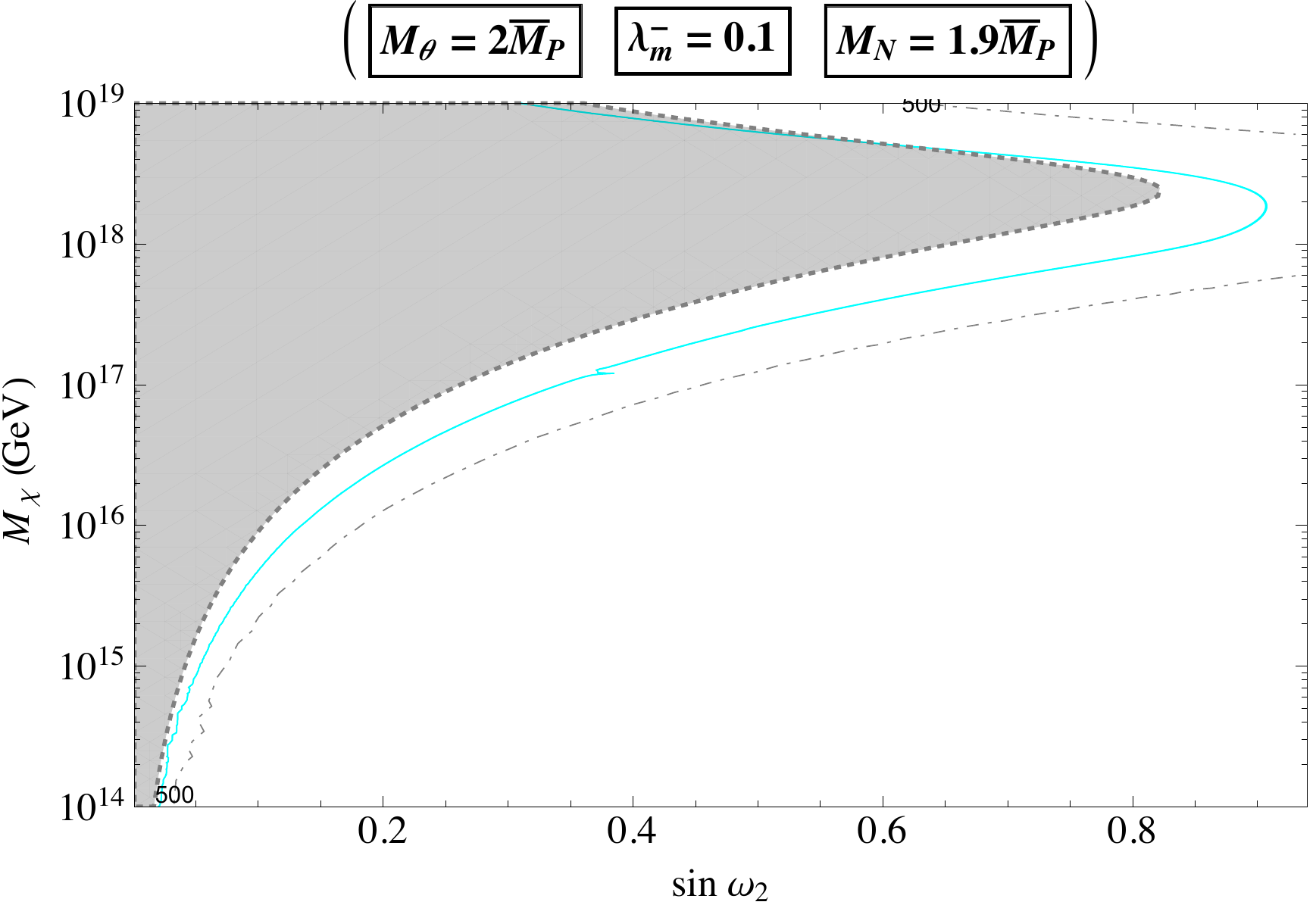}
\includegraphics[width=.49\textwidth]{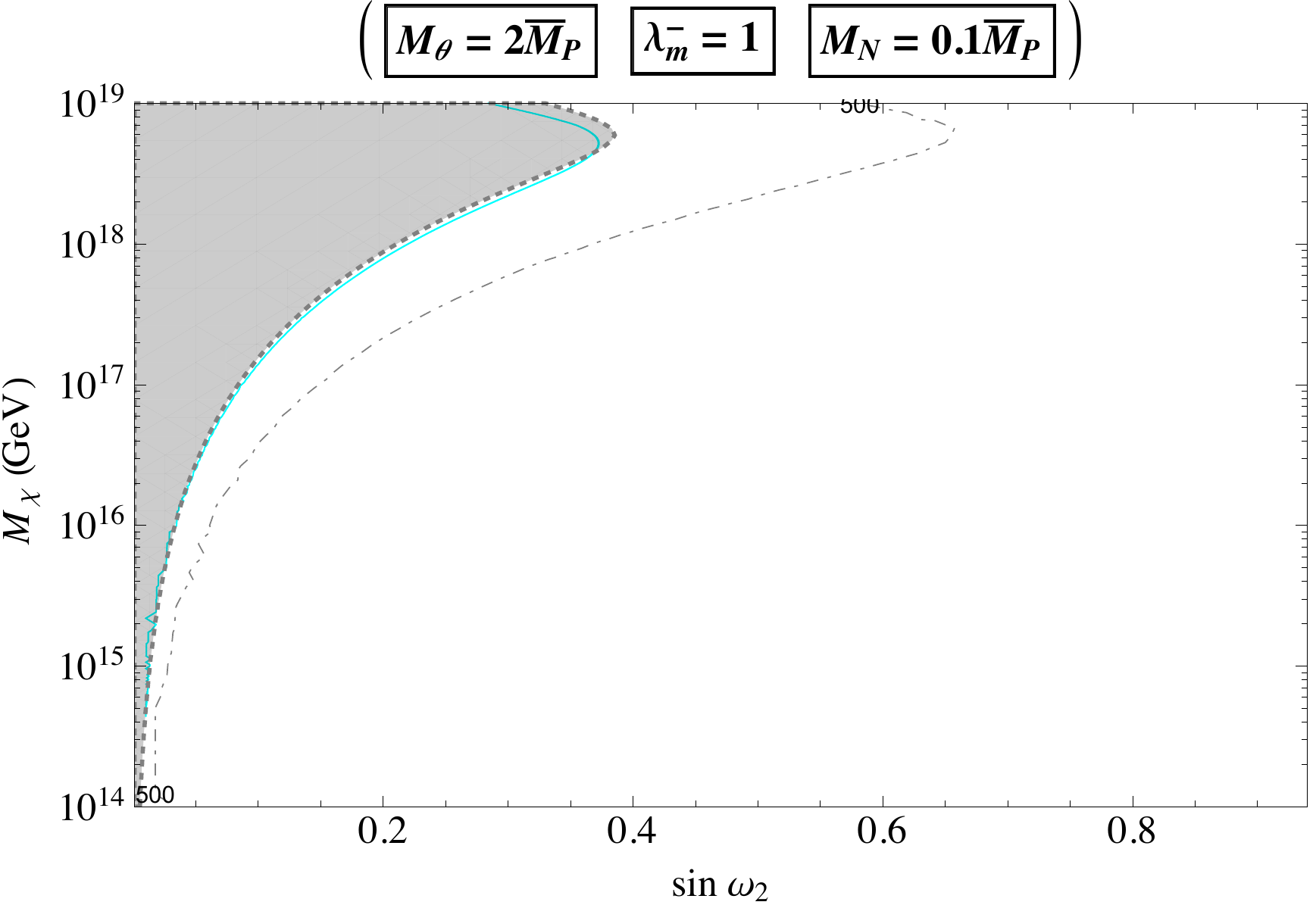}
\includegraphics[width=.49\textwidth]{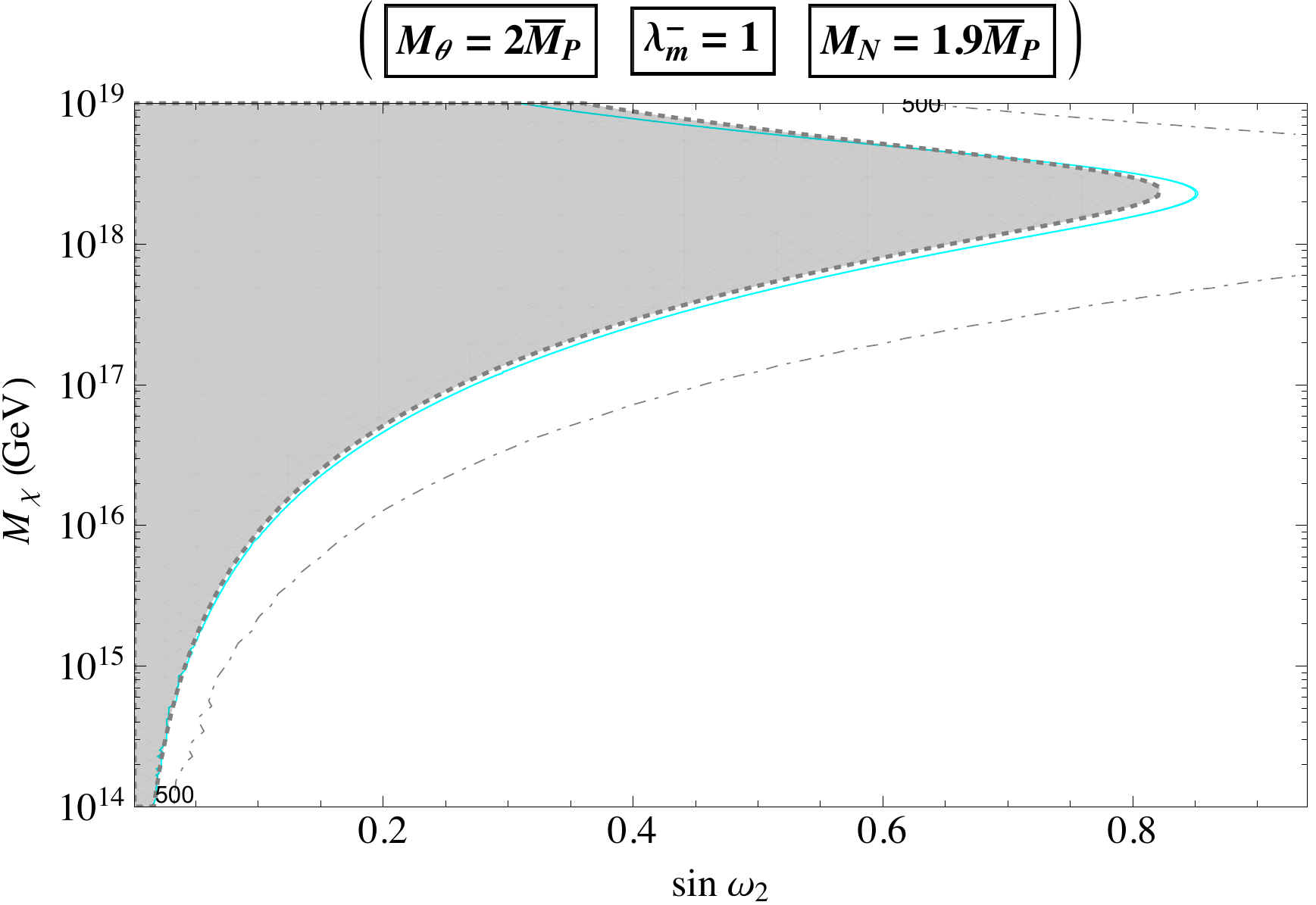}
\caption{Dark matter relic abundance curve (solid blue) as reported by the Planck Collaboration \cite{Ade:2015lrj}, with a $1\sigma$~uncertainty thickness, in the $\sin\omega_{2} - M_{\chi}$~plane, for $M_{\theta} = 2\bar{M}_\text{P}$, and several (small and large) values of the free parameters $\lambda_{m}^{-}$ and $M_{\mathcal N}$. Universal values of the remaining free parameters $\xi_{H} = \xi_{\chi} =1$ and $M_{\kappa} = 0.1 \bar{M}_\text{P}$ are selected for illustration. (see the caption of Fig.~\ref{relicplots1} for details)}
\label{relicplots2}
\end{figure}

The regions of the parameter space where the observed value of the relic abundance can be accommodated within the current framework are demonstrated in Figs.~\ref{relicplots1}~and~\ref{relicplots2}, in the $\sin\omega_{2} - M_{\chi}$~plane, for $M_{\theta} = \bar{M}_\text{P}$ and $M_{\theta} = 2\bar{M}_\text{P}$, respectively. The panels further illustrate the dependence on the small and large benchmark values of the input parameters $\lambda_{m}^{-}$ and $M_{\mathcal N}$. A dependence on the remaining formally relevant free parameters (i.e., $M_{\kappa}, \xi_{H}, \xi_{\chi}$) is negligible within the allowed region, since the $W^{\pm},Z$ and $h$~pairs form the dominant channels (Fig.~\ref{DMann}). The viable region is confined by requiring the WIMPzilla dark matter mass not to exceed the $2000\, \Treh$~boundary \cite{Chung:1998rq}. One observes that a larger $\lambda_{m}^{-}$~coupling, a heavier $\theta$~LW~graviton, and a lighter right-handed Majorana neutrino favor larger WIMPzilla masses and/or smaller values of the mixing angle. In addition, the figures demonstrate that the non-relativistic approximation remains valid throughout, for the WIMPzilla masses compatible with the abundance curve.

\section{Discussion}\label{Disc}

\begin{figure}
\includegraphics[width=.49\textwidth]{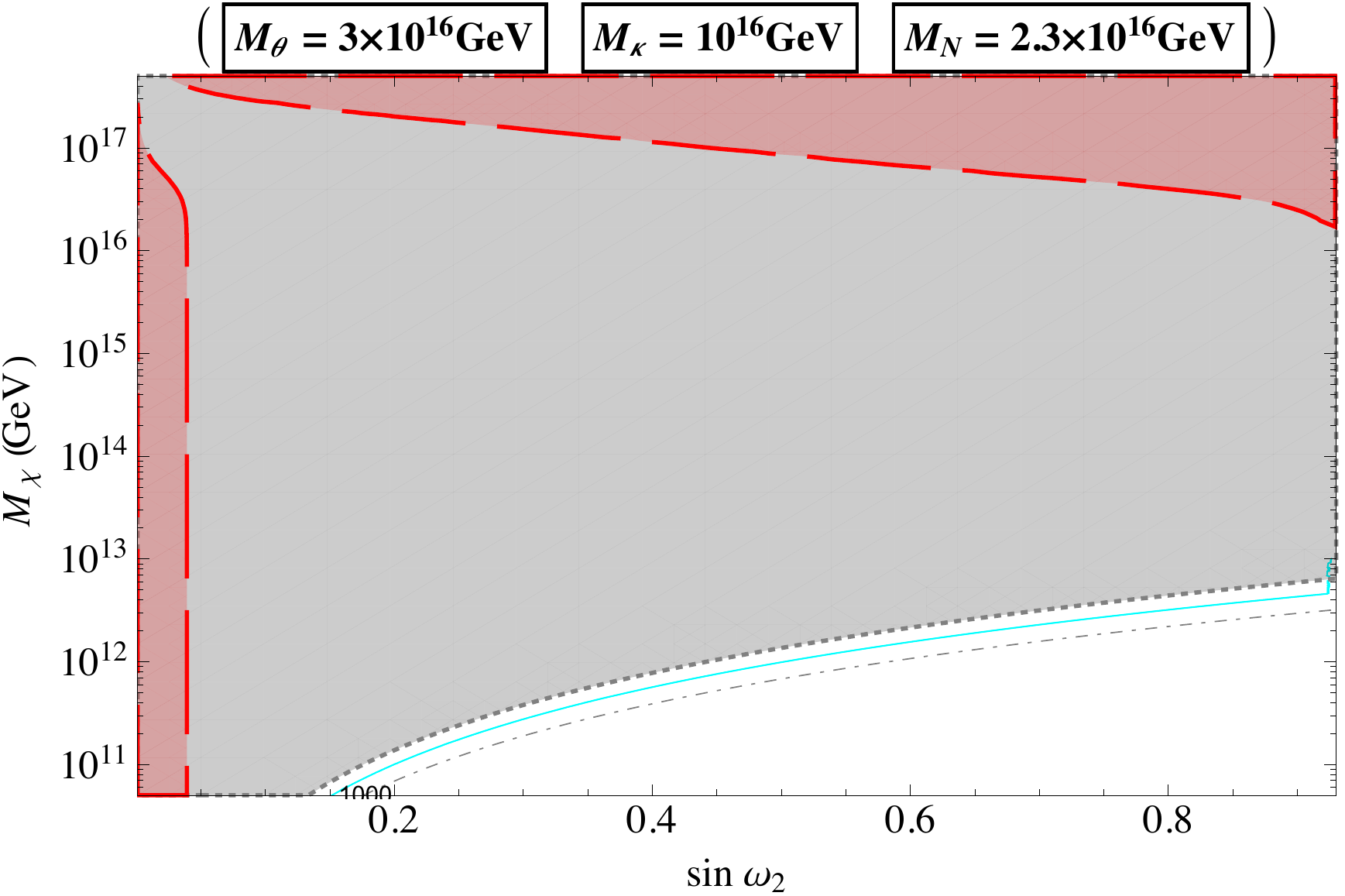}
\includegraphics[width=.49\textwidth]{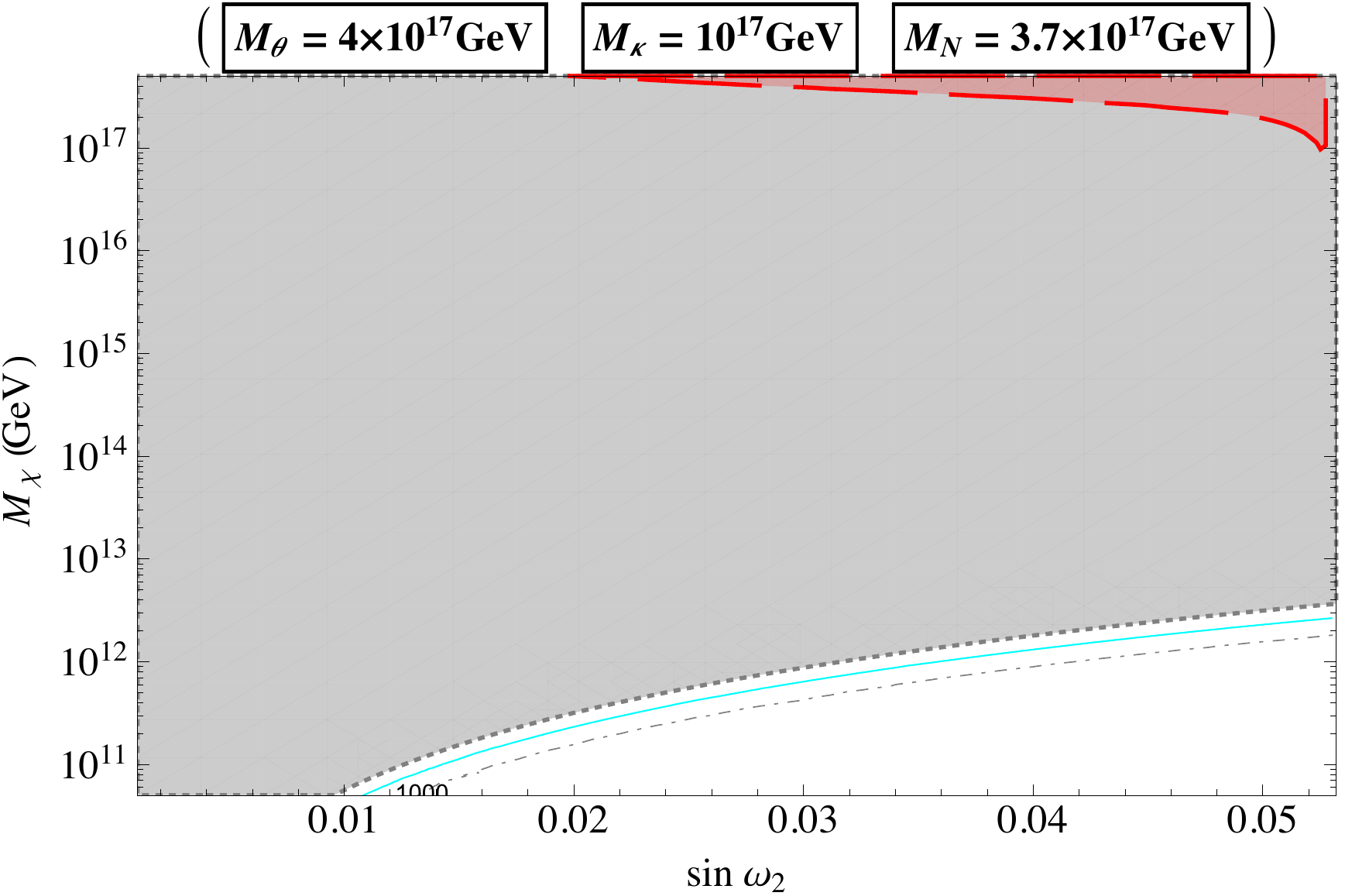}
\caption{Examples of the viable regions of the parameter space in the $\sin\omega_{2} - M_{\chi}$~plane for smaller (left) and larger (right) benchmark values of the remaining mass parameters. Universal values of $\lambda_{m}^{-} = 0.5$ and $\lambda_{\chi} = \xi_{H} = \xi_{\chi} =1$ are selected for illustration. The entire displayed regions are compatible with the perturbativity and vacuum stability requirements, as elaborated in Sec.~\ref{VST}. Inflationary constraints from the Planck TT+LowP at 95\%~C.L. (dashed red line) result in an upper bound on the dark matter mass, as well as lower and upper limits on the mixing angle (depicted range). The WIMPzilla relic abundance curve (solid blue), with a $1\sigma$~uncertainty thickness, resides in between $M_{\chi} / \Treh = 1000$ (dot-dashed line) and $M_{\chi} / \Treh = 2000$ (dotted line). $M_{\chi} / \Treh$ values exceeding $\sim2000$ are excluded by demanding the dark matter to constitute a WIMPzilla candidate (gray region). All colored regions are excluded.}
\label{comboplots}
\end{figure}

Finally, let us combine the findings from the previous sections, in order to investigate the viability of the model in accordance with the perturbativity and vacuum stability, inflation, and the dark matter constraints. Examples of the viable $\sin\omega_{2} - M_{\chi}$~regions, compatible with all aforementioned requirements, are demonstrated in Fig.~\ref{comboplots}, for benchmark values of the remaining input parameters. For each benchmark plot, the inflationary constraints from the Planck TT+LowP at 95\%~C.L. are satisfied within a certain range of the mixing angle values,\footnote{See Fig.~\ref{fig:cnt1} for comparison. Note, however, that in that figure the colored regions correspond to the \textit{viable} parameter space, whereas in Fig.~\ref{comboplots} all colored regions are \textit{excluded}.} in addition to imposing an upper limit on the dark matter mass. Specifically, heavier masses of the remaining bosonic and fermionic degrees of freedom, $\theta, \kappa, \mathcal N$, significantly reduce the range of the allowed mixing angles.

Furthermore, one observes that the WIMPzilla mass, compatible with the dark matter relic abundance, is confined below $\sim 10^{13}$~GeV. This (rather general) result is a direct consequence of the constraints imposed by the inflation, combined with the WIMPzilla nature of the dark matter. In particular, the small amplitude of the primordial scalar perturbations, $A_{s} \sim 10^{-9}$, implies an accordingly small amplitude for the inflaton potential \eqref{V01}, $\mathcal M \lesssim 0.1 \bar{M}_\text{P}$, within vast regions of the parameter space (c.f. Fig.~\ref{fig:cnt1}). This, in turn, introduces a relatively light inflaton \eqref{msigfin}, with a mass several orders of magnitude below the reduced Planck mass, which results in moderate reheating temperatures \eqref{Treh}. Within the WIMPzilla paradigm, the dark matter mass satisfies the condition $M_{\chi} \lesssim 2000\, \Treh$, and hence may not be too heavy for moderate reheating temperatures. In addition, small values of $\mathcal M$, as favored by the inflationary constraints, require the masses of the bosonic and fermionic degrees of freedom to reside in the relative vicinity of one another (c.f. \eqref{Bmodel}); in particular, the masses of the LW~graviton and the right-handed Majorana neutrinos play the dominant role in this respect, given their large internal degree of freedom coefficients.

In summary, one concludes that within a classically scale invariant framework in which the pseudo-Nambu-Goldstone boson of the (approximate) scale symmetry is identified with the inflaton,\footnote{Alternatively, one may consider the $\kappa$ (or the $h$) scalar as the inflaton of the framework, with a non-vanishing tree-level potential away from the flat direction. Such analysis, although interesting, is outside the scope of the current treatment.} the WIMPzillas satisfying the observed relic abundance are generically much lighter than the reduced Planck scale,\footnote{WIMPzilla masses much lighter than the reduced Planck scale may require (unnaturally) small values of the $\lambda_{\eta\chi}$~quartic coupling within certain regions of the parameter space (c.f.~\eqref{couplings}).} due to the observational constraints imposed by the inflation, and in particular the small amplitude of the primordial scalar perturbations.

\section{Conclusion}\label{Concl}

In this treatment, we have introduced a unified framework which simultaneously tackles several open problems of particle physics and cosmology consistently; namely, the hierarchy problem, neutrino masses, dark matter, and inflation. The model concerns the minimal addition of one complex gauge singlet to the SM content, in a scale- and $CP$-symmetric manner. By embedding the model within a scale symmetric and renormalizable theory of gravity---the Agravity---the dynamically-generated nonzero VEV of the $CP$-even singlet scalar (utilizing the Coleman-Weinberg mechanism) may induce the Planck scale (via the scalar non-minimal couplings), as well as the weak scale (via the Higgs portal couplings). The classical scale symmetry guarantees the absence of a quadratic sensitivity between the scales, whereas the $CP$-symmetry prevents the pseudoscalar singlet from decaying, rendering it a dark matter candidate. In addition, nonzero masses for the SM neutrinos are induced by the scale-symmetric see-saw mechanism, which includes three flavors of the right-handed Majorana neutrinos. We have, thus, captured and addressed several important open issues, faced by the contemporary physics, within a single consistent framework.

We have demonstrated that the consistency of the introduced framework can be verified up to substantial trans-Planckian energies ($\mu > 250 \bar{M}_\text{P}$), where the scalar potential is stable and the couplings remain perturbative. Identifying the pseudo-Nambu-Goldstone boson of the (approximate) scale symmetry with the inflaton field, we exhibited the viability of the slow-roll inflationary paradigm within the framework's parameter space, in accordance with the latest observational values, as well as the reheating of the Universe due to the inflaton decay. Within the introduced framework, the dark matter candidate can assume the role of the WIMPzilla, whose relic abundance is compatible with the observations within vast regions of the parameter space.

We have identified the viable regions of the parameter space, which simultaneously accommodate all the aforementioned requirements, and demonstrated the results in various benchmark exclusion plots. In particular, we have reached the important (and rather general) conclusion that within a classically scale invariant framework in which the pseudo-Nambu-Goldstone boson of the (approximate) scale symmetry is identified with the inflaton, the masses of the inflaton and the WIMPzilla, as well as the resulting reheating temperature are (much) smaller than the reduced Planck scale, while satisfying the inflationary and the relic abundance observational values.

\section*{Acknowledgment}

We are grateful to Ki-Young Choi for valuable discussions and input in the course of the completion of this study. We thank Hyun Min Lee and Raphael Flauger for interesting comments. A.F. also thanks Alessandro Strumia for useful correspondence, and the APCTP Focus Research Program for its hospitality while parts of this work were completed. The work of A.F. was supported by the IBS under the project code IBS-R018-D1.


\begin{thebibliography}{99}

\bibitem{Bardeen}
W.\ A.\ Bardeen, ``On Naturalness in the Standard Model'',
FERMILAB-CONF-95-391-T, ``Beyond Higgs'', FERMILAB-CONF-08-118- T;
H.~Aoki and S.~Iso,
  Phys.\ Rev.\ D {\bf 86}, 013001 (2012)
  [arXiv:1201.0857 [hep-ph]].
  

\bibitem{Farzinnia:2013pga} 
  A.~Farzinnia, H.~J.~He and J.~Ren,
  Phys.\ Lett.\ B {\bf 727}, 141 (2013)
  [arXiv:1308.0295 [hep-ph]].
    
  \bibitem{Coleman:1973jx}
S.~R.~Coleman and E.~J.~Weinberg,
Phys.\ Rev.\ D {\bf 7} (1973) 1888.

 \bibitem{LHCnew}
G.~Aad {\it et al.,}  [ATLAS Collaboration],
Phys.\ Lett.\ B\,{716} (2012) 1 [arXiv:1207.7214 [hep-ex]];
S.~Chatrchyan {\it et al.,}  [CMS Collaboration],
Phys.\ Lett.\ B\,{716} (2012) 30 [arXiv:1207.7235 [hep-ex]].

   \bibitem{seesaw}
P.~Minkowski, Phys. Lett. B {\bf 67} (1977) 421;
T.~Yanagida, in \emph{Proceedings of the Workshop on the Unified
Theory and the Baryon Number in the Universe} (O.~Sawada and
A.~Sugamoto, eds.), KEK, Tsukuba, Japan, 1979, p.\,95;
M.~Gell-Mann, P.~Ramond, and R.~Slansky, \emph{Supergravity} (P.~van
Nieuwenhuizen et al., eds), North Holland, Amsterdam, 1979, p.\,315;
R.~N. Mohapatra and G.~Senjanovi{\'c}, Phys.\ Rev.\ Lett.\ {\bf 44} (1980) 912;
J.~Schechter and J.~W.~F.~Valle,
  Phys.\ Rev.\ D {\bf 22}, 2227 (1980).
  
\bibitem{Farzinnia:2014xia} 
  A.~Farzinnia and J.~Ren,
  Phys.\ Rev.\ D {\bf 90}, no. 1, 015019 (2014)
  [arXiv:1405.0498 [hep-ph]].
  
\bibitem{Farzinnia:2015uma} 
  A.~Farzinnia,
  Phys.\ Rev.\ D {\bf 92}, no. 9, 095012 (2015)
  [arXiv:1507.06926 [hep-ph]].
  
\bibitem{Farzinnia:2014yqa} 
  A.~Farzinnia and J.~Ren,
  Phys.\ Rev.\ D {\bf 90}, no. 7, 075012 (2014)
  [arXiv:1408.3533 [hep-ph]].
  
\bibitem{Salvio:2014soa} 
  A.~Salvio and A.~Strumia,
  JHEP {\bf 1406}, 080 (2014)
  [arXiv:1403.4226 [hep-ph]].
  
\bibitem{Ade:2015lrj}
  P.~A.~R.~Ade {\it et al.} [Planck Collaboration],
  arXiv:1502.02114 [astro-ph.CO].
  
\bibitem{SIinfl}
A.~D.~Linde,
  Phys.\ Lett.\ B {\bf 108}, 389 (1982);
   A.~Albrecht and P.~J.~Steinhardt,
  Phys.\ Rev.\ Lett.\  {\bf 48}, 1220 (1982);
  M.~U.~Rehman, Q.~Shafi and J.~R.~Wickman,
  Phys.\ Rev.\ D {\bf 78}, 123516 (2008)
  [arXiv:0810.3625 [hep-ph]];
  A.~Cerioni, F.~Finelli, A.~Tronconi and G.~Venturi,
  Phys.\ Lett.\ B {\bf 681}, 383 (2009)
  [arXiv:0906.1902 [astro-ph.CO]];
   V.~V.~Khoze,
  JHEP {\bf 1311}, 215 (2013)
  [arXiv:1308.6338 [hep-ph]];
    G.~Barenboim, E.~J.~Chun and H.~M.~Lee,
  Phys.\ Lett.\ B {\bf 730} (2014) 81
  [arXiv:1309.1695 [hep-ph]];
  J.~Joergensen, F.~Sannino and O.~Svendsen,
  Phys.\ Rev.\ D {\bf 90}, no. 4, 043509 (2014)
  [arXiv:1403.3289 [hep-ph]];
   K.~Kannike, A.~Racioppi and M.~Raidal,
  JHEP {\bf 1406}, 154 (2014)
  [arXiv:1405.3987 [hep-ph]];
  G.~Ballesteros and J.~A.~Casas,
  Phys.\ Rev.\ D {\bf 91}, 043502 (2015)
  [arXiv:1406.3342 [astro-ph.CO]];
   C.~Csaki, N.~Kaloper, J.~Serra and J.~Terning,
  Phys.\ Rev.\ Lett.\  {\bf 113}, 161302 (2014)
  [arXiv:1406.5192 [hep-th]];
   T.~Inagaki, R.~Nakanishi and S.~D.~Odintsov,
  Astrophys.\ Space Sci.\  {\bf 354}, no. 2, 2108 (2014)
  [arXiv:1408.1270 [gr-qc]];
   S.~Iso, K.~Kohri and K.~Shimada,
  Phys.\ Rev.\ D {\bf 91} (2015) 4,  044006
  [arXiv:1408.2339 [hep-ph]];
   G.~Barenboim and W.~I.~Park,
  Phys.\ Rev.\ D {\bf 91}, no. 6, 063511 (2015)
  [arXiv:1501.00484 [hep-ph]]
  K.~Kawana,
  PTEP {\bf 2015}, 073B04 (2015)
  [arXiv:1501.04482 [hep-ph]];
  D.~Croon, V.~Sanz and J.~Setford,
  JHEP {\bf 1510}, 020 (2015)
  [arXiv:1503.08097 [hep-ph]];
  K.~Kannike, A.~Racioppi and M.~Raidal,
  arXiv:1509.05423 [hep-ph].
  
\bibitem{Kannike:2015apa} 
  K.~Kannike, G.~H\"utsi, L.~Pizza, A.~Racioppi, M.~Raidal, A.~Salvio and A.~Strumia,
  JHEP {\bf 1505}, 065 (2015)
  [arXiv:1502.01334 [astro-ph.CO]].
     
\bibitem{Chung:1998rq} 
  D.~J.~H.~Chung, E.~W.~Kolb and A.~Riotto,
  Phys.\ Rev.\ D {\bf 60}, 063504 (1999)
  [hep-ph/9809453].

\bibitem{AlexanderNunneley:2010nw} 
  L.~Alexander-Nunneley and A.~Pilaftsis,
  JHEP {\bf 1009}, 021 (2010)
  [arXiv:1006.5916 [hep-ph]].

\bibitem{LW} 
  T.~D.~Lee and G.~C.~Wick,
  Nucl.\ Phys.\ B {\bf 9}, 209 (1969);
  Phys.\ Rev.\ D {\bf 2}, 1033 (1970).
  
  \bibitem{acausal} 
  R.~E.~Cutkosky, P.~V.~Landshoff, D.~I.~Olive and J.~C.~Polkinghorne,
  Nucl.\ Phys.\ B {\bf 12}, 281 (1969);
  B.~Grinstein, D.~O'Connell and M.~B.~Wise,
  Phys.\ Rev.\ D {\bf 79}, 105019 (2009)
  [arXiv:0805.2156 [hep-th]].
  
\bibitem{Chivukula:2010kx} 
  R.~S.~Chivukula, A.~Farzinnia, R.~Foadi and E.~H.~Simmons,
  Phys.\ Rev.\ D {\bf 82}, 035015 (2010)
  [arXiv:1006.2800 [hep-ph]].
  
\bibitem{Salvio:2015gsi} 
  A.~Salvio and A.~Strumia,
  arXiv:1512.01237 [hep-th].
  
\bibitem{Kaiser:2010ps} 
  D.~I.~Kaiser,
  Phys.\ Rev.\ D {\bf 81}, 084044 (2010)
  [arXiv:1003.1159 [gr-qc]].
  
\bibitem{Gildener:1976ih}
E.~Gildener and S.~Weinberg,
Phys.\ Rev.\ D {\bf 13} (1976) 3333.

\bibitem{Karam:2015jta} 
  A.~Karam and K.~Tamvakis,
  Phys.\ Rev.\ D {\bf 92}, no. 7, 075010 (2015)
  [arXiv:1508.03031 [hep-ph]].

\bibitem{GenBeta1} 
  M.~E.~Machacek and M.~T.~Vaughn,
  Nucl.\ Phys.\ B {\bf 222}, 83 (1983);
  Nucl.\ Phys.\ B {\bf 236}, 221 (1984);
  Nucl.\ Phys.\ B {\bf 249}, 70 (1985);
  M.~x.~Luo, H.~w.~Wang and Y.~Xiao,
  Phys.\ Rev.\ D {\bf 67}, 065019 (2003)
  [hep-ph/0211440].
  
\bibitem{GenBeta2} 
D.~J.~Gross and F.~Wilczek,
  Phys.\ Rev.\ D {\bf 8}, 3633 (1973);
  H.~D.~Politzer and G.~G.~Ross,
  Nucl.\ Phys.\ B {\bf 75}, 269 (1974);
  I.~Jack and H.~Osborn,
  J.\ Phys.\ A {\bf 16}, 1101 (1983).
  
  \bibitem{SMvac} 
 G.~Degrassi, S.~Di Vita, J.~Elias-Miro, J.~R.~Espinosa, G.~F.~Giudice, G.~Isidori and A.~Strumia,
  JHEP {\bf 1208}, 098 (2012)
  [arXiv:1205.6497 [hep-ph]];
  D.~Buttazzo, G.~Degrassi, P.~P.~Giardino, G.~F.~Giudice, F.~Sala, A.~Salvio and A.~Strumia,
  JHEP {\bf 1312}, 089 (2013)
  [arXiv:1307.3536 [hep-ph]].
  
  \bibitem{StabPlanck} 
  V.~Branchina and E.~Messina,
  Phys.\ Rev.\ Lett.\  {\bf 111}, 241801 (2013)
  [arXiv:1307.5193 [hep-ph]];
  V.~Branchina, E.~Messina and A.~Platania,
  JHEP {\bf 1409}, 182 (2014)
  [arXiv:1407.4112 [hep-ph]];
  V.~Branchina, E.~Messina and M.~Sher,
  Phys.\ Rev.\ D {\bf 91}, 013003 (2015)
  [arXiv:1408.5302 [hep-ph]];
  V.~Branchina and E.~Messina,
  arXiv:1507.08812 [hep-ph].
    
\bibitem{Giudice:2014tma} 
  G.~F.~Giudice, G.~Isidori, A.~Salvio and A.~Strumia,
  JHEP {\bf 1502}, 137 (2015)
  [arXiv:1412.2769 [hep-ph]].
  
\bibitem{Schrempp:1996fb} 
  See e.g., B.~Schrempp and M.~Wimmer,
  Prog.\ Part.\ Nucl.\ Phys.\  {\bf 37}, 1 (1996)
  [hep-ph/9606386].
  
  \bibitem{Buchbinder:1989jd} 
  I.~L.~Buchbinder, O.~K.~Kalashnikov, I.~L.~Shapiro, V.~B.~Vologodsky and J.~J.~Wolfengaut,
  Phys.\ Lett.\ B {\bf 216}, 127 (1989);
  I.~L.~Buchbinder, S.~D.~Odintsov and I.~L.~Shapiro,
  Bristol, UK: IOP (1992) 413 p.
		      
  \bibitem{MCMC}
   N. Metropolis, A.W. Rosenbluth, M.N. Rosenbluth, A.H. Teller, and E. Teller, 
   J.\ Chem.\ Phys.\ {\bf 21}, 1087 (1953);
   W.K. Hastings, 
   Biometrika {\bf 57}, 97 (1970).
   
\bibitem{Abrahamse-etal-2008}
  A.~Abrahamse, A.~Albrecht, M.~Barnard and B.~Bozek,
  Phys.\ Rev.\ D {\bf 77}, 103503 (2008)
  [arXiv:0712.2879 [astro-ph]].
  
  \bibitem{DMrelic} 
  M.~Srednicki, R.~Watkins and K.~A.~Olive,
  Nucl.\ Phys.\ B {\bf 310}, 693 (1988);
P.~Gondolo and G.~Gelmini,
  Nucl.\ Phys.\ B {\bf 360}, 145 (1991).
  
  \bibitem{Kolb:book}
  E.~W.~Kolb and M.~S.~Turner,\emph{ The Early Universe}, Frontiers in physics (Westview Press, New York, 1994), ISBN 9780813346458.

  
   


\end{thebibliography}
\end{document}